\documentclass[a4paper,12pt, epsfig]{article}
\usepackage{epsfig}
\usepackage{amssymb}
\usepackage{amsfonts}
\usepackage{amsmath}
\usepackage{euscript}
\usepackage{verbatim}
\usepackage{latexsym}
\usepackage{graphicx}

\newskip\humongous \humongous=0pt plus 1000pt minus 1000pt

\newif\ifdtup

\catcode`\@=11

\@addtoreset{equation}{section}

\def\@normalsize{\@setsize\normalsize{15pt}\xiipt\@xiipt
\abovedisplayskip 14pt plus3pt minus3pt%
\belowdisplayskip \abovedisplayskip
\abovedisplayshortskip \z@ plus3pt%
\belowdisplayshortskip 7pt plus3.5pt minus0pt}

\def\small{\@setsize\small{13.6pt}\xipt\@xipt
\abovedisplayskip 13pt plus3pt minus3pt%
\belowdisplayskip \abovedisplayskip
\abovedisplayshortskip \z@ plus3pt%
\belowdisplayshortskip 7pt plus3.5pt minus0pt
\def\@listi{\parsep 4.5pt plus 2pt minus 1pt
     \itemsep \parsep
     \topsep 9pt plus 3pt minus 3pt}}

\relax

\catcode`@=12

\catcode`\@=11

\def\section{\@startsection{section}{1}{\z@}{3.5ex plus 1ex minus
   .2ex}{2.3ex plus .2ex}{\large\bf}}

\def\SymBoxes#1#2#3#4{\newdimen\un@t \un@t#3%
\raisebox{#1}{\rule{#2\un@t}{#4}\hskip-#2\un@t
\@tempdimb\un@t \advance\@tempdimb by-#4\@tempcntb#2\relax%
\@whilenum{\@tempcntb>0}\do{
\rule{#4}{\un@t}\hskip\@tempdimb \advance\@tempcntb by\m@ne}%
\hskip-#2\un@t \rule[\un@t]{#2\un@t}{#4}%
\rule[\un@t]{#4}{#4}\hskip-#4
\rule{#4}{\un@t}}\hskip-#4}                

\begin{document}

\newcommand{\beq}{\begin{equation}}
\newcommand{\eeq}{\end{equation}}
\newcommand{\bea}{\begin{eqnarray}}
\newcommand{\eea}{\end{eqnarray}}
\newcommand{\beas}{\begin{eqnarray*}}
\newcommand{\eeas}{\end{eqnarray*}}
\newcommand{\defi}{\stackrel{\rm def}{=}}
\newcommand{\non}{\nonumber}
\newcommand{\nn}{\nonumber}
\newcommand{\bquo}{\begin{quote}}
\newcommand{\enqu}{\end{quote}}
\renewcommand{\(}{\begin{equation}}
\renewcommand{\)}{\end{equation}}
\def \eqn#1#2{\begin{equation}#2\label{#1}\end{equation}}
\def\IZ{{\mathbb Z}}
\def\IR{{\mathbb R}}
\def\IC{{\mathbb C}}
\def\IQ{{\mathbb Q}}
\def\de{\partial}
\def\Tr{ \hbox{\rm Tr}}
\def\H{ \hbox{\rm H}}
\def\HE{ \hbox{$\rm H^{even}$}}
\def\HO{ \hbox{$\rm H^{odd}$}}
\def\K{ \hbox{\rm K}}
\def\Im{ \hbox{\rm Im}}
\def\Ker{ \hbox{\rm Ker}}
\def\const{\hbox {\rm const.}}
\def\o{\over}
\def\im{\hbox{\rm Im}}
\def\re{\hbox{\rm Re}}
\def\bra{\langle}\def\ket{\rangle}
\def\Arg{\hbox {\rm Arg}}
\def\Re{\hbox {\rm Re}}
\def\Im{\hbox {\rm Im}}
\def\exo{\hbox {\rm exp}}
\def\diag{\hbox{\rm diag}}
\def\longvert{{\rule[-2mm]{0.1mm}{7mm}}\,}
\def\a{\alpha}
\def\dag{{}^{\dagger}}
\def\tq{{\widetilde q}}
\def\p{{}^{\prime}}
\def\W{W}
\def\N{{\cal N}}
\def\cH{{\cal H}}
\def\cI{{\cal I}}
\def\hsp{,\hspace{.7cm}}
\newcommand{\C}{\ensuremath{\mathbb C}}
\newcommand{\I}{\ensuremath{\mathbb I}}
\newcommand{\Z}{\ensuremath{\mathbb Z}}
\newcommand{\R}{\ensuremath{\mathbb R}}
\newcommand{\rp}{\ensuremath{\mathbb {RP}}}
\newcommand{\cp}{\ensuremath{\mathbb {CP}}}
\newcommand{\vac}{\ensuremath{|0\rangle}}
\newcommand{\vact}{\ensuremath{|00\rangle}                    }
\newcommand{\oc}{\ensuremath{\overline{c}}}
\begin{titlepage}
\begin{flushright}
\end{flushright}
\def\thefootnote{\fnsymbol{footnote}}

\begin{center}
{\Large {\bf
Quantum Field Theory, Black Holes and Holography\\
\vspace{0.3cm}
}}
\end{center}

\bigskip
\begin{center}
{\large  Chethan
KRISHNAN}\\
\end{center}

\renewcommand{\thefootnote}{\arabic{footnote}}

\begin{center}
{\em  { SISSA,\\
Via Bonomea 265, 34136, Trieste, Italy\\
{\rm {\texttt{krishnan@sissa.it}}}\\}}

\end{center}

\noindent
\begin{center} {\bf Abstract} \end{center}
These notes are an expanded version of lectures given at the Croatian School on Black Holes at Trpanj, June 21-25, 2010. The aim is  to provide a practical introduction to quantum field theory in curved spacetime and related black hole physics, with AdS/CFT as the loose motivation. 

\begin{center}
\end{center}

\vfill

\end{titlepage}
\bigskip

\hfill{}
\bigskip

\tableofcontents


\setcounter{footnote}{0}
\section{\bf Not Quantum Gravity}

The subject matter of these lectures is the general topic of quantum field theory on black hole spacetimes. On top of serving as the paradigmatic example, black holes are also of intrinsic theoretical interest because they bring out the tension between general  relativity and quantum field theory in a maximally revealing way. It is often said that black holes provide the kind of workhorse for quantum gravity that the Hydrogen atom provided for quantum mechanics in its infancy. 
Quantum theory deals with microscopic things while general relativity (which is the natural setting for curved spacetimes\footnote{A curved spacetime can be thought of as arising from putting a lot of gravitons in the same state (``a coherent background").}) is usually relevant only at macroscopic 
length scales. So before we start, it behooves us to explain why it is worthwhile to consider the two in conjunction.

A usual first observation is that $\hbar, c$ and $G$ can together be used to construct units of length, time and energy, as first noticed long ago by Planck. These natural units are
\begin{eqnarray}
{\rm Planck\ Length,} \ L_P=\Big(\frac{G \hbar}{c^3}\Big)^{1/2}, \ \  {\rm Planck\  Time,}\ T_P=\frac{L_p}{c}\\ {\rm  and\ \ Planck\ Mass,} \  M_p=\frac{\hbar}{L_P^2 T_P^{-1}}. \hspace{1in} \nonumber
\end{eqnarray}
One can also define a Planck energy scale by $E_P=M_P c^2$, which comes out to about $10^{19} GeV$. This means that in physical phenomena that probe beyond the Planck scale (eg.: a transPlanck-ian collision between two particles), one will need a theory that takes account of both gravity and quantum mechanics simultaneously. Remarkably, it is an experimental fact that the scales of particle physics happen to be far above/below the Planck length/energy. This is the reason why gravity is utterly negligible at the scales relevant for particle physics. We would never have noticed gravity at all if we were only to do particle experiments. But despite this, and again remarkably, gravity is in fact visible in the deep IR (i.e., energies far below the Planck scale) and was the first fundamental interaction to be noticed by humans: this is because of a specific dynamical feature of gravity, namely that large amounts of matter contribute constructively to the total gravitational field.  Said differently, we feel gravity because we live close to large, massive objects; the universe feels it because the Hubble scale captures the total amount of matter-energy in the Universe.

The above paragraph is supposed to convince the reader that above some scale gravity must be quantized. But quantizing gravity, as is well-known, is beset with many conceptual and technical difficulties. One oft-stated problem is that since the coupling constant $G$ is dimensionful, we expect more and more counter-terms to be necessary as we go to higher and higher orders in perturbation theory, resulting in a lack of predictivity. Note that this is a problem when the typical energies of the processes involved are Planckian. For energies $E$ far below the Planck scale, one can work to whatever order in $E/E_P$ as one wants by truncating the perturbation expansion at that order. Once one does a finite number of experiments to fix the counter-terms up to that order, quantum gravity (up to that order and up to that energy scale) is a perfectly predictive {\em effective} field theory. But at the Planck scale, the ``small" parameter $E/E_P$ is order one: so the effective field theory fails and we lose all control. The usual expectation from particle physics for such breakdown is that new degrees of freedom become relevant at the Planck scale. This is analogous to the breakdown of the Fermi theory of weak interactions at the weak scale, $M_W$: the new degrees of freedom there were the weak gauge bosons whose mass was at the weak scale. Below $M_W$ one could ``integrate out" these gauge bosons from the path integral and work with Fermi theory as the effective theory of weak interactions, but above $M_W$ we needed a different theory, namely electro-weak theory.

Another question is the meaning of observables in quantum gravity: since Einstein's general relativity is a diffeomorphism invariant theory, it seems that spacetime coordinate is a meaningless quantity in gravity. So if we take diffeomorphisms seriously as a gauge redundancy, only integrals over all spacetime can arise as well-defined quantum observables. This integration is the continuum analogue of the tracing over gauge indices that one does in non-Abelian gauge theories to construct gauge invariant observables. Trouble is that even if we were able to successfully come up with such a setup for gravity, it is not clear how spacetime locality can emerge from a description with no spacetime coordinates\footnote{This is the situation, for example, in the AdS/CFT correspondence, where we believe that we have a consistent theory of quantum gravity in an asymptotically anti-de Sitter spacetime in terms of a Yang-Mills theory in a different spacetime without gravity. The diffeomorphisms of the original spacetime are ``solved" by the Yang-Mills theory, and as a result it is not at all clear how locality of the original gravity theory in AdS is encoded in the Yang-Mills theory.}. Another related question, if one wants to canonically quantize, is that of the choice of $t=$ const. spatial slices where we can define our canonical commutation relations: such slices are not respected by diffeomorphisms. In a theory where time-reparametrization is a gauge invariance, the generator of time translations (the Hamiltonian) vanishes classically. In the quantum theory, the analog of this statement is that one should impose that the Hamiltonian annihilates the physical states! The precise meaning of time evolution in such a quantum system is sometimes called the ``Problem of Time". Yet more puzzles (entropy, information loss, unitarity, ...) show up if one tries to understand black holes and horizons in a quantum theory, which we will discuss in some detail later. The presence of singularities in classical general relativity is another suggestion that something has to give: in a full theory of nature, one does not expect that regular initial value data can evolve into singular configurations where the theory breaks down.

But even at a very basic level, it is not clear that quantizing gravity frontally in perturbation theory is the correct way to go. This is because typically, when we do quantization, we go to an energy scale where the theory decouples into free theories with no interactions. For QED this happens at low energies where the theory splits into free Maxwell field and free electrons, while for QCD this happens in the UV where we have free quarks and gluons. From there, we can perturbatively add interactions between the various players. But the fundamental fact, due to the principle of equivalence, that gravity couples to everything that has energy {\em including itself} means that at any energy scale where the non-linearities of gravity are important, the contribution from matter will also be important (and vice versa), and it is inconsistent to quantize pure gravity first and then add matter later. So before one can consistently start to quantize gravity,
one needs to know {\em all} physics from here to the Planck scale: that is, we need to have a consistent and compelling UV completion for gravity\footnote{One reason why string theory is attractive is because it provides an essentially unique (and therefore compelling) way to do this UV completion {\em provided} one demands ``worldsheet conformal invariance" in the theory. 
}. As a corollary, this questions the validity of certain attempts to quantize pure gravity as a stand-alone theory, without worrying about matter.

But we might learn a few things about quantum gravity even without trying to quantize it fully. These lectures deal with one such technology, namely quantum field theory in curved space. The subject of QFT in curved space is best thought of as the propagation of quantized matter fields in a {\em classical} background containing a large number of gravitons. This is not quite the limit where individual scattering events become Planckian. In this setting, quantum gravity is still not significant when computing particle cross sections. In other words, we are still not in the regime where the UV of the quantum field theory gets modified due to gravity, even though its IR (long-distance) is different from the usual flat spacetime. This can be a consistent regime to consider because the field theories that we write down at low energies are insensitive to high scales except through the RG runnings of the coupling constants. In other words, we expect that the complications in the UV are going to be identical in both curved and flat spacetimes, because at short distances the length scale introduced by the spacetime curvature is irrelevant\footnote{Note that to make this precise, we will have to define a consistent notion for the renormalized expectation value of the stress tensor in curved spacetime, because we don't want uncontrolled backreaction on the geometry. This is a technical subject and we will discuss it in a later section.} and spacetime looks effectively flat. These lectures then are concerned with the IR modification of quantum field theory due to a curved geometry. We expect this to be a consistent thing to do, because we know that flat space quantum field theory works, even though we are ignoring gravity: all the complications from the UV are captured via the renormalization group by a few coupling constants. 
Our experience with flat space quantum field theory tells us that there indeed exists a notion of classical spacetime in which matter fields propagate quantum mechanically.

In fact, we can go a bit further and even talk about the classical backreaction of quantum matter. We can define a renormalized stress tensor for the fields in the curved background. This tensor can act as a source for the {\em classical} gravitational field according to
\bea \label{exp}
G_{\mu\nu}=-8 \pi G \langle T_{\mu\nu}\rangle,
\eea
which we will refer to as the expectation value form of the Einstein equations. This stress tensor can serve as a source for the backreaction of the quantum field on the classical geometry. The stress tensor is quadratic in the basic fields and therefore requires a systematic renormalization procedure to define it. 
The importance of the stress tensor also lies in the fact that as we will see, the notion of a particle is ill-defined in curved space and therefore it is better to work with local observables constructed from the fields.

But this is as far as we can go. If one tries to go one more step in the iteration, namely to use the backreacted metric again to determine the modification to the quantum field (in some semblance of perturbation theory), one runs into logical difficulties. The time evolution of the field depends on the backreacted metric, but this backreaction is non-linear because gravity is non-linear. So the time evolution becomes non-linear, which is not something one expects in a sensible quantum theory. Of course, this just means that in a full theory we need to treat gravity, and not just matter, quantum mechanically. This could be taken as an argument why quantizing gravity is a necessity, not an option, in an otherwise quantum world.

\subsection{Quantum Gravity at One Loop}

As explained before, 
the basic non-linearity of gravity suggests that it is not possible to decouple matter (alone) from the metric at any scale. One could then think that in a gravitational background where quantum effects of matter are significant, we might also expect quantum effects of gravity to also be significant\footnote{In the sense, for example, that we expect a black hole to Hawking radiate not just scalars, spinors and vectors, but also gravitons.}. 
Indeed, if one splits the metric into a background classical piece and a graviton fluctuation\footnote{This is morally analogous to treating photon exchange quantum mechanically while treating the background electric field as classical.},
\bea
g_{\mu \nu}=g_{\mu \nu}^c+h_{\mu\nu}, \label{back}
\eea
and the fluctuation is treated as a quantum field, then it contributes with the same strength as the rest of the matter fields at one loop in the quantum effective action.
That is, the contribution to the quantum effective action from gravitons is {\em not} suppressed by $M_{\rm Planck}$ at one-loop, and is equally important as that of ordinary fields. The reason for this is the trivial fact that at one loop we only have simple bubble diagrams and there are no internal vertices\footnote{We are talking about the effective action. Green functions with external vertices arise by differentiation of the effective action with respect to fields.}. It is only through vertices that the $G_N \sim 1/M_{\rm Planck}^2$ (or for that matter any coupling constant) can show up. When expanded around $g_{\mu \nu}^c$, the free part containing $h_{\mu \nu}$ is the only piece in the gravitational action that contributes at one loop.

To one loop, therefore, we can treat gravitons $h_{\mu\nu}$ as just another (free) quantum matter field in the QFT-in-curved-spacetime language. One expects that (\ref{back}) might be a reasonable approximation because this sort of a split between background and fluctuation is how we deal with gravitational waves\footnote{There is indirect, but very strong, quantitative evidence for gravitational waves from observing the binary pulsar PSR 1913+16. With the LIGO and LISA observatories, we expect to detect gravitational waves directly.}, which are the prototype waves whose quanta we believe are gravitons\footnote{To whatever extent we believe they are fundamental.}. This also means that the graviton contribution to the loop corrected Einstein equation can be taken to the right side of (\ref{exp}) and interpreted as part of the contribution to the expectation value of the stress tensor. We will see in a later section that the one loop corrected Einstein equations involve a renormalization of the couplings of the Einstein tensor (i.e, $G_N$), the metric (i.e., the cosmological constant $\Lambda$) and couplings of two other higher order curvature tensors. This will be demonstrated by computing the renormalization to the stress tensor of a (free) quantum scalar field in a fixed background. By the above arguments, the one loop contribution from the metric fluctuation will be identical. In other words, the stress tensor computation will also capture the one-loop divergence structure of gravity\footnote{An interesting special case is when one ignores all matter fields except the metric and its fluctuation. In this case it turns out that one can use the {\em background} equations of motion (i.e., vacuum Einstein equations) to do a field redefinition that
re-absorbs all divergences. That is, there is no need to renormalize the coupling constants and the theory is finite at one loop. This result is special to $D=4$ because it uses some special properties of the Gauss-Bonnet curvature in four dimensions. Of course, pure gravity does have divergences at two-loop. A review of these matters can be found in \cite{Deser}.} \cite{tHooftV}.

To summarize: In practice, by quantum field theory in curved space we will mean (1) propagating free quantum fields in a fixed background, or if we go one step further (2) quantum gravity coupled to matter truncated at one loop in the above sense. We will work with free matter fields, even though the assumption of freedom can be (perturbatively) relaxed to some extent, see p. 6 and chapter 15 of \cite{Birrell}. Note that for free fields, the one loop effective action is exact, because there are no interaction vertices. Of course, such a statement is not true for the metric fluctuation, so for them the one-loop truncation is truly a truncation.


In the rest of the introduction, we will try to give a flavor of the various things we will discuss in these lectures. The presentation is necessarily rather sketchy, so the reader should not panic.

One essential feature of curved space quantum field theory is that there is often no canonical notion of a particle. This is intimately tied to the fact that unlike in Minkowski space, there is often no preferred set of coordinates in which one can quantize the fields and identify the basis modes as particles. What looks like the vacuum in one frame can look like a state with particles in another. One of the most dramatic ways in which this phenomenon manifests itself is in the Unruh effect: an accelerating observer in empty flat space will see an isotropic flux of hot radiation.

We will explore some of these effects in the context of black hole spacetimes. It turns out that one of the effects of spacetimes with horizons is that they act like thermal backgrounds for quantum fields. This is impressive on two counts: from a classical gravity perspective, this means that black holes (despite their names) can radiate. There is a sensible way to treat them as thermodynamical objects, which is the subject of black hole thermodynamics. A remarkable result of black hole thermodynamics is that the area of a black hole captures its entropy. That raises the question: what are the microstates of the black hole which add up to its entropy? This is certainly a problem that lies outside the regime of general relativity where black holes are essentially structure-less. One needs a theory of quantum gravity to even begin to address this problem. Remarkably, it turns out that for certain special kinds of black holes, string theory has managed to provide an explanation for the microstates in terms of D-brane states, resulting in some rather detailed matches between microscopic and macroscopic entropy. 

Another interesting consequence of these ideas is that the thermal nature of black holes is a signal of an apparent loss of unitarity in the quantum evolution of the field. What started out as a pure state in the far past looks like a thermal state (a mixed state, a density matrix) at late times. This seeming loss of unitarity has been a thorny problem for decades, but again, with developments like AdS/CFT we now believe that we have a qualitatively correct picture of how unitarity is preserved.

Some of the recent applications of these web of ideas has been in the AdS/CFT (holographic) correspondence. Many old-school features of black holes have very natural re-interpretations in AdS/CFT. String theory has offered some plausible answers for the puzzles raised by black holes, and in fact AdS/CFT seems to suggest that one should certainly take the thermal interpretation of black holes seriously. It turns out that much of the recent developments in applied AdS/CFT relating it to condensed matter (eg., \cite{cond}), fluid dynamics (eg., \cite{fluid}) and heavy ion QCD (eg., \cite{RHIC}) are all dual versions of black hole physics. The fact that black hole thermodynamics and AdS/CFT mesh together beautifully, should be taken as evidence for both.

The purpose of these lectures is to explore parts of the above web of phenomena. In particular, we will try to emphasize the aspects of quantum field theory in curved space that are ``practical" and ``useful". We will start with a description of quantum field theory in flat Minkowski spacetime, but with a curved space outlook. The purpose of this section is to clarify the nature of the generalizations involved, when we migrate to curved space in later sections. Section 3 will develop the basics of quantum field theory in curved spacetime. 
In section 4 we will turn to an example that is quintessential to the subject, namely QFT on Rindler space and the Unruh effect. Then we turn to Green functions in complex plane which connect up Euclidean, Lorentzian and Thermal quantum field theory. In subsequent sections we discuss the Euclidean approach to quantum gravity, quantum field theory on black hole backgrounds and the rather technical subject of stress-tensor renormalization.  The final section is a selection of topics from the AdS/CFT correspondence which are related to black hole physics. 

My major influences in preparing the classical parts of these lecture notes have been Hawking's original papers (with various collaborators) cited throughout, Unruh's remarkable paper \cite{Unruh}, Birrll\&Davies \cite{Birrell} and Preskill's lectures \cite{Preskill} (which are in a class of their own). I have also consulted Kay\&Wald \cite{Kay}, Mukhanov\&Winitzki \cite{Mukhanov}, Ross \cite{Ross}, Wald \cite{WaldQFT} and Weinberg \cite{Weinberg}. The parts on AdS/CFT are mostly adapted from MaldacenaI \cite{MaldacenaI} and II \cite{MaldacenaII}, WittenI \cite{WittenI} and II \cite{WittenII}, MAGOO \cite{MAGOO} and Aharony \cite{Aharony}. Since the subject matter is vast and historical, it is impossible to give credit everywhere that it is due, so I apologize in advance for the numerous inevitable omissions. It is often hard to remind oneself that a piece of lore that is considered standard now, was the fruit of struggle, sweat and tears for the last generation.

\section{QFT in Flat Spacetime}

For us, field theory will mean free scalar fields. The idea is to probe curved space with the simplest probe. Free fermionic and Maxwell fields in curved space have also been studied quite a bit and is of relevance in some situations; fermionic modes are relevant for example in the study of certain instabilities of spinning black holes as well as in recent holographic constructions of non-Fermi liquids. But we will not consider higher spin fields at all in these lectures. We will forget about interactions as well: the main results of thermal black holes, like Hawking radiation,  have been found to be robust against the presence of interactions\footnote{Most of this is for weakly coupled interactions. Strongly coupled theories, like QCD, in black hole backgrounds are a different story and hardly anything is known. There is no notion of particle/mode here because quantizing around a weakly coupled fixed point (like the UV fixed point of QCD) as one usually does in QFT is not useful because the states one finds that way are not the asymptotic states useful for defining, say, S-matrices. But see \cite{Rang1} for a dual description of certain black hole vacuum states using the AdS/CFT correspondence, for large-$N$ gauge theories, when they are strongly coupled.} and mostly only add technical complications. 

\subsection{One-particle Hilbert Space}

We start by reviewing the basic notions of quantizing fields in flat space, so that we can generalize what can be generalized to curved space later. Flat space quantum field theory is about making quantum mechanics consistent with special relativity. What this means is that we want to construct a quantum theory such that
\begin{itemize}
\item Physics is Lorentz invariant, i.e., frame independent in all inertial frames.
\item Information does not propagate faster than speed of light (``relativistic causality").
\end{itemize}
The problem as it is posed is non-trivial and in fact requires introducing some auxiliary notions like ``fields". One cause for worry is that uncertainty causes wave-packets to spread, but for relativistic causality to work, we want them to {\em not} spread so fast that they get outside the lightcone. Another point of view might be that in general we expect things to get simpler when we add more symmetry, so adding Lorentz invariance should make quantum mechanics simpler, not more complicated. But we know that quantum field theory is more complicated than quantum mechanics. Both these points have interesting resolutions. The first is tied to the remarkable fact that for a relativistic theory, for points outside the lightcone, forward and backward propagation amplitudes (remarkably) cancel\footnote{This requires that integer spin fields are quantized using commutators and half-integer spins using anti-commutators, and is the origin of the spin-statistics theorem.}. The second is tied to the fact that in a relativistic theory, one cannot work with only a finite number of particles as in quantum mechanics: so the seeming simplicity of an added symmetry comes at the expense of an infinite number of degrees of freedom. This results in the complications in perturbation theory having to do with renormalization.

The basic goal is to construct a quantum theory of non-interacting relativistic particles. Two strategies:
\begin{itemize}
\item Strategy A. First construct a Hilbert space of relativistic particle states. Then introduce ``fields" as a tool for implementing a notion of spacetime locality for observables acting on this Hilbert space.
\item Strategy B. Start with a relativistic classical field theory, and quantize these fields canonically. The spectrum contains states that are naturally interpreted as particles.
\end{itemize}
Remarkably, both lead to the same theory in the end. The first strategy is more natural if one wants to stay close to the notion of a particle as one goes from non-relativistic to relativistic physics (this was our original motivation), while the second is better if one wants to stay close to symmetries and causality. The particle notion is fuzzy in curved space because it is not a canonical choice in a non-inertial frame, so we will often prefer the latter approach. 

But we start with some comments on strategy A. A natural definition of a a relativistic particle is as a unitary irreducible representation of the Poincare group on a Hilbert space. The Poincare group is the (semi-direct) product of the translation group (the normal subgroup in the semi-direct product) and the Lorentz group. We work with the proper Lorentz group, so parity and time-reversal are excluded. So what we we consider are transformations $(\Lambda, a)$ such that
\bea
{\rm Lorentz}:& \ \ \Lambda: x^\mu \rightarrow \Lambda^{\mu}_{\ \nu}x^\nu, \ \ \eta_{\mu\nu}\Lambda^{\mu}_{\ \sigma}\Lambda^{\nu}_{\ \rho}=\eta_{\sigma \rho}, \ \ \ \Lambda^0_{\ 0} > 0, \ \ {\rm det}\Lambda =+1. \\
{\rm Trnsaltion}:& a: x^\mu \rightarrow x^\mu+a^\mu, \\
{\rm Poincare}:& (\Lambda, a): x \rightarrow \Lambda x +a, \nonumber \ \ {\rm with\ the \ product  \ rule}\\ & ( \Lambda_1, a_1).(\Lambda_2, a_2)=(\Lambda_1\Lambda_2, \Lambda_1a_2+a_1).
\eea
To construct a quantum theory that has Poincare invariance as a symmetry, we need to introduce a Hilbert space in which there is a unitary action of the Poincare transformations. By definition, this means that states in the Hilbert space transforms as
\bea
(\Lambda, a): |\psi\rangle = U (\Lambda, a)| \psi \rangle, \ \ {\rm with} \ \ \langle \psi|\phi\rangle = \langle U \psi| U\phi \rangle,
\eea
so that
\bea
U( \Lambda_1, a_1).U(\Lambda_2, a_2)=U(\Lambda_1\Lambda_2, \Lambda_1a_2+a_1).
\eea
These two are part of what it means to have a symmetry in a quantum system. The first step in strategy A is to explicitly construct such a representation.

This is done in gory detail in chapter 2 of Weinberg (for example). The basic idea is to introduce plane wave states $| k \rangle$ which are eigenstates of translation (in fact they form an ir-rep of translation) labeled by momenta, and then look for conditions so that these states will form representations of the Lorentz group as well. One helpful fact is that the eigenvalues of translation on physical particles have to satisfy $p^\mu p_\mu \ge 0$ and $p^0 \ge 0$. We can use this to go to a frame where the problem of finding the representations of the full Lorentz group simplifies (the ``little group"). For spin zero which we are concerned with, the problem is quite a bit simpler than the general approach of Weinberg, but since we do not need these details we will not present them. The only fact that we will quote is that since these plane wave states form a complete set, the following relation holds:
\bea
{\mathbb I} = \int d\mu(k) |k \rangle \langle k |, \ {\rm where} \ d\mu(k)= \frac{d^3 k}{(2 \pi)^3 2 k^0} \label{measure}
\eea
is the Poincare invariant measure. This can be used to show that
\bea
\langle k|k' \rangle=(2 \pi)^3 (2 k^0) \delta^3({\bf  k}-{\bf k}'). \label{norm}
\eea
The one-particle Hilbert space is fixed by the transformation law that $U(\Lambda) |k\rangle  = | \Lambda k\rangle$ and the above norm on the Hilbert space.

Once we have the one-particle Hilbert space, we can form tensor products to form (reducible) multi-particle Hilbert states, and form a Fock space as a direct sum of such $n$-particle Hilbert spaces. At this point the Fock space just sits there: but once we introduce fields which can create and annihilate particles, this becomes the natural arena for dynamics.

\subsection{Fields: Locality in Spacetime}

At this stage we have outlined the first step of strategy A. But a quantum theory is defined not just by the states, but also by the operator algebra (the algebra of observables) acting on the states\footnote{Since all (infinite-dimensional) Hilbert spaces are isomorphic, it is the operator algebra that breathes life into a theory.}. So far we have only defined the Hilbert space representation of the Poincare group. This describes the kinematics, but to complete the picture we need to introduce (local) operators which act 
on this representation space. Our physical prejudice is that these operators should capture some notion of locality: flat spacetime is after all merely a gadget for capturing locality, Lorentz invariance and (relativistic) causality. A field accomplishes precisely that. A field is an operator valued distribution, i.e., a field $\phi(x)$ can be used to construct an operator $\int d^4 x \phi(x) f(x)$ that acts on the Hilbert space of the theory for any suitable test function $f(x)$. Note that the definition of the field is local. In practice we imagine (in the free theory limit that we are working with) that the effect of operating with fields is to emit or absorb particles and thereby mix up the different $n$-particle Hilbert spaces. This also provides a natural setup for introducing interactions as local operators built out of the basic fields.

We will see in the next section that Lagrangian quantization provides a nice book-keeping device for dealing with the various fields and their interactions that we would like to introduce in the theory. In the simplest case of a scalar field, we introduce the basic field operator $\phi(x)$ from which all other (local) operators can be constructed as an operator with the following properties:
\begin{enumerate}
\item $\phi$ is a superposition of creation and destruction operators\footnote{\label{a} Creation operators are defined between multi-particle Hilbert spaces as $
a(k)^\dagger|k_1,...,k_N\rangle \sim | k, k_1,...,k_N\rangle$.
This determines its matrix elements in the Fock space, and by Hermitian conjugation those of the annihilation operators as well. Note that they are defined in momentum space. When we fix the normalization (the ``$\sim$") in a relativistically invariant way, this is in fact an incredible amount of information and can be used to fix $\phi$ uniquely using properties 2 and 3.} acting on the $n$-particle Hilbert spaces.
\item $\phi=\phi^\dagger$, because we are working with real (uncharged) fields.
\item $\phi$ is a Lorentz scalar: $U(\Lambda, a)^{-1} \phi(x) U(\Lambda, a)=\phi(\Lambda x+a)$.
\end{enumerate}
The first property might seem a bit ad-hoc. The point is that creation and annihilation operators serve a dual purpose. Physically, they are introduced as a concrete way of realizing the possibility that particles can be created or destroyed. Technically, they are useful because {\em any} operator acting on the Fock space can be expressed as a sum of products of creation and annihilation operators. This might seem surprising at first, but in fact is trivial: any operator is fully defined by its expectation values on every state, so it is possible to simulate any such expectation value by means of an appropriately constructed combination of $a$'s and $a^\dagger$'s. See Weinberg chapter 4.2 for an induction-based proof. In any event, it is natural that one constructs the basic field operator as a linear combination of the two basic operations on the Fock space.

As mentioned in footnote \ref{a}, these three conditions are in fact enough to fix the form of the operator (essentially) uniquely: one can write down $\phi$ as a general linear combination of $a$  and $a^\dagger$ and then apply the conditions systematically to massage it to the familiar form:
\bea
\phi(x)=\int \frac{d^3 k}{(2 \pi)^{3/2} (2 k^0)^{1/2}} \Big(e^{-ik.x}a(k)+e^{ik.x}a(k)^\dagger\Big).
\eea
In arriving at this, on top of Lorentz invariance, we have also used the defining properties of ``particles" (like $p^\mu p_\mu=m^2$). Note that in position space, this latter condition translates to the fact that $\phi(x)$  satisfies the Klein-Gordon wave equation. The scalar field is nothing but a linear combination of positive frequency and negative frequency solutions of the KG equation. It is possible to check by direct computation that
\bea
[\phi(x), \phi(y)]=0, \ {\rm when} \ (x-y)^2<0,
\eea
and therefore relativistic causality is protected outside the light-cone as necessary. This uses the fact that the creation and annihilation operators satisfy their usual algebra\footnote{This is where the spin-statistics connection enters.}, which can be shown to hold as an operator relation on the Fock space using their definition (footnote \ref{a}).

\subsection{Lagrangian Quantization}

After our brief overview of strategy A, now we turn to strategy B, which is the one that is more immediately useful for generalizations to curved spacetime.
The basic observation is that there is a one-to-one map between states in the one-particle Hilbert space and positive frequency solutions of the Klein-Gordon equation. Positive frequency solutions are solutions which have a Fourier decomposition in terms of purely positive frequency modes. Since an arbitrary state in the one-particle Hilbert space can be expanded (by definition) as
\bea
| \tilde f \rangle = \int \frac{d^3 k}{(2 \pi)^3 2 k^0} \tilde f(k) | k \rangle,
\eea
this state can be mapped to the positive energy solution $\langle 0 | \phi(x) | \tilde f$ via
\bea
| \tilde f \rangle \rightarrow \langle 0 | \phi(x) | \tilde f \rangle=\int \frac{d^3 k}{(2 \pi)^3 2 k^0} \tilde f(k) e^{-i k.x}.
\eea
 This manifestly contains only positive frequency modes. Since the only information contained on both sides of the map are the Fourier coefficients $\tilde f(k)$, it is clear that one side contains precisely the same information as the other, and therefore this is an equivalence. Of course, there is nothing special about positive frequency, and one can find such an isomorphism with negative frequency solutions as well by conjugating some of the ingredients.

This isomorphism gives us an alternative construction of the Hilbert space as the space of positive frequency solutions of the Klein-Gordon equation.  The solutions of the classical Klein-Gordon equation
\bea
(\partial_\mu \partial^\mu +m^2) \phi(x)=0
\eea
can be expanded in the basis
\bea
u_k(x)=e^{-ikx}, \ \ u_k(x)^*=e^{ikx}, \ \ {\rm with} \ \ k^2-m^2=0,\  k^0>0 \label{basis}
\eea
where $kx \equiv \omega t- {\bf k}.{\bf x}$. The $u_k$ are called positive energy modes and $u_k^*$ are negative energy modes. If we declare that under a Lorentz transformation
\bea
u_k(x)\rightarrow u_k(\Lambda^{-1}x),
\eea
then the positive and negative frequency solutions do not mix and form irreducible representations of Lorentz group. It is immediately checked that
\bea
\Lambda : (u_k (x), u_k(x)^*) \rightarrow (u_{\Lambda k} (x), u_{\Lambda k} (x)^*).
\eea
It is trivial to check that translations also retain the positive/negative frequency nature of the mode, and therefore this construction gives an ir-rep of the Poincare group and therefore is a good candidate for the one-particle Hilbert space.  This construction in terms of solutions of the KG equation does generalize reasonably straightforwardly to situations where the particle concept does not.

Another equivalence that will be useful to remember is the relation between positive (negative) frequency solutions and initial value data. The basic idea is that when restricted to only modes of positive(negative) frequency, we only need one piece of data to evolve the initial data of a second order ODE (Klein-Gordon) into the future. Therefore
\bea
\{\text{pos. frequency solns.}\} \sim \{\text{neg. freq. solns.}\}  \sim \nonumber\hspace{0.55in} \\ \sim \{\text{initial value data on a hypersurface}\} \sim \{\text{functions on} \ {\mathbb R}^3\}.
\eea

Finally, we need to introduce a norm on the Hilbert space as defined in this new way, so that it matches with the norm defined in (\ref{norm}). This is accomplished via the Klein-Gordon inner product between two {\em solutions} $f(x)$ and $g(x)$ :
\bea
(f,g)=i \int_{\Sigma} d\Sigma \ [f(x)^* \partial_t g(x)-(\partial_t f(x)^*) g(x)],
\eea
the integral is over a constant time slice at $t$ (which we call $\Sigma$). This might seem like a non-covariant choice, but the same expression when written covariantly looks like
\bea
(f,g)=i\int_{\Sigma}d^3 x \ n^{\mu} \ [f(x)^* \partial_\mu g(x)-(\partial_\mu f(x)^*) g(x)].
\eea
where $n^\mu$ is the normal to the surface and defines a local direction for time.  If we perturb the spacelike hypersurface $\Sigma$, then we can compute the difference between the initial and final values of the norm and the result can be written using the divergence theorem as
\bea
(f,g)_{\Sigma_1}-(f,g)_{\Sigma_2}=i\int_X d^4 x \ \partial^\mu[f(x)^* \partial_\mu g(x)-(\partial_\mu f(x)^*) g(x)].
\eea
Here $X$ is the spacetime cylinder bounded by the two spacelike slices: $\partial M= \Sigma_1-\Sigma_2$ (the sign is to keep track of orientation). But on the solutions of the Klein-Gordon equation the integrand of the last expression vanishes, so the norm is indeed independent of the choice of hypersurface.

Using the definition of the Klein-Gordon norm it is directly checked that
\bea
(u_k, u_{k'})=(2 \pi)^3 (2 k^0) \delta^3({\bf  k}-{\bf k}'), \ (u_k, u_{k'}^*)=0, \ (u_k^*, u_{k'}^*)=-(2 \pi)^3 (2 k^0) \delta^3({\bf  k}-{\bf k}'). \label{basisnorm}
\eea
The first of these is the solution space analogue of (\ref{norm}). The last relation shows that the norm is not positive definite when it is defined outside positive frequency solutions\footnote{But note that  the norm $-(f,g)$ is positive definite on negative frequency solutions and negative definite on positive frequency solutions.}.

These considerations demonstrate that we have every right to use this ``one-particle" Hilbert space to construct our Fock space.  We can identify the coefficients in the expansion of the field in terms of the positive and negative frequency solutions, as the creation and annihilation operators. The quantization follows from imposing the canonical commutation relations on them in the usual way. As usual, imposing canonical commutators on the fields will be equivalent to imposing the creation-annihilation commutation relations on the creation-annihilation operators. This is what we will do, and when we move on in the next sections to the curved spacetime, we will find that this path is essentially indispensable.

\section{QFT in Curved Spacetime}

Since Poincare symmetry is no longer a global symmetry of spacetime when we move on to curved space, working with ir-reps of Poincare group to construct one-particle states is not useful in curved space. Said another way, there is no reason why the Hilbert space of the quantum theory built on a curved background should respect Poincare as a symmetry group. But of course locally, when the wavelengths are small compared to the curvature scale of the geometry, particles become an approximately useful notion (this is why particle physics works after all). It is also useful to keep in mind that because of the causal structure of light cones in curved geometry, causality is preserved globally if it is preserved locally.

The basic goal is that we want to construct a ``one-particle" Hilbert space such that the fields (maps from spacetime to operators) acting on this Hilbert space are causal. We will generalize the ``positive frequency" approach to quantization in order to get some traction in curved space. So we work with the solution space of the Klein-Gordon equation in curved spacetime. Start with the covariantized (i.e., minimally coupled) action\footnote{Sometimes we will also consider an action of the form
\bea
S=\int d^4 x \ \sqrt{g} \frac{1}{2} [g^{\mu \nu} \partial_\mu\phi \partial_\nu \phi-m^2 \phi^2-\xi R \phi^2], \label{confcupLag}
\eea
where $\xi$ is a parameter and $R$ is the Ricci scalar of the background metric. The discussion here is essentially unchanged by the addition of this term, because the scalar still appears as a quadratic. The transformation properties of this action under local rescalings of the metric will be of interest to us in the discussion on the conformal anomaly in a later section.}
\bea
S=\int d^4 x \ \sqrt{g} \frac{1}{2} [g^{\mu \nu} \partial_\mu\phi \partial_\nu \phi-m^2 \phi^2].
\eea
It is invariant under general coordinate transformations (diffeomorphisms):
\bea
x \rightarrow x'(x), \ \text{if $\phi$ is a scalar:} \ \phi(x) \rightarrow \phi(x').
\eea
The equation of motion is the curved space Klein-Gordon equation:
\bea
(\nabla^\mu \nabla_\mu +m^2) \phi=0
\eea
We want to work with solution space of this equation parallel to the flat space case. For this to be successful strategy, we need to have globally well-defined solutions for KG in the spacetime geometry. A sufficient condition for this is that the spacetime is globally hyperbolic. Roughly speaking, global hyperbolicity is the condition that the entire history can be determined by the data one puts on one spatial slice. Such a spatial slice is called a global Cauchy surface. A slightly more detailed discussion of this idea can be found in the Appendix.

\subsection{Canonical Quantization}

Global hyperbolicity guarantees that in curved space, analogous to flat space,
\bea
\text{Space of global Klein-Gordon solns.} \sim \text{Initial value data on a Cauchy surface.}
\eea
We want to keep things as closely related to flat space as possible, while allowing interesting generalizations. There are many interesting globally hyperbolic geometries which are not flat, so we will restrict ourselves to globally hyperbolic spacetimes in what follows, and proceed with the quantization. We first introduce a generalization of the flat space KG inner product.

For two solutions $f$ and $g$, the Klein-Gordon inner product is defined by
\bea
(f,g)=i\int_{\Sigma}d^3 x \ \sqrt{h}\  n^{\mu} \ [f(x)^* \partial_\mu g(x)-(\partial_\mu f(x)^*) g(x)].
\eea
Here $h_{ij}$ is defined as the induced metric on $\Sigma$ and $n^\mu$ is the future-pointing unit normal. Essentially by a repetition of the flat space case, we can show the slice-independence of this inner product.

Now we want to impose canonical equal-time commutators (CCR). The covariant version is
\bea
[\phi(x), n^\mu \partial_\mu \phi(y)]_{\Sigma}=\frac{i}{\sqrt{h}}\delta^3 ({\bf x}-{\bf y}), \ \
[\phi(x),  \phi(y)]_{\Sigma}=0, \ \ [n^\mu \partial_\mu \phi(x), n^\nu \partial_\nu \phi(y)]_{\Sigma}=0.
\eea
The normalization of the delta function is such that it multiplies correctly with $\int d^3 x \sqrt{h}$. Note also that since the CCR is defined on a spacelike surface (the Cauchy slice), the second relation $[\phi(x),  \phi(y)]_{\Sigma}=0$ immediately has the consequence that causality is preserved for spacelike separations.

The good thing is that once these CCRs are imposed on one spacelike slice, they are automatically satisfied on any other, as long as $\phi$ satisfies the KG equation. To show this, we use the fact that if $\phi$ satisfies CCR on $\Sigma$, then for {\em any} $f, \ g$ that are solutions to KG equation,
\bea
[(f,\phi)_{\Sigma},(g,\phi)_{\Sigma}]=-(f,g^*)_{\Sigma}. \label{trick}
\eea
This is easy to prove by direct computation of LHS and using the fact that $\phi$ satisfies CCR. In fact the converse is also true: if arbitrary KG solutions $f, \ g$ satisfy (\ref{trick}), then the $\phi$ satisfies CCR on $\Sigma$. This is because $f$ and $g$ and their normal derivatives are {\em arbitrary} choices one can make on $\Sigma$: the statement that CCR holds (as an operator relation) is precisely the statement that it holds on $\Sigma$ no matter what it is applied to. In other words, the statement that a KG solution $\phi$ satisfies the CCR is equivalent to the statement that for arbitrary KG solutions $f, g$, eqn. (\ref{trick}) is satisfied. But since we already know that $(\ ,\ )_{\sigma}$ is slice-independent, eqn.(\ref{trick}) is also slice-independent. Which in turn means that CCR is also slice-independent. QED. 

In the flat space case there was a complete basis (the plane wave basis $u_k(x)=e^{-i k.x}$ and its conjugate, see (\ref{basis}) and (\ref{basisnorm})) in which we could expand any solution of the Klein-Gordon equation. That this was possible was a consequence of the fact that a Fourier basis is guaranteed in flat space. Assuming that the spacetime is approximately flat in the far past, we can argue for the existence of a complete basis even in curved space as follows. Since we have shown the slice independence of the inner product and the CCR in curved space, we can start with a slice $\Sigma$ in the past where all solutions can be expanded in terms of an (approximately) flat space basis. The evolution forward of these basis solutions will give us the required complete basis on any $\Sigma$.

For such a basis, analogous to (\ref{basisnorm}), we have
\bea
(u_i, u_j)_{\Sigma}=\delta_{ij}, \ \ (u_i, u_j^*)_{\Sigma}=0, \ \ (u_i^*, u_j^*)_{\Sigma}=-\delta_{ij}. \label{pseudonorm}
\eea
We have used a discrete schematic notation. In this basis, we can expand
\bea
\phi=\sum_i (u_i a_i+u_i^*a_i^\dagger),
\eea
where the coefficients $a_i, a_i^\dagger$ can be computed as
\bea
a_i=(u_i, \phi), \ \ a_i^\dagger=-(u_i^*, \phi) \ \ \text{{\bf (on any slice)}}. \label{defa}
\eea
The canonical commutation relations imply the usual creation-annihilation algebra (a quick way to derive this is to use (\ref{trick})):
\bea
[a_i, a_j]=0, \ \ [a_i^\dagger, a_j^\dagger]=0, \ \ [a_i, a_j^\dagger]=\delta_{ij}.
\eea
By (\ref{defa}), the creation-annihilation algebra (CAA) is precisely the statement that (\ref{trick}) holds for $f$ and $g$ replaced by a basis element. Since the basis elements span the entire solution space, this means that (\ref{trick}) holds for all KG solutions, which is equivalent (as discussed earlier) to the condition that the CCR holds. So CAA is equivalent to CCR.

This completes our quantization of the scalar field in curved spacetime: we start with a complete basis $\{u, u^*\}$ of solutions to Klein-Gordon and then construct the ``one-particle" Hilbert space ${\cal H}^{(1)}$ as the space spanned by the ${u_i}$. Taking direct sums of direct products of ${\cal H}^{(1)}$, we arrive at the Fock space ${\cal H}$. The creation and annihilation operators are naturally defined on ${\cal H}$, and using them we can construct the field operator $\phi$ on ${\cal H}$. The observables in the resulting quantum theory respects causality because $[\phi(x), \phi(y)]=0$ as a consequence of the CCR when $x$ and $y$ lie on a spacelike slice.

Till this point, nothing drastically different from Minkowski space QFT has happened. But now we introduce the crucial point which makes curved space QFT different from its flat space cousin.

\subsection{Bogolubov Transformations and S-matrix}

We argued that in curved spacetime, fairly generally, we expect the existence of a basis of KG solutions that satisfies the pseudo-orthonormality relations (\ref{pseudonorm}). But unlike in flat space, here there is an ambiguity involved in the choice of basis which has no canonical resolution: there are many ways to choose the $\{u_i\}$ with positive KG norm. For example, if $u, \ u^*$ satisfies
\bea
(u,u)=1, \ \ (u,u^*)=0, \ \ (u^*,u^*)=-1,\label{SHOB1}
\eea
then so does $u', \ u'^*$ defined by
\bea
u'=\cosh \theta u+\sinh \theta u^*, \ \ u'^*=\sinh \theta u + \cosh \theta u^*. \label{SHOB2}
\eea

This feature is quite general. The construction of the one-particle Hilbert space and therefore the Fock space is not unique in curved spacetime. Since modes (or creation-annihilation operators) are immediately connected to particles, this means that the notion of particles is coordinate dependent in curved spacetime. When we worked with flat Minkowski space, this ambiguity was resolved because there was a natural choice of time coordinate (the Minkowski time), which enabled a specific choice of poitive energy/norm\footnote{The relation between positive energy and positive norm solutions is part of the focus of the next section.} as the definition of the one-particle Hilbert space. A related statement is that we used the global symmetry of the spacetime to demand that our vacuum (or equivalently, the 1-particle Hilbert space built on it) be Poincare invariant. In general spacetimes, we have neither of these, and the choice of the vacuum is exactly that: a choice. Often, we will have some specific physical question in mind which will dictate the choice of the vacuum, we will see this for example in defining vacuum states on black holes. Another scenario encountered in cosmological settings is that often there is a well-defined time translation invariance symmetry in some parts of the spacetime and we can use it to define a set of positive energy/norm modes. We will discuss this briefly in the next section.

Consider two sets of complete orthonormal sets of modes in a spacetime both of which have been split into positive and negative norm subspaces: $u_i$ and $v_i$. Then we can expand the scalar $\phi$ as
\bea
\phi=\sum_i (u_i a_i+ u_j^* a_j^\dagger),
\eea
with the $a$, $a^\dagger$ satisfying the usual algebra and
\bea
(u_i, u_j)=\delta_{ij}, \ \ (u_i, u_j^*)=0, \ \ (u_i^*, u_j^*)=-\delta_{ij}.
\eea
as before. We have suppressed the spatial slice on which the KG inner product is defined because it is invariant under that choice. The vacuum state on which this quantization is built is characterized by
\bea
a_i|0\rangle_a=0, \ \ \forall \ i.
\eea
But we can consider an entirely analogous construction in terms of $v_i$:
\bea
\phi=\sum_i (v_i b_i+ v_j^* b_j^\dagger),
\eea
with the $b$, $b^\dagger$ satisfying the usual algebra and the new vacuum state is given by
\bea
b_i|0\rangle_b=0, \ \ \forall \ i.
\eea
Since both sets are complete, we can expand the second basis modes in terms of the first:
\bea
v_j=\alpha_{ij}u_j+\beta_{ij}u_j^*, \ \ v_j^*=\beta_{ij}^*u_j+\alpha_{ij}^*u_j^* \label{Bog}
\eea
This is what is called a Bogolubov transformation. Plugging these into the expansion in terms of the $v_i$ and comparing coefficients of $u_i$ and $u_i^*$, we find that
\bea
a_i=\sum_j(\alpha_{ji}b_j+\beta_{ji}^*b_j^\dagger), \label{Bog2}\ \ b_j=\sum_i(\alpha_{ji}^*a_j-\beta_{ji}^*a_j^\dagger).
\eea
The orthonormality of the modes results in conditions on the Bogolubov coefficients $\alpha$ and $\beta$:
\bea
\sum_k(\alpha_{ik}\alpha_{jk}^*-\beta_{ik}\beta_{jk}^*)=\delta_{ij}, \ \ \sum_k (\alpha_{ik}\beta_{jk}-\beta_{ik}\alpha_{jk})=0.
\eea

The basic point we need to remember is that when the $\beta$ are non-zero, i.e., when the positive norm modes are not preserved under a Bogolubov transformation, then the vacuum state of one basis looks populated with particles in the other. For example,
\bea
a_i|0\rangle_b=\sum_j\beta_{ji}^* |1_j\rangle_b \neq 0.
\eea
An explicit demonstration that there are particles, can be seen by taking the expectation value of the number operator of $u_i$ modes in the $a$-basis in the $b$-vacuum:
\bea
\langle 0| N^{(a)}_i |0\rangle_b =\sum_{j}|\beta_{ij}|^2.
\eea
where the superscript $(a)$ stands for the vacuum in which the number operator is defined: $N^{(a)}_i=a_i^\dagger a_i$. So the $a$-``vacuum" contains $b$-particles.

A useful object that we will consider is the ``S-matrix" that relates two such bases. The idea is that even though the notion of vacuum and notion of creation-annihilation operators get mixed up, the theory is unitary, so there exists a unitary operator that can map one vacuum to the other.  We will refer to such an operator as the S-matrix even though the notion of scattering is not necessarily built into it\footnote{In a spacetime that is approximately time-independent in the far future and the far past, the past-vacuum as it evolves into the future, can look full of particles of the future-vacuum. In this context, it is meaningful to talk about cosmological particle production and S-matrices. See the next subsection.}. The object we seek is $U$, defined by
\bea
U|\psi_b \rangle = |\psi_a\rangle
\eea
for any Fock space state $|\psi_{a,b} \rangle$. The index denotes the basis (i.e., built on $a$-vacuum or $b$-vacuum). Since matrix elements are invariant, we have
\bea
U^{-1}a_iU=b_i
\eea
(and its Hermitian conjugate, $U^{-1}a_i^\dagger U=b_i^\dagger$).
By rewriting this as
\bea
U^{-1}a_iU=\sum_i(\alpha_{ji}^*a_j-\beta_{ji}^*a_j^\dagger)
\eea
(and its conjugate), we finally have an operator equation for $U$ entirely in terms of the $a$-Fock space. The basic thing to remember is that any operator in the $a$-Fock space can be written in terms of $a$ and $a^\dagger$. By trying out an ansatz, we can obtain an explicit expression for $U$ in terms of $a$ and $a^\dagger$. The derivation essentially is about the Harmonic oscillator algebra writ large, but involves keeping track of many indices, so we will skip it\footnote{A toy version of the general problem  is the special case of a single harmonic oscillator: with the Bogolubov transformations taking the form (\ref{SHOB1}, \ref{SHOB2}). This is a simple exercise.}. The problem is solved with all its indices in section 10.2.3 and the appendices of the book by Frolov and Novikov \cite{Frolov}. A transparent derivation in a convenient matrix notation, together with some slick and instructive tricks can be found in Preskill's lecture notes \cite{Preskill}. We will adopt his notation and write the final result as
\bea
U={1\over{\rm Det}[\alpha^\dagger\alpha]^{1/4}}:\exp\Big[\frac{1}{2}a(\alpha^{-1}\beta)a+a(\alpha^{-1}-\I)a^{\dagger}+\frac{1}{2}a^\dagger(-\beta^*\alpha^{-1})a^\dagger\Big]: \label{S-matrix}
\eea
The result is normal-ordered and there is an overall phase ambiguity that we haven't fixed and will not be important. Here, the $\alpha$ and $\beta$ are defined as matrices so that the expressions in (\ref{Bog}) are written as
\bea
v=\alpha u+\beta u^*, \ \ v^*=\beta^* u+\alpha^* u^*
\eea
In what follows, we will repeatedly use this expression for the operator $U$ relating various bases.

\subsection{Stationary Spacetimes and Asymptotic Stationarity}

In a generic spacetime, there is no {\em canonical} split into positive and negative norm states. But we will mostly consider spacetimes with a timelike Killing vector. These are what are called stationary spacetimes. Black holes and Robertson-Walker cosmology are examples. Note that static is not the same as stationary: spinning black holes are stationary, but not static. On a stationary spacetime, one has a canonical choice of positive norm modes: the positive frequency modes, like the ones we discussed in the flat space case.

But before we proceed to explain this canonical split on stationary spacetimes, we want to emphasize a couple of subtleties that are hardly ever emphasized, but are crucial for the discussion. Stationarity guarantees that there is a timelike Killing vector, but it does not mean that this vector is well-defined globally. This is the case for instance for static black holes: the light-cone flips over inside the horizon. But this turns out {\em not} to be a problem in quantizing fields on static black holes. This is because what we mean by quantizing fields on black holes, is to quantize fields in the region outside the horizon: we treat the outside region as a full spacetime in itself. The reason why it is not unreasonable to do this, is because the region outside the horizon of a static black hole is by itself globally hyperbolic (even though it is not geodesically complete). To define the one-particle Hilbert space and proceed with a consistent quantization, what matters is (only) that the spacetime is globally hyperbolic as we discussed above. So static black holes are fine. On the other hand, on spinning black holes\footnote{Spinning black holes are stationary, but not static.} the problem is even worse, because there the usual timelike Killing vector associated to the Schwarzschild-like time flips over at the ergosphere. The ergosphere begins outside the horizon, but the globally hyperbolic region begins right at the horizon\footnote{There are related problems having to do with classical and quantum super-radiance and related issues, but these are deeper waters which we need not get into at the moment.}. How does one proceed in such a situation? One possible approach that is natural from many perspectives is to work with a frame co-rotating with the horizon and use the time coordinate in it as the timelike Killing vector. But unfortunately in flat space, such a rigid co-rotation will become superluminal far away from the black hole and such an approach cannot work. In fact, it is not clear if there is a consistent quantization of spinning black holes in Minkowski space. Interestingly, in asymptotically anti-de Sitter spaces, such a co-rotation construction is legitimate because the warp factor in the geometry ameliorates the possibility of faster-than-light rotation. Some recent directions in understanding rotating black holes can be found in \cite{rot, AdSrot, Krishnan}.

Onward to the quantization on stationary spacetimes! On stationary spacetimes, there is a canonical split of the space of solutions into positive and negative norm. The idea is to  work with eigen-basis of the time-translation generators, i.e., an energy/frequency basis. This means that we can take KG solutions in the form
\bea
u_k(t,x)=e^{-i \omega_k t} f_k(x).
\eea
Time invariance implies in particular that
\bea
0=\frac{d}{dt}(u_k,u_j) \sim (\omega_k-\omega_j) (u_k,u_j)
\eea
So solutions of distinct frequency are orthogonal. On the slice $t=0$ (and therefore on any slice) the solutions are positive definite (no sum over $k$ below):
\bea
(u_k,u_k)=2 \omega_k\int_{t=0} d^3 x \ |f_k(x)|^2
\eea
Since we work with normalizable solutions, then the basis can therefore be orthonormalized with a positive definite subspace (and a negative definite subspace, together they sum up to the whole space):
\bea
(u_i,u_j)=\delta_{ij}, \ \ (u_i, u_j^*)=0, \ \ (u_i^*, u_j^*)=-\delta_{ij}.
\eea
This is the positive frequency/energy basis and from this point on defining the creation/annihilation operators etc. proceed as usual.

In such a spacetime if one defines two different sets of modes, both of which are expanded in terms of the positive energy modes alone, then the two observers will agree on their notion of particles and vacuum. In particular, the $\beta_{ij}$ from the end of last subsection is zero and $a$-vacuum remains $b$-vacuum.

There is a generalization of the idea of stationarity that is useful when the spacetime is not necessarily stationary globally, but approaches a stationary spacetime in the far past or the far future. We can refer to this situation as asymptotic stationarity. In this case bases which tend to positive frequency modes in the future or past can be useful.  In the special case that the spacetime is stationary {\em both} in the far past and the far future while being time-dependent at intermediate times, the vacuum defined by the late-time positive frequency modes need not coincide with the vacuum defined by the early-time positive frequency modes. In such a situation, the S-matrix of the last section can be used to compute cosomological\footnote{Cosmology is a word used to capture any time-dependent background.} particle production.

To conclude this rather formal chapter, we illustrate by means of an example, what it means to construct a vacuum in a curved spacetime.

\subsection{Example: Inflationary Cosmology and Bunch-Davies Vacuum}

It is an observational fact that the large scale structure of the Universe (starting at the Big Bang to today) can be captured by the Robertson-Walker metric:
\bea
ds^2=-dt^2+a^2(t)\Big(\frac{dr^2}{1-k r^2}+r^2 d\Omega^2\Big),
\eea
for appropriate choices of the scale factor $a(t)$ and the spatial curvature $k$ ($=-1,0$ or $1$). One of the tasks of cosmology is to explain the time evolution of metric and matter by coupling this metric to appropriate forms of matter. In the last decade, a fairly substantial amount of observational evidence has accumulated showing that the Universe underwent an early era of expansion when the scale factor underwent an exponential increase with time. This is what is called inflation. Apart from providing a natural set up where the homogeneity and spatial flatness of the Universe emerge more or less naturally, inflation has the great virtue that it provides a mechanism for formation of inhomogeneities (and structure) in the post-inflationary universe. The idea is that large scale structure (and CMB anisotropies) were the result of inflation stretching quantum fluctuations to macroscopic scales. This means that studying quantum field theory in an inflating Universe is of great interest in getting predictions for, say, CMB anisotropies. As an exercise in some of the things that we have introduced in the previous section, we will do this in this section. Our task will merely be to define a canonical vacuum state that is often used in this context: the so-called Bunch-Davies vacuum. We will not deal with the various important applications of these ideas. This subsection is not self-contained, but we will explicitly warn the reader  in the places where leaps of faith are required.

We will restrict our attention to spatially flat ($k=0$) inflating Friedman-Robertson-Walker universes:
\bea
ds^2=a^2(\eta)\left[-d\eta^2+d\vec{x}^2\right].
\eea
We have introduced conformal time,
\bea
\eta=\int\frac{dt}{a(t)}.
\eea
Conformal time runs from $\eta=-\infty$ in the asymptotic past to $\eta=0$ in the asymptotic future. In the simplest models of inflation, exponential expansion is implemented by coupling a scalar field to this metric. We will assume that inflation is driven by a single  scalar  $\varphi$ ``slowly rolling" down an almost flat potential. In this case, the inflationary stage can be characterized by a so-called (dimensionless) slow-roll parameter
that is proportional to the derivative of the Hubble ``constant" $H$. Hubble constant is defined by
\bea
H=\frac{\dot a}{a}=\frac{a'}{a^2}
\eea
where dot denotes derivative with respect to the usual FRW time and the prime denotes derivatives with conformal time.
The only fact we will need is that during slow-roll inflation, the slow-roll parameter is almost zero (and constant). 
As a first approximation then, we can treat $H$ as approximately constant. A spacetime of zero spatial curvature with a constant Hubble parameter is called a de Sitter space, and our aim therefore is to describe a vacuum for quantum field theory on de Sitter space. Note that de Sitter space is the prototype for inflation, because a constant Hubble parameter means that the scale factor is changing exponentially. De Sitter space constructions in string theory are relevant both for the inflationary phase in the early Universe \cite{inflation}, as well as in the current era where dark energy is beginning to dominate \cite{KKLT}. De Sitter is a maximally symmetric space, even though this is not immediately clear from the way we have written down the metric. The robustness of inflationary de Sitter against inhomogeneities has been investigated recently in \cite{CKTX2}. In the context of the string landscape, which stipulates that the fundamental theory has many vacua of varying vacuum energies, each of these vacua is expected to give rise to a de Sitter Universe. In this context, transitions between such vacua via decompactification was studied in \cite{CKTX1}.

To describe perturbations in such an almost-de Sitter space, we first need to choose a gauge. This is because diffeomorphism invariance introduces arbitrariness in the choice of coordinates (including time). One standard choice is to choose constant-$\varphi$ slices as time foliation: note that $\varphi$ is constant on each spatial slice if the universe was perfectly homogeneous. When there are perturbations in fields, one way to choose a time-foliation is via equal-$\varphi$ slices. In this case, fluctuations in the scalar field $\delta \varphi$, are (by definition) zero. A non-trivial fact is that the metric perturbations in such an appropriately chosen gauge can be brought to the form
\bea
ds^2\equiv a^2(\eta)\left[-(1+2\delta N)d\eta^2+\exp (2\zeta) \delta_{ij} (dx^i+a \, \delta N^i  d\eta)(dx^j+a \, \delta N^jd\eta)\right].
\eea
It turns out that the variables $\delta N$ and $\delta N^i$ are constrained, and  can be expressed in terms of $\zeta$ and its derivatives \cite{MaldacenaInflation}. To linear order, $\zeta$ describes the curvature perturbation in slices of constant  scalar field. For reasons that we will not discuss, we do not need to consider tensor perturbations for the minimalist discussion of this section. We can expand the curvature perturbation in Fourier modes,
\bea
\zeta(\eta, \vec{x})\equiv
\sum_{\vec{n}\in \mathbb{Z}^3}
\zeta_{\vec{k}}(\eta) \exp(i\, \vec{k} \cdot \vec{x}),
\quad \text{with} \quad \vec{k}=\frac{2\pi}{L}  \vec{n} .
\eea
By doing this, we have put the spatial slice of the universe in a box, we can send it to infinity at the end of the calculation. In this notation $\zeta_{\vec{k}}$ is dimensionless. Reality conditions of the Fourier decomposition imply that $\zeta_{\vec{k}}=\zeta^*_{-\vec{k}}$.

Substituting the perturbed metric into the Einstein-Hilbert action and keeping terms up to quadratic order, one obtains the action
\bea
S= \frac{V}{2}  \sum_{\vec{k}} \int d\eta \left[ \hat{v}_{\vec{k}}' \, \hat{v}_{-\vec{k}}'
-\left(k^2-\frac{a''}{a}\right)\hat{v}_{\vec{k}}\, \hat{v}_{-\vec{k}}\right],
\eea
where we have introduced the Mukhanov variable  $\hat{v}_{\vec{k}}$ given by
\bea
\hat{v}_{\vec{k}}\equiv \sqrt{2 \epsilon} \, a \, M_P \,  \hat{\zeta}_{\vec{k}}
\eea
and $M_P$ is the reduced Planck mass, $M_P^2=(8\pi G)^{-1}$. Note that $V \equiv L^3$. The wave equation takes the form
\bea
v_{\vec{k}}''+\left (k^2-\frac{a''}{a}\right)v_{\vec{k}}=0,
\eea
and the Klein-Gordon inner product on these modes reduces to $ (v_{\vec{k}} v_{\vec{k}}^*{}'-v^*_{\vec{k}} v_{\vec{k}}')V=i$.

Choosing a set of positive energy modes fixes the vacuum, as we have discussed at length. The Bunch-Davies vacuum corresponds to choosing Minkowski space behavior $v_{\vec{k}} \propto e^{-i k \eta}$ in the asymptotic past, so that
\bea
v_{\vec{k}}=\frac{e^{-i k \eta}}{\sqrt{2 k V}}  \left(1- \frac{i}{k\eta}\right)
\eea
are the positive energy modes. This vacuum is usually taken as the vacuum state for perturbations in the inflationary era. Note that Bunch-Davies vacuum happens only to be one of the many vacua for scalar fields that one can define in de Sitter space: its importance is that it matters for inflationary perturbation theory, the details of which are not our concern here.

\section{QFT in Rindler Space}

Rindler space is (a patch of) flat spacetime in the coordinate system of an accelerating, and therefore non-inertial, observer. By the principle of equivalence, this system can capture aspects of genuinely curved geometries and this is the reason to study it. In fact Unruh effect in Rindler space is a phenomenally useful prototype for many of the generic features of black holes and horizons in general.

\subsection{The Uniformly Accelerated Observer/Detector}

By a uniformly accelerated observer what we mean is {\em not} that the observer is accelerating at a uniform rate in some {\em fixed} inertial reference frame. That is ill-defined in special relativity because in the $x-t$ plane, the particle will traverse the trajectory
\bea
x \sim \frac{1}{2}t^2,
\eea
and will move outside the light cone\footnote{One can of course consider curves in the geometry that get outside the light cone, but in special relativity, we believe that such curves cannot be the time evolutions of physical particles. We want to look at the physically interesting notion of constant acceleration.}. By uniform acceleration, we mean that the particle is accelerating at a constant rate in the frame in which its instantaneous velocity is zero. In other words it is the co-moving acceleration, and it is being measured in a different inertial frame at each instant.

We would first like to know what the worldline of such an observer is. That is, what does a a fixed (lab) observer see? The reason we care about this is that we plan to construct Rindler space by foliating Minkowski space with such worldlines. In what follows, for simplicity, we will work with two dimensional Minkowski spacetime. As long as the acceleration of the particle is linear (and not centrifugal etc.) this captures everything that is of essence because the extra coordinates just add a dynamically irrelevant sphere to each point in the 2D spacetime.

The most direct way to determine the worldline is to use Lorentz transformations. Let's first forget about our problem and determine the relativistic relative acceleration formula, analogous to the famous relative velocity formulae. We know the relative velocity formula, i.e., the velocity of an object as seen in a frame moving with velocity $V$:
\bea
\frac{dx'}{dt'}=\frac{dx/dt-V}{1-V (dx/dt)}\ , \ \ \text{where} \ \ x'=\gamma(x-V \ t), \ \ t'=\gamma (t- V \ x ),
\eea
and $\gamma=\sqrt{1-V^2}$. To find the relative acceleration formula we need just to differentiate this and determine $d^2 x'/dt'^2$ and write it in terms of the $x - t$ coordinates. The result simplifies to
\bea
\frac{d^2 x'}{dt'^2}=\frac{(1-V^2)^{3/2}}{(1-V (dx/dt))^3}\frac{d^2 x}{dt^2}
\eea
With this at hand we can go back to our problem of determining the trajectory (worldline) equation in the lab frame for a uniformly accelerating (in co-moving frame) particle. So we have constant acceleration $a$ in the primed frame, and the velocity $V$ of the primed frame is the instantaneous velocity $dx/dt$ of the particle. In other words,
\bea
V=\frac{dx}{dt}, \ \ \text{and} \ \  \frac{d^2 x'}{dt'^2}=a.
\eea
So the equation we need to solve is:
\bea
\frac{d^2 x}{dt^2}=\Big(1-(\frac{dx}{dt})^2\Big)a.
\eea
Despite being a second order non-linear ODE, this is easy to integrate: one first defines a velocity $v \equiv dx/dt$ and notices that $d^2 x/dt^2=v \ dv/dx$ to exchange the independent variable from $t$  to $x$. Assuming $v=0$ at $t=0$ and shifting the origin of $x$ appropriately, the end result is
\bea
x^2-t^2=\frac{1}{a^2},
\eea
so the uniformly accelerating particle lives on a hyperboloid in Minkowski spacetime.

It is instructive to parametrize this worldline using the proper time of the accelerating observer. This means we want to find $x(\tau), \ t(\tau)$ such that
\bea
x(\tau)^2-t(\tau)^2=\frac{1}{a^2}, \ \ \text{and} \ \ dt^2-dx^2=d \tau^2,
\eea
These two equations can be simultaneously solved with an appropriate initial condition to give
\bea
x(\tau)=\frac{1}{a} \cosh(a \tau), \ \ t(\tau)=\frac{1}{a} \sinh(a \tau)
\eea
From this structure it is easy to see that the time translation as seen by the accelerating detector are the boosts $\beta$ of conventional Minkowski coordinates:
\bea
\left( \begin{array}{c}
t(\tau)\\
x(\tau)
\end{array}
\right)
\rightarrow
\left(
\begin{array}{cc}
\cosh\beta & \sinh\beta \\
\sinh\beta & \cosh\beta
\end{array}
\right)
\left(
\begin{array}{c}
\frac{1}{a} \cosh(a \tau) \\
 \frac{1}{a} \sinh(a \tau)
\end{array}
\right)
=\left( \begin{array}{c}
t(\tau+\beta)\\
x(\tau+\beta)
\end{array}
\right)
\eea
which just shifts the proper time by the boost parameter.

\subsection{Rindler Geometry}

Now, we turn to a description of (a patch of) Minkowski space in terms of a coordinate system natural from the perspective of uniformly accelerated observers. This is the Rindler geometry. Basically, we will foliate this patch with the worldlines of uniformly accelerated observers from the last section (with different accelerations).

What is a natural coordinate system from the perspective of accelerating particles?
\begin{figure}
\begin{center}
\includegraphics[
height=0.38\textheight
]{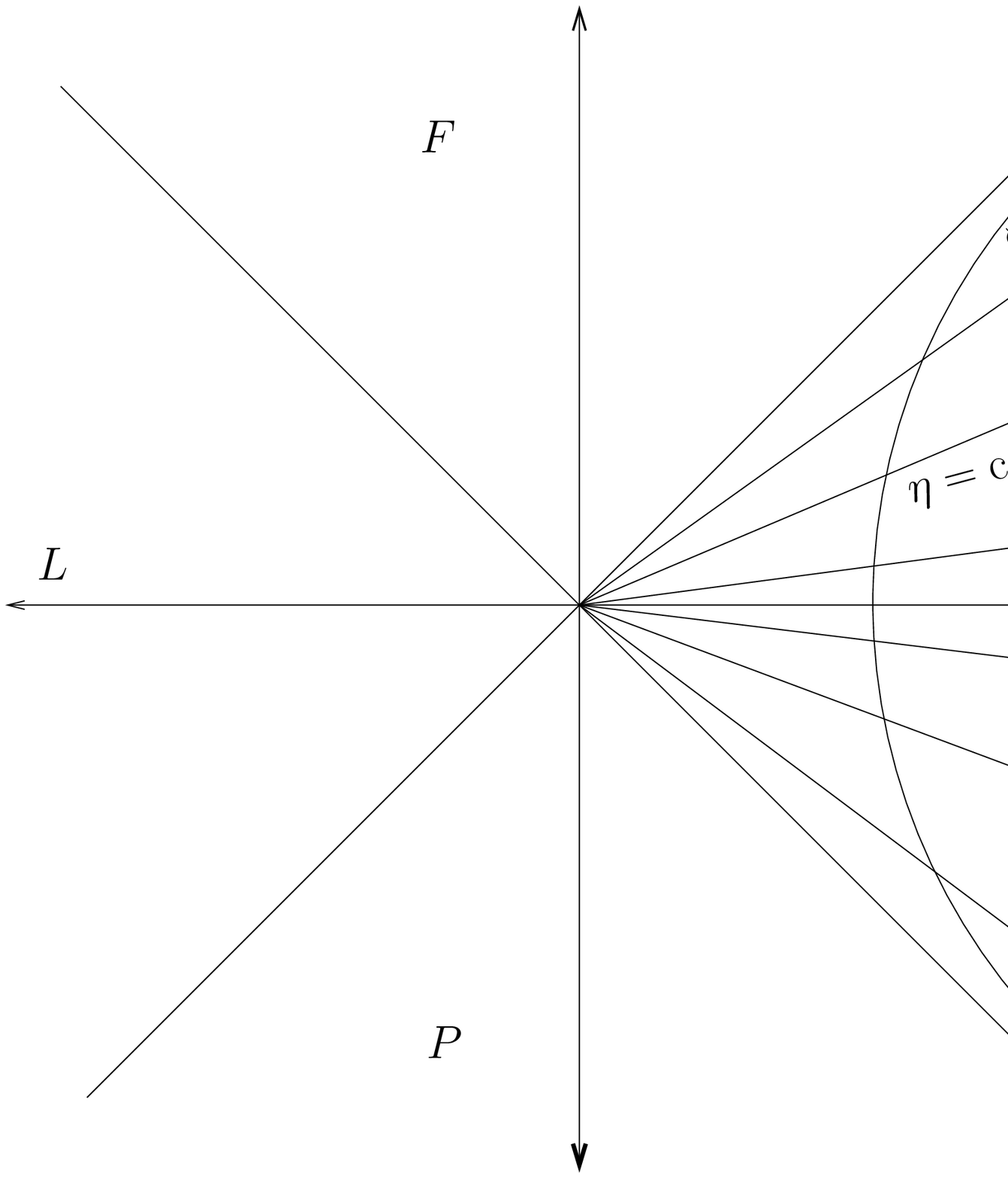}
\caption{Rindler geometry is (the right wedge of) Minkowski space, as foliated by constant acceleration trajectories. The hyperboloid denotes a constant acceleration foliation and captures the Rindler spatial coordinate $\xi$. Rindler time along it is measured by $\eta$.}
\label{curves}
\end{center}
\end{figure}
It is clear that on any worldline the natural time is the proper time. The acceleration is fixed on any worldline, so that works as a spatial coordinate\footnote{Note that in a natural coordinate system adapted to the observer, the position coordinate doesn't change. The proper acceleration has precisely this property.}. So we choose
\bea
x=\xi \cosh \eta, \ \ t = \xi \sinh \eta.
\eea
Note that $\xi$ has the interpretation of inverse proper acceleration, and $\eta$ looks like proper time on the worldline with that acceleration. The flat metric in these coordinates is
\bea
ds^2=\xi^2 d\eta^2-d\xi^2.
\eea
Note that for $\xi$ positive, which corresponds to positive norm for the acceleration, this spans only the patch $u<0, \ v>0$ of Minkowski space, where $u$ and $v$ define the lightcone of flat space:
\bea
u= t-x = -\xi e^{-\eta}, \ \ \text{and} \ \ v=t+x=\xi e^{\eta}.
\eea
Note that one can also define such a null coordinate system $(U,V)$ in terms of Rindler variables:
\bea
R: \ \ u=-e^{-U}, v=e^{V}, \ \ \text{so that}, \ \ ds^2=e^{V-U}dUdV.
\eea
This exponential relation between two coordinate systems is a standard relation when the geometry has a horizon and in fact the thermal features of black hole horizons also ultimately can be traced back to this kind of an exponential: there, the exponential relates Schwarzschild coordinates to Kruskal coordinates, here it relates Rindler to Minkwoski\footnote{We mention that our use of $(u,v)$ and $(U,V)$ are out of sync. between the black hole case and Rindler case. If we were to stick to the typical black hole notation for Rindler as well, it would have been more logical to call $t-x$ as $U$ instead of $u$, for example. Since this seems fairly standard, we will do this anyway at the risk of some confusion.}. There the horizon is the event horizon of the black hole while here, it is the acceleration horizon associated to the fact that the regions beyond the null rays are causally disconnected. 

As we saw, the Rindler patch/wedge does not span all of Minkowski space. This in itself is not a problem for quantization since the Rindler wedge is actually globally hyperbolic: the surfaces $\eta$=const. are Cauchy. But we could ask the question how the vacuum state of Minkowski looks in terms of Rindler states. Note that this is a physically meaningful question (``What does an accelerating particle see?") so we want to formulate the Rindler problem in a way that this problem has an answer. The trouble is that this problem cannot be answered satisfactorily, the way we have posed it, because modes defined on the Rindler patch do not form a complete set in which we can expand a general solution of the Minkowski Klein-Gordon equation. This is obvious because the $\eta=$const. line, while being a full Cauchy surface for the Rindler wedge is not a Cauchy surface of Minkowski. So to answer this problem, we need to find a ``natural" way to extend the Rindler wedge to other regions of Minkowski space, and define appropriate modes there so that together these modes can form a complete set for Minkowski KG solutions.

The hint is provided by the fact that even though the Rindler wedge is globally hyperbolic, it is not geodesically complete. The way to see this is to note that a vertical line connecting $v$ and $u$ axes has finite proper time (this is easily checked in the flat, i.e. non-Rindler, coordinates and should be the same in Rindler as well because it is {\em proper} time). So just like an exterior Schwarzschild observer can figure out that there are regions inside the horizon by computing the proper time of in-falling trajectories and finding that they are finite, the Rindler observer also knows that there is life beyond Rindler. So he can analytically extend the coordinates to these other regions. The coordinates that are regular under such continuations are the $u, v$ coordinates defined before: these are simply Minkowski in a light-cone form, and nothing special is happening at the axes except that they are passing through zero. As we emphasized, Minkowski for Rindler is the analogue of Kruskal for Schwarzschild. 
On the contrary, Rindler coordinates $(\eta, \xi)$ cease being useful at the horizons $u=0$ and $v=0$ because $\eta \rightarrow \pm\infty$ and $\xi \rightarrow \infty$. From our knowledge of the signs of Minkowskian $u$ and $v$ in various regions of the lightcone, we can write down the appropriate extension of the Rindler patches as
\bea
F: \ \  (u,\ v)=(e^{-U}, e^V), \ \ L: \ \ (u,\ v)=(e^{-U}, -e^V), \ \ P: \ \ (u,\ v)=(-e^{-U}, -e^V),
\eea
with
\bea
U=\eta-\ln \xi, \ \ V=\eta+\ln \xi
\eea
in all four regions $R, F, L$ and $P$. Together these cover the entire Minkowski space.

An important feature of this analytic extension that we will use later is that time runs ``backward" in the region $L$: from the above expressions it is easily seen that $t \sim -\xi \sinh \eta$ in $L$. This means that on a fixed $\xi$ foliation, $\eta$ increases in the opposite direction as that of the Minkowskian future direction.

\subsection{Unruh's Analytic Continuation Argument}

Now we turn to the quantization on Rindler/Minkowski space such that we can ask the question alluded to in the previous section. What does the Minkowski vacuum look in terms of Rindler states? The question requires us to expand a general Minkowski state (expanded in the usual flat space basis)  in terms of a full basis constructed in terms of Rindler coordinates. Then, using the Bogolubov transformation to relate the expansions, we can draw conclusions about (say) one vacuum in terms of the other.

The Minkowski part of this story is easy, and we already know the answer. The general KG solution can be expanded in the usual plane wave modes in terms of the creation/annihilation operators defined on the Minkowski vacuum $| 0 \rangle_M$. We will call these modes $u_k^M$ and the creation/annihilation operators $a$ and $a^\dagger$.

Rindler has a timelike Killing vector $\eta$, so the general solution can be taken in the form
\bea
v_k^R
\sim e^{-i \omega_k \eta}. 
\eea
We won't explicitly solve for the $\xi$-dependent part 
because we won't need it.
The superscript stands for the right wedge, we will momentarily introduce modes on the left wedge as well. Together they span the whole Minkowski spacetime. On the left wedge, the modes are similar, but with the crucial difference that the positive frequency modes are
\bea
v_k^L
\sim e^{i \omega_k \eta} 
\eea
This is because propagation is towards the future in Minkowski $t$, which is directed oppositely to $\eta$ in the left wedge.
In other words:
\bea
v^{R}_{k} \sim \left\{ \begin{array}{ll}
                    e^{-i\omega_k \eta}
&\quad \mbox{R}\\
          0 & \quad \mbox{L}
                \end{array}\right. \\
v^{L}_{k} \sim \left\{ \begin{array}{ll}
                    0
&\quad \mbox{R}\\
          e^{+i\omega_k \eta} & \quad \mbox{L}
                \end{array}\right.
\eea
The scalar can be expanded either as
\bea
\phi=\sum_k (u_k^M a_k + u_k^{M*} a_k^\dagger) \label{Mbasis}
\eea
in terms of the Minkowski modes or as
\bea
\phi=\sum_k (v_k^R b_k^R + v_k^{R*} b_k^{R\dagger}+v_k^L b_k^L + v_k^{L*} b_k^{L\dagger}) \label{Rbasis}
\eea
in terms of the combined left and right Rindler modes. Creation/annihilation operators on Rindler, we denote by $b, b^\dagger$. Comapring the structure of the Minkowski and Rindler expansions, we note the important point: there are two separate Hilbert spaces in the Rindler picture. They correspond to the left wedge and the right wedge, characterized by the fact that they are annihilated by $b_k^L$ and $b_k^R$ respectively. As we noted before, we need both to construct a general solution in the full Minkowski space because a Cauchy slice of Minkowski goes through both. This means that our Bogolubov transformation will relate the Minkowski vacuum $|0\rangle_M$ to $|0\rangle_R \equiv |0\rangle^R \otimes |0\rangle^L$.  The upper $R$ (on the RHS) stands for the right hand wedge, while the lower $R$ (on the LHS) stands for Rindler, this should not cause any confusion in what follows. Note in particular that this means that (for example) $b_k^L$ should be understood as $ \I^R \otimes b_K^L$, where $\I^R$ is the identity operator acting on the right Hilbert space. We have suppressed this above and in what follows, to avoid clutter.

The Bogolubov transformation that we are looking for relating Minkowski and (the doubled) Rindler can therefore be written adapting (\ref{Bog2}) as
\bea
a_i=\sum_j (\alpha_{ij}^R b_j^R+\beta_{ij}^{R*}b_{j}^{R\dagger}+\alpha_{ij}^L b_j^L+\beta_{ij}^{L*}b_{j}^{L\dagger}), \ \ {\rm etc.}
\eea
where
\bea
\alpha^{R*}_{ji}=(u_i^M,v^R_j), \ \ \text{and so on.}
\eea
These are essentially Fourier transforms of the $v^R$ and $v^L$ because the Minkowski modes are plane waves. This is a complicated integral and it might seem that this is going to get messy, if at all the integrals are doable\footnote{As it turns out, the integrals are indeed messy, but doable.}.

But Unruh has shown that indirect arguments can go a long way. Since this argument introduces many useful ingredients for our understanding of similar physics on black holes, this is the path we will pursue. The basic idea is that we do not need to work with $u_k^M$ necessarily: we can work with any complete set of positive frequency modes defined with respect to Minkowski time, and the Minkowski vacuum would be the same. Now, any positive energy solution has an expansion as an integral over the momenta of the form
\bea
\int \frac{d^3k}{(2 \pi)^3 2 k^0} \tilde f (k) e^{-ik. x}. \label{pos}
\eea
If one thinks in the complex plane, this means that an equivalent definition of positive frequency solution is as a solution that is analytic and bounded in ${\rm Im} \ t <0$.

So if we can construct positive energy Minkowski modes as linear combinations of Rindler modes (possibly both positive and negative frequency), then we have our Bogolubov transformation. In other words, we want to construct modes that are regular everywhere in the lower half $t$ plane. One helpful intermediate step is to note that an equivalent notion of positive energy is to have regularity for both ${\rm Im} \ u<0$ and ${\rm Im} \ v<0$. This is because we can write
\bea
e^{-i(\omega t - k x)}=e^{-\frac{i}{2} (\omega+k)u}e^{-\frac{i}{2} (\omega-k)v}
\eea
Note that both $(\omega+k)$ and $(\omega-k)$ have the same sign\footnote{Note: $\omega=\sqrt{k^2+m^2}$, and that $(\omega+k)$ and $(\omega-k)$ stand for $(\omega+k_z)$ and $(\omega-k_z)$ and so ${\rm sign}(\omega)={\rm sign}(\omega+k_z)={\rm sign}(\omega-k_z)$. In the massless case, it is useful to remember that $k$ is actually $\sum k_i^2$ over the transverse coordinates, so semi-definite signs can show up (think 1+1 dimensional case as an example), but this can be thought of as a limit where one adds a mass term and then takes the mass $\rightarrow 0$ limit at the end of the computation. 
}, in particular they are both positive for positive frequency modes. From the fact that the integral over $(\omega+k)$ and $(\omega-k)$ treated as independent variables in (\ref{pos}) has to converge, we know that for positive frequency modes, we need regularity in both ${\rm Im} \ u<0$ and ${\rm Im} \ v<0$.

\begin{figure}
\begin{center}
\includegraphics[
height=0.3\textheight
]{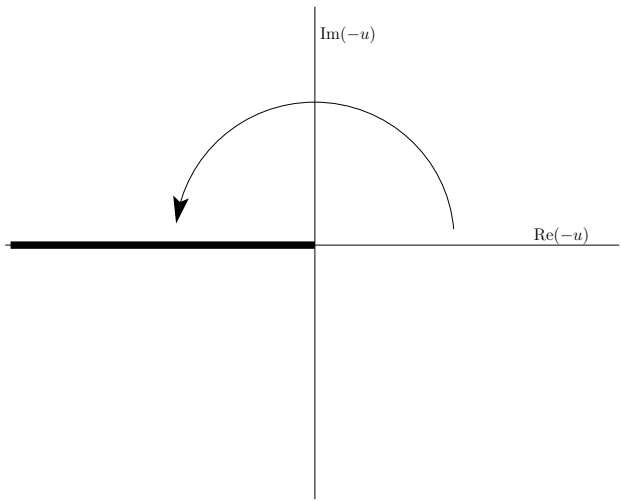}
\caption{Unruh's analytic continuation in the $-u$ plane. Solid line is the branch cut of $\ln(-u)$.}
\label{UnruhC}
\end{center}
\end{figure}

With these preliminary comments, we first make the observation that if there was an analytic continuation of the positive frequency Rindler R-modes through the lower half $u$-plane (and $v$-plane) to the positive frequency Rindler L-modes, then that would mean that they are both comprised entirely of Minkowski positive frequency modes. But this is not the case as we now show. In region R ($u<0, v>0$), from our earlier construction of the analytic extension of Rindler
\bea
\eta=\frac{1}{2}(\ln \ v-\ln (-u))
\eea
while in region L ($u>0, v<0$),
\bea
\eta=\frac{1}{2}(\ln(-v)-\ln (u))
\eea
Lets focus first on the analytic continuation in $u$. 
In the R-wedge (which has $u<0$), the positive energy Rindler mode is
\bea
e^{-i\omega \eta} \sim e^{\frac{i}{2}\omega \ln(-u)}.
\eea
Since the log is well-defined for positive arguments, there is no ambiguity in evaluating this on the R-wedge. But on the other hand, in the L-wedge, the precise analytic continuation will affect the value because the value of the log will depend on whether it is evaluated above or below its branch cut. Since what we want to do is to 
make sure that the result is positive frequency in Minkowski modes, that means that we have to think of the result as the real boundary value of a function analytic in ${\rm Im} \ u < 0$, or which is the same, ${\rm Im}(-u) > 0$. This means  that the log should be approached from above its branch cut: $\ln(-u)=\ln (u)+i\pi$. See figure \ref{UnruhC}.
Therefore the contribution from the $u$-piece to the positive energy Minkowski mode restricted to the L-wedge is
\bea
e^{-\frac{\pi \omega}{2}}e^{\frac{i}{2}\omega \ln u} 
\eea
A similar factor arises from the analytic continuation in $v$.
Together then the positive frequency piece on L-wedge takes the form
\bea
e^{-\pi \omega}(e^{-i\omega\eta})^* \sim e^{-\pi \omega} v^{L*}_{k},
\eea
where in the last step we have re-expressed the result in terms of Rindler L-modes.
All this means that we can write new positive energy modes for Minkowski vacuum in terms of Rindler modes as
\bea
U_k^I=\sqrt{\frac{e^{\pi \omega_k}}{2 \sinh \pi \omega_k}}
\Big(v^R_k+e^{-\pi \omega_k}v^{L*}_k\Big),
\eea
where we have normalized it appropriately using the Klein-Gordon inner product of the Rindler modes. Note, as we observed in a previous section, that while $(u,u) \sim +1$, for negative frequency modes $(u^*,u^*) \sim -1$. This is crucial in getting this specific form for the normalization: we have used $\sinh x=\frac{e^x-e^{-x}}{2}$.

It is important to note that this is not yet a full basis of Minkowski positive frequency modes. This is because we got these modes by combining $v^R_k$ and the analytic continuation of  $v^R_k$ (even though we expressed the latter in terms of $v^{L*}_k$), while from (\ref{Mbasis}, \ref{Rbasis}) and the fact that Minkowski is spanned by the two wedges together, we should expect a doubling. This is basically the statement that by Rindler vacuum, we mean the tensor product of the vacua on the two wedges. To get these other positive modes one can start from the L-modes and do the analytic continuation just as we did for R-modes. The result is
\bea
U_k^{II}=\sqrt{\frac{e^{\pi \omega_k}}{2 \sinh \pi \omega_k}}
\Big(v^L_k+e^{-\pi \omega_k}v^{R*}_k\Big).
\eea
Together, these span the positive frequency Minkowski KG solutions, and we have the Bogolubov transformation. The $U_k^I$ and $U_k^{II}$ together are equivalent to the $u_k^M$ discussed at the beginning of this section. The negative frequency solutions just follow by complex conjugation and add no extra information.

\subsection{Emergence of Thermality: The Density Matrix}

Comparing the structure of these Bogolubov transformations with those presented in Section 3, we can write down the ``S-matrix" that relates the vacua.  First we write the Bogolubov transformations of the last subsection in a matrix form
\bea
\left(
\begin{array}{c}
U_k^{I}\\
U_k^{II}\\
U_k^{I*}\\
U_k^{II*}
\end{array}
\right)=
\sqrt{\frac{e^{\pi \omega_k}}{2 \sinh \pi \omega_k}}
\left(
\begin{array}{cccc}
1&0&0&e^{-\pi \omega_k} \\
0&1&e^{-\pi \omega_k}&0 \\
0&e^{-\pi \omega_k}&1&0 \\
e^{-\pi \omega_k}&0&0&1
\end{array}
\right)
\left(
\begin{array}{c}
v_k^R\\
v_k^{L}\\
v_k^{R*}\\
v_k^{L*}
\end{array}
\right)
\eea
We want to write the S-matrix (really, the U-matrix) as
\bea
|0\rangle_M=U  |0\rangle_R.
\eea
as before. Note that in the notation of section 2, it means that it is more convenient to think of the Bogolubov transformations above as acting on the left. So we have the $\alpha$ and $\beta$ matrices as
\bea
\alpha_{\omega_k}^{\dagger}=\sqrt{\frac{e^{\pi \omega_k}}{2 \sinh \pi \omega_k}}
\left(
\begin{array}{cc}
1&0\\
0&1
\end{array}
\right), \ \ \
\beta_{\omega_k}^{\rm T}=\sqrt{\frac{e^{-\pi \omega_k}}{2 \sinh \pi \omega_k}}
\left(
\begin{array}{cc}
0&1\\
1&0
\end{array}
\right),
\eea
Plugging this into (\ref{S-matrix}), one finds
\bea
|0\rangle_M&=&\prod_{k'}\sqrt{2 e^{-\pi \omega_k'}\sinh \pi \omega_k'}
 \exp\Big(\sum_{k}e^{-\pi \omega_{k}}b_{k}^{R \dagger} b_{k}^{L \dagger}\Big)  |0\rangle_R \\
&=&\prod_{k}
\sqrt{2 e^{-\pi \omega_k}\sinh \pi \omega_k}
\sum_{n_k=0}^\infty e^{-\pi \omega_{k}n_k} |n_k\rangle^R \otimes  |n_k\rangle^L
\eea
where we have used the fact that Rindler vacuum is the tensor product of the left and right vacua as discussed before. An observer on the right wedge has no access to the left wedge, so he is appropriately described by a density matrix where the left wedge states are traced over:
\bea
\rho_R=\sum_L |0\rangle_M \langle 0|_M=\prod_{k}
2 e^{-\pi \omega_k}\sinh \pi \omega_k 
 \sum_{n_k=0}^\infty e^{-2 \pi \omega_{k}n_k} |n_k\rangle^R  \langle n_k|^R
\eea
Note that a density matrix for the canonical ensemble is of the form
\bea
\rho \sim \sum e^{-\beta E_k} |E_k\rangle \langle E_k|,
\eea
so what we have here is a ${\rm Tr} \rho =1$ normalized thermal density matrix at a temperature $T=1/\beta= 1/2 \pi$. But note that this temperature is with respect to the Rindler coordinate $\eta$ which is related to the proper time via $\eta = a\tau$ as we noted earlier. So the temperature with respect to the proper time is $a/(2\pi)$. In terms of the coordinate $\xi$ which captures the inverse acceleration, this becomes $1/(2\pi \xi)$.

The conclusion  therefore is that a particle with a proper acceleration $a$ in empty flat space should see an isotropic flux of thermal radiation at a temperature $a/2 \pi$. This is the Unruh effect and it captures the essence of quantum field theory in curved spacetime. Note that the tracing over the unobservable Hilbert space was crucial to get the thermal density matrix, so in any situation where horizons are important, we expect thermality to arise.

\section{The Unreasonable Effectiveness of the Complex Plane}
\label{Green}

The thermality of Rindler is ultimately a consequence of the fact that it had a horizon. By the principle of equivalence, we might expect to see thermal features in general spacetimes with horizons. This indeed seems to be the case, even though we won't explore this in much detail (see Gibbons-Hawking paper for a general discussion of thermodynamics of horizons like de Sitter.) What we will do is to give more perspective on the remarkable emergence of thermality in what seemed at the beginning to be zero temperature quantum field theory. To do this, we first investigate some general features of propagators in a thermal environment. We will see that the emergence of thermality has a connection with Euclideanizing spacetime (at least for static spacetimes).

\subsection{KMS condition}

We first start with correlation functions of a Harmonic oscillator at finite temperature. That is, we consider the quantum harmonic oscillator in the canonical ensemble.  The expectation value of the occupation number $N=a^\dagger a$ takes the form
\bea
\langle n \rangle_\beta =\frac{{\rm Tr}(e^{-\beta H}N)}{{\rm Tr}(e^{-\beta H})}\equiv \frac{\sum_{n} n\ e^{-\beta n\omega}}{\sum_{n} e^{-\beta n \omega} } = \frac{1}{e^{\beta \omega}-1}.\label{occu}
\eea
We ignore the zero point energy in the Hamiltonian so that $H=N \omega$.
This is the thermal expectation value of the harmonic oscillator occupancy at a temperature $T= 1/\beta$ in the canonical ensemble. (This basically follows from the definition of the canonical partition function as a Boltzmann-weighted sum over states.)

Now, lets look at the Green function for the Harmonic oscillator at finite temperature. This is given by
\bea
G_+^\beta(t)=\langle x(t) x(0) \rangle_\beta
\eea
where we have used time-translation invariance to shift the origin of time. As a Heisenberg operator
\bea
x(t)=\frac{1}{\sqrt{2 \omega}}[e^{-i\omega t} a+e^{i\omega t} a^\dagger].
\eea
The normalization is a convention for the Fourier transforms, which generalizes nicely when we go to relativistic field theory. By explicit computation, then, this thermal correlator takes the form
\bea
& G_+^\beta(t)=\frac{1}{2\omega}\langle e^{-i\omega t} a a^\dagger+e^{i\omega t} a^\dagger a \rangle_\beta \\ &=\frac{1}{2\omega}[e^{-i\omega t} \langle n+1 \rangle_\beta+e^{i\omega t}\langle n \rangle_\beta ]=\frac{e^{\beta \omega}}{2 \omega (e^{\beta \omega}-1)}(e^{-i\omega t}+e^{-\beta \omega}e^{i\omega t}).
\eea
We have used the creation-annihilation algebra as well as (\ref{occu}). This captures the time correlations of a harmonic oscillator at finite temperature.

A basic observation is that
\bea
G_+^{\beta}(t-i\beta)=G_+^\beta(-t) \equiv G_-^\beta(t).
\eea
This is called the Kubo-Martin-Schwinger (KMS) condition. We did this for the Harmonic oscillator because there are not too many irrelevant indices. But the statement is quite general, and can (obviously) immediately be applied to perturbative field theory as well by writing the harmonic oscillator in fancier ways. More interestingly, KMS condition can actually be shown to hold {\em formally} for {\em any} time invariant quantum mechanical system in thermal equilibrium, without even resorting to a weak-coupling oscillator description. Consider the definition of the thermal expectation value, applied to the two point operator $A(t) B(0)$:
\bea
\langle A(t) B(0) \rangle_\beta \equiv \frac{1}{Z}{\rm Tr} (e^{-\beta H}A(t) B(0))
\eea
From the Heisenberg equation of motion $A(t)= e^{+i Ht} A(0) e^{-iHt}$ and the cyclicity of the trace, it is straightforward to show that
\bea
\langle A(t) B(0) \rangle_\beta = \langle A(0) B(-t) \rangle_\beta, \ \ \langle A(t-i\beta) B(0) \rangle_\beta = \langle  B(-t) A(0) \rangle_\beta.
\eea
The first result means that in thermal equilibrium there is time-translation invariance and the second result is the general KMS condition. Note that the KMS  condition is a formal thing because we applied the Heisenberg EOM to define the operator at a complex time: $A(t-i\beta)$.

In practice, we will often take the fact that an operator expectation value satisfies the KMS condition as indicative of the fact that the expectation value is being taken in a thermal state. Effectively what we are saying is that a thermal state is defined as a state $| 0\rangle_\beta$ where
\bea
\langle 0| X | 0\rangle_\beta=\frac{1}{Z}{\rm Tr} (e^{-\beta H}X)
\eea

\subsection{Green Functions}

Note that a trivial generalization of the harmonic oscillator to the scalar field can be defined:
\bea
G_+^\beta(t,{\bf x},{\bf y})=\langle \phi(t, {\bf x}) \phi(0,{\bf y}) \rangle_\beta
\eea
Plugging in the mode expansion from section 3 for a system with time-translation invariance (we assume it), we end up with an analogous structure to the harmonic oscillator:
\bea
G_+^\beta(t,{\bf x},{\bf y}) =\sum_k \ {\rm (stuff)} (e^{-i\omega_k t}+e^{i\omega_k t}e^{-\beta \omega_k})
\eea
we have suppressed the spatial dependence.
Assuming that the convergence is dictated by the exponential pieces in the sum\footnote{While this is indeed a good assumption for many real-world systems, it is not a guarantee: if the number of states diverges sufficiently fast (i.e., faster than exponential) the sum need not converge. This is basically the Hagedorn transition and happens for example in string theory where there are lots of states at high levels.}, this converges only in the strip $-\beta < {\rm Im}(t) < 0$. Note that the usual field theory two-point function converges for all ${\rm Im}(t) < 0$, because it is the zero temperature limit where $\beta \rightarrow \infty$. In this limit $G_+^\beta(t)$ reduces to a sum over positive energy modes, and for this reason $G_+^\beta(t)$ is called the positive frequency Wightman correlator.  One can also define the negative frequency Wightman correlator as
\bea
G_-^\beta(t,{\bf x},{\bf y})=\langle \phi(0, {\bf x}) \phi(t,{\bf y}) \rangle_\beta=G_+^\beta(-t,{\bf x},{\bf y})
\eea
which is convergent in the the region $0< {\rm Im}(t) < \beta$. 

An important ingredient at this point is the observation that by direct computation
\bea
G_+^\beta(t,{\bf x},{\bf y})-G_-^\beta(t,{\bf x}, {\bf y})= [\phi(t,{\bf x}),\phi(0,{\bf y})]
\eea
which is independent of $\beta$, and therefore is the same as at zero temperature. But we know that the zero-temperature commutator is zero outside the light cone.
\begin{figure}
\begin{center}
\includegraphics[
height=0.3\textheight
]{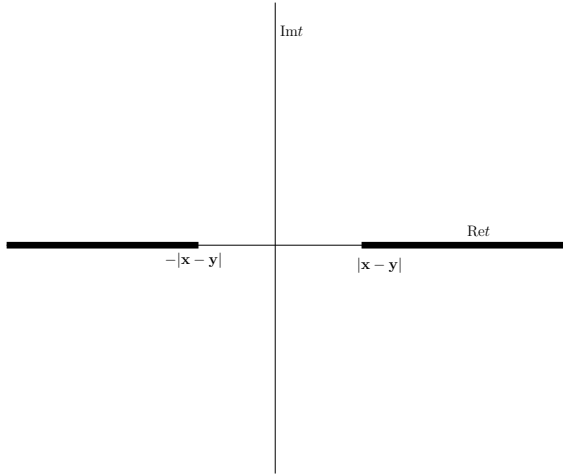}
\caption{Analyticity structure of zero-temperature Green functions. $G_-(t,{\bf x}, {\bf y})$ is analytic in the upper half $t$-plane and $G_+(t,{\bf x}, {\bf y})$ is analytic in the lower half plane. The difference between the two on the real $t$-axis is the commutator (see main text). Therefore when the commutator is non-zero, there is a branch cut on the real $t$ axis. But when it is zero, one can analytically continue $G_+$ to $G_-$. This happens when $|t| < |{\bf x}-{\bf y}|$, corresponding to correlators computed in the region outside the lightcone.}
\label{Unruh}
\end{center}
\end{figure}
This means that if we think of these Green functions as functions of $t$ at fixed ${\bf x}$ and ${\bf y}$ there is some range of small enough $t$ where the spacelike separation between ${\bf x}$ and ${\bf y}$ is bigger than $t$ and we are outside the light cone. That is, on the real $t$ axis, near the origin there is a strip where $G_+^\beta(t,{\bf x},{\bf y})=G_-^\beta(t,{\bf x},{\bf y})$. For bigger values of real $t$, there is a branch cut and the difference is captured by the commutator (which is non-vanishing now since we are inside the light cone). So, $G_-^\beta(t,{\bf x},{\bf y})$ is the analytic continuation to the upper strip of $G_+^\beta(t,{\bf x},{\bf y})$ on the lower strip. Together with the KMS condition, which says
\bea
G_+^\beta(t-i \beta)=G_-^\beta(t)
\eea
this means that we can continue $G_+^\beta(t)$ to everywhere on the complex plane, except for the periodic cuts in the imaginary direction. (We will suppress the spatial coordinates when it causes no confusion.)
\begin{figure}
\begin{center}
\includegraphics[
height=0.3\textheight
]{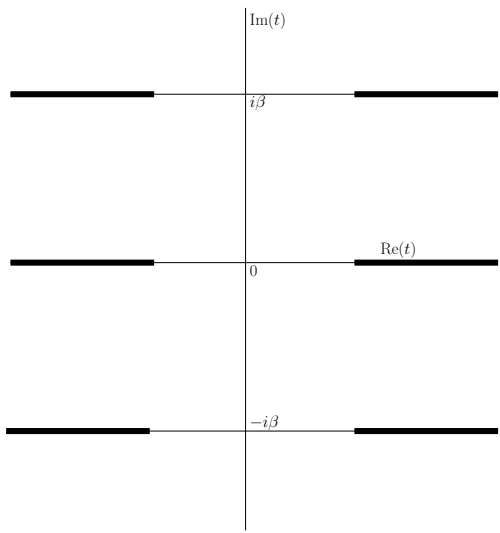}
\caption{The finite temperature analog of the previous figure. At finite temperature, i.e., $\beta < \infty$, there are new branch cuts that have moved in from infinity, and they are periodic according to the KMS condition. Close to the real axis, the analyticity structure and the continuations are as in the zero temperature case: the strips of analyticity of $G_+^\beta (t)$ and $G_-^\beta (t)$ alternate from there. All of these can be determined as the boundary values close to the cuts of the a unique function that is defined on the imaginary $t$ axis.}
\label{Unruh}
\end{center}
\end{figure}

The real time Wightman functions are boundary values of this analytically continued function on the complex plane. In fact we can also think of the Feynman Green functions in this way. From the definition
\bea
i G_F^\beta(t)=\theta(t) G_+^\beta(t)+\theta(-t)G_-^\beta(t)
\eea
it is immediate that it corresponds to the specific boundary value of the complex Green function where we approach the real axis from below on the left half plane and from above on the right half plane.
Note also that all these constructions work as long as KMS condition holds and the operators commute at spacelike separations.

What we have discussed above is often called real-time thermal field theory. The interpretation is that the ``time" has a real and imaginary part: the real part captures the dynamics, in a thermal background captured by the imaginary part. The idea is that a field theory is coupled to a canonical thermal bath, and we want to compute its dynamics.

\subsection{Euclidean Field Theory: An Alternate Derivation of the Unruh Effect}

In the last subsection we found that the complexified two point function carries a lot of information about time dependence and thermal equilibrium. The crucial point was that we were continuing around the cuts rather than under them to another sheet.

Since the analytic continuation is unique, we can define the function on the imaginary line and then analytically continue from there. It turns out, remarkably, that a Euclidean Green function defined for the Euclidean Klein-Gordon equation is the appropriate object that when continued to the real line, gives rise to the correct Lorentzian Green functions. In fact the Green functions of Euclidean theory are unique: if one specifies that the solution falls off at Euclidean infinity, then the Green function is fixed. From there, one can analytically continue to Lorentzian  space. At zero temperature, this is the familiar fact that one can define the Lorentzian Green functions by Wick rotating from flat Euclidean space.

What is impressive is that this construction works also at finite temperature. The difference is that now the geometry in which the Euclidean Klein-Gordon equation is defined is the Euclidean cylinder, and not flat space. This is because the time direction has gotten compactified as $\tau \sim \tau +\beta$. The claim then is that the finite temperature Wightman functions (for example) that we can compute are defined as the analytic continuation of the the asymptotically dying Green functions of Klein-Gordon on this cylinder. That this works for any static spacetime is something that we can check explicitly by writing down the Green functions in Lorentzian and Euclidean spaces  directly and analytically continuing the latter appropriately: we will not do this in detail.
This approach works for any static spacetime.

In particular, now this gives us a simple derivation of the the fact that the Unruh detector sees a thermal bath at temperature $1/2 \pi$. First, note that Euclideanized Minkowski space (which is nothing but Cartesian Euclidean space, of course) is periodic in Euclideanized Rindler time:
\bea
ds^2=-dt^2+dz^2, \ \ {\rm with} \  \ z=\xi \cosh \eta, \ \ t=\xi \sinh \eta, \ \ \\ \text{goes to} \ \ ds^2=d\tau^2+dz^2, \ \ {\rm with} \ \ z=\xi \cos \eta_E, \ \ \tau= \xi \sin \eta_E,
\eea
when we define $t=i \tau$ and $\eta=i\eta_E$. So the Euclideanized Minkowski coordinates are periodic with period $2\pi$ in $\eta_E$.
This means that the Minkowski and Rindler Green functions, when Euclideanized are related by
\bea
G_E(\tau, z)\equiv G_E(\eta_E,\xi)=G_E(\eta_E+2 \pi,\xi)
\eea
The last relation is a consequence of the periodicity above. Therefore, after the Wick rotation the Minkowski correlator has an interpretation as a thermal correlator (in Rindler time) with $\beta=2 \pi$ or temperature $1/2 \pi$. In other words, Minkowski vacuum looks like a thermal state in terms of Rindler time.

This is a structure that we will again see in the case of black holes. Two coordinate systems; the Green function in one being thermal in terms of the time of the other.

\section{QFT on Black Holes}

Now we turn to a topic that is an application of the ideas presented so far, while also being a topic of fundamental significance in quantum gravity: black holes. The defining characteristic of a black hole in classical general relativity is the presence of an event horizon. As we have already seen, quantum field theory on spacetimes with horizons leads to curious physical phenomena, but so far we saw that in the context of observer dependent horizons. In black hole spacetimes, global event horizons are expected  to be fundamental (at least in classical relativity), and the phenomena will have more drastic consequences. Among these will be the conclusion that we need to associate an entropy with black holes (which is troublesome because in relativity black holes are hairless), and that quantum evolution seemingly ceases to be unitary in black hole backgrounds. Both these problems can be thought of as hints regarding the quantum nature of gravity and one of the strongest arguments for string theory as a correct quantum gravity is that it provides plausible resolutions (of varying degrees of completeness) to these problems. Some useful facts about black holes, causal structures, geodesics and Penrose diagrams are collected in the Appendix.

\subsection{Potential Barriers: Reflection and Transmission}

\begin{figure}
\begin{center}
\includegraphics[
height=0.3\textheight
]{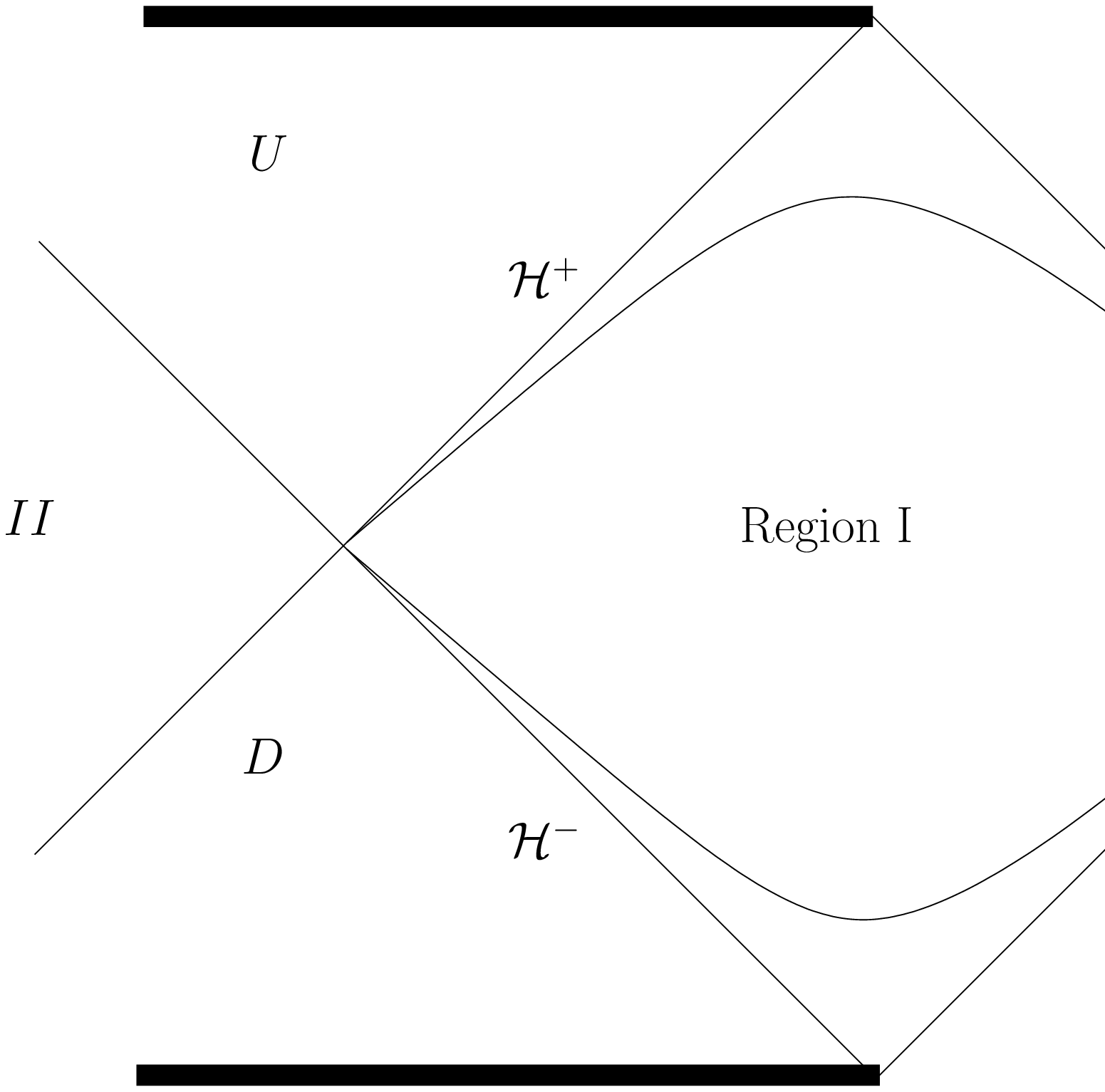}
\caption{For massless fields, region I is globally hyperbolic and its Cauchy surface can be taken to be $\cH^- \cup \cI^-$ or $\cH^+ \cup \cI^+$. They can be taken as limits of the usual spacelike Cauchy surfaces (like the ones depicted in the figure).}
\label{distortCauchy}
\end{center}
\end{figure}

We start this section with something boring and straightforward, but necessary: we discuss the wave equation in the Schwarzschild geometry and the nature of its solutions for use in the next section. The wave equation in tortoise coordinates (see Appendix) takes the form
\bea
\left(\frac{\partial^2}{\partial r_*^2}+\omega^2-V_l(r_*)\right)u_{l,\omega}(r_*)=0
\eea
where we have decomposed the scalar $\phi$ as
\bea
\phi=\sum_{l,m}\frac{u_{l,\omega}(r_*)e^{-i\omega t}}{r} Y_{lm}(\theta,\phi).
\eea
Note that the spacetime is stationary and therefore we are working with an energy basis. The spherical symmetry of the geometry ensures that only the radial part is non-trivial. We will work with the  massless scalar to keep things simple. Another advantage  of the massless case is that one can take the Cauchy surfaces of region I of Kruskal geometry to be $\cH^- \cup \cI^-$. See figure \ref{distortCauchy}. It is straightforward to write down the explicit form of the potential, but we will only need the asymptotics. The potential goes to zero both at the horizon and at infinity with a peak in the middle.
\bea
V_l(r_*)_{r_* \rightarrow -\infty} \sim \exp(r_*/2M), \ \ V_l(r_*)_{r_* \rightarrow \infty} \sim \frac{l(l+1)}{r^2}.
\eea
The wave equation in this Schrodinger form can be solved in terms of waves propagating from/to the horizon ($r_*\rightarrow -\infty$) and from/to infinity ($r_*\rightarrow \infty$). The solution in these two asymptotic regions are simple sinusoids because the potential vanishes:
\bea
r_* \rightarrow \infty: \ \  u(r_*) = A^{+}e^{i\omega r_*}+B^{+}e^{-i\omega r_*} \nn \\
r_* \rightarrow -\infty: \ \ u(r_*) = B^{-}e^{i\omega r_*}+A^{-}e^{-i\omega r_*}. \label{asympeq}
\eea
We can think of the $A$ as outgoing waves at the horizon $-$ and the boundary (i.e., infinity) $+$. When we include the time dependence back in, it is clear that they are outgoing waves in the future. Similarly, the $B$ are incoming waves (in the past). We suppress all inessential indices from now on. Now, note that by linearity, the solution that tends to  $e^{-i\omega r_*}$ (say) at the horizon should go to a linear combination of  $e^{i\omega r_*}$ and $e^{-i\omega r_*}$ at the other end. This means that there should exist a linear relation between the coefficients of the form
\bea
\left(
\begin{array}{c}
A^{+}\\
A^{-}
\end{array}
\right)=S\left(
\begin{array}{c}
B^{-}\\
B^{+}
\end{array}
\right)\equiv
\left(
\begin{array}{cc}
s_{11}&s_{12} \\
s_{21}&s_{22}
\end{array}
\right)
\left(
\begin{array}{c}
B^{-}\\
B^{+}
\end{array}
\right) \label{smatrixS}
\eea
where we will call the $2 \times 2$ matrix an S-matrix: the $B$ coefficients capture the propagation towards the potential barrier and the $A$ coefficients capture the scattered wave. Since the complex conjugate of the Schrodinger equation is also satisfied, the complex conjugate of the solutions (and therefore their asymptotic forms (\ref{asympeq})) should remain solutions. This implies the following replacements on (\ref{asympeq}):
\bea
B^{\pm} \rightarrow A^{\pm *}
\eea
Reinterpreting (\ref{smatrixS}) in terms of the new variables, we get
\bea
\left(
\begin{array}{c}
A^{-}\\
A^{+}
\end{array}
\right)=(S^*)^{-1}\left(
\begin{array}{c}
B^{-}\\
B^{+}
\end{array}
\right)
\eea
Using the fact that the S-matrix is unitary, $SS^\dagger=1$ (we will show it momentarily), this yields the relation
\bea
s_{11}=s_{22}
\eea
between its matrix elements.

Now we show the unitarity of the S-matrix. Again using the complex conjugate equation, one can show directly that
\bea
\frac{\partial}{\partial r_*}\Big(f\frac{\partial f^*}{\partial r_*}-f^*\frac{\partial f}{\partial r_*}\Big)=0.
\eea
This implies unitarity of the S-matrix:
\bea
|B^{+}|^2+|B^{-}|^2=|A^{+}|^2+|A^{-}|^2.
\eea

When one of the four coefficients $B^{\pm}, A^{\pm}$ is zero, the elements of the S-matrix are usually interpreted in terms of reflection and transmission coefficients.
We will take $B^{+}=0$ and normalize the solution so that $B^{-}=1$ in the following discussion. The natural interpretation of $A^{+}$ and $A^{-}$ in this context is as transmission and reflection coefficients respectively, and we will call them $t$ and $r$. We can write these solutions schematically as
\bea
u_{\rm past}^{\rm horizon}=t \ u_{\rm future}^{\rm boundary}+r \ u_{\rm future}^{\rm horizon}
\eea
From the definition of the S-matrix this fixes $s_{11}=t$ and $s_{21}=r$. Using unitarity and the $s_{11}=s_{22}$, we can fully fix the S-matrix to be
\bea
S=\left(
\begin{array}{cc}
t&\frac{-t \ r^*}{t^*} \\
r&t
\end{array}
\right)
\eea
Once the S-matrix is fixed using this specific choice of solution, it is fixed forever, so we can use it for computing the scattering of the solution where $B^{-}$ is turned off to zero, and $B^{+}$ is non-zero. For notational convenience, we will adopt the convention that $B^{+}=t^*/t$ instead of 1. Note that this is just a redefinition of the phase. Plugging into (\ref{smatrixS}), this gives us $A^{-}=t^*$ and $A^{+}=-r^*$. In terms of the previous schematic notation, we have
\bea
u_{\rm past}^{\rm boundary}=t^* \ u_{\rm future}^{\rm horizon}-r^* \ u_{\rm future}^{\rm boundary}.
\eea

\subsection{Hawking Radiation}

Astrophysical black holes (whether they be stellar or galactic) are believed to have formed by gravitational collapse.
\begin{figure}
\begin{center}
\includegraphics[
height=0.3\textheight
]{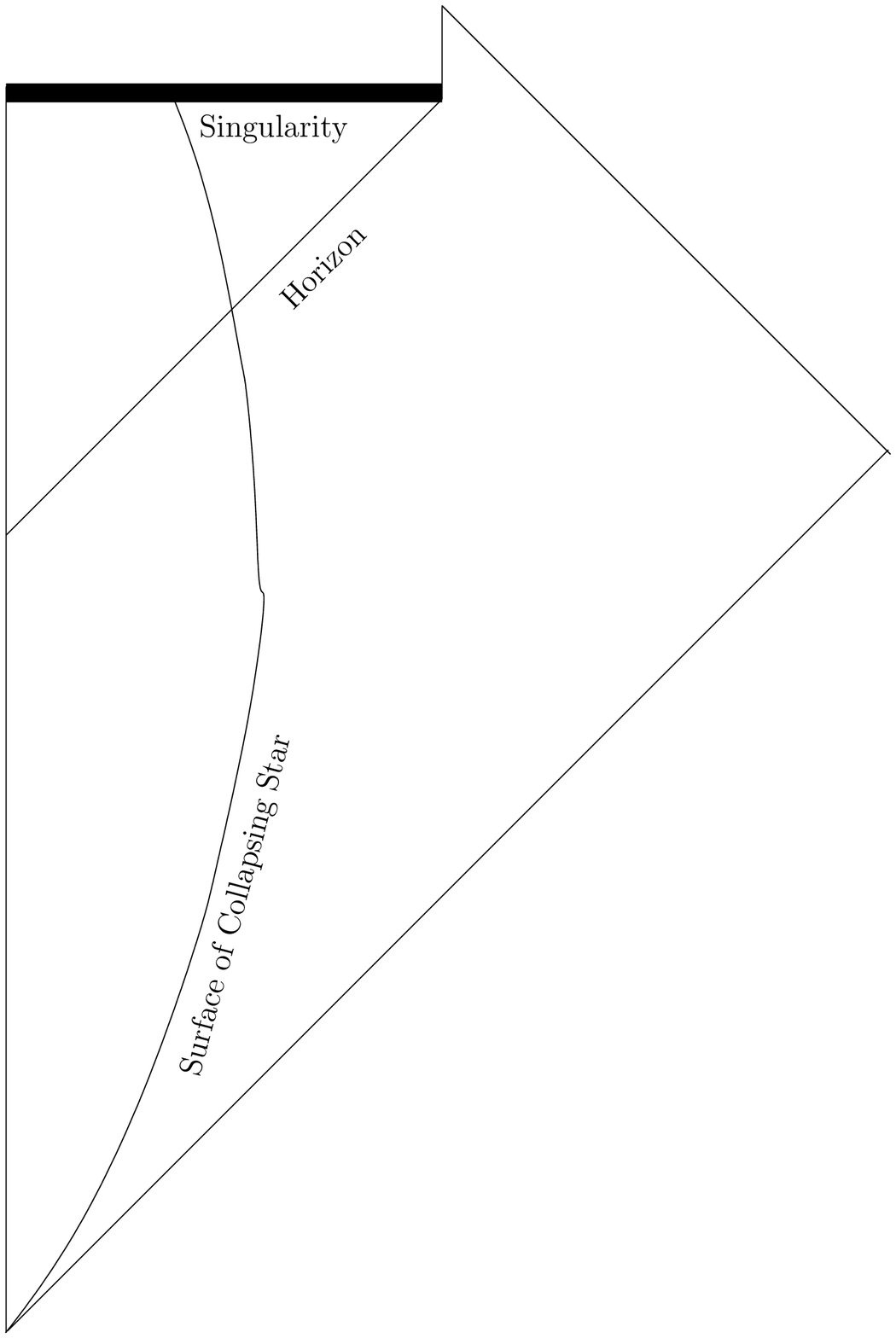}
\caption{Penrose diagram of collapsing dust.}
\label{curves}
\end{center}
\end{figure}
This means that the horizon was not eternal, and that it formed at a certain stage of the collapse. But quantum field theory on the realistic and messy context of gravitational collapse is difficult, so mostly we will be studying quantum field theory in the ``eternal" Kruskal geometry\footnote{As we will discuss, black holes in flat space radiate and lose mass, have negative specific heat, etc. so an eternal black hole is not a truly meaningful concept in flat space. 
But for the moment, we are not considering thermodynamical stability. By eternal, we basically just mean the Kruskal manifold. In asymptotically anti-de Sitter spaces, it turns out that there does exist a notion of a truly eternal black hole, see Section \ref{adscft}.}. By quantization in Kruskal, what we really mean is that we quantize fields in region I of the Kruskal geometry. Note that this is a perfectly acceptable globally hyperbolic patch of the full Kruskal manifold, just like the Rindler wedge. As it happens, we are not losing much by focusing on the eternal black hole: it turns out that it is reasonable to make a certain identification between the basis modes  in the collapsing geometry and those in Kruskal, so that we can understand many aspects of the former by studying the latter. We will explain this connection shortly.

Our aim, 
is to investigate realistic black holes with a gravitational collapse in their dark past. So we want to write down a vacuum state in the black hole spacetime in the far past (before the collapse created a black hole), then Bogolubov analyze it using the positive frequency modes that are natural in the far future (where now there is a black hole). A priori, this problem seems horrendously difficult because the collapse is time dependent. There is matter and (at least radial) inhomogeneities in the original spacetime, so following a state before, through and after the collapse seems essentially intractable. But Hawking managed to solve the problem in his original paper by realizing that the modes emerging from a black hole long after the collapse are highly red-shifted.
\begin{figure}
\begin{center}
\includegraphics[
height=0.4\textheight
]{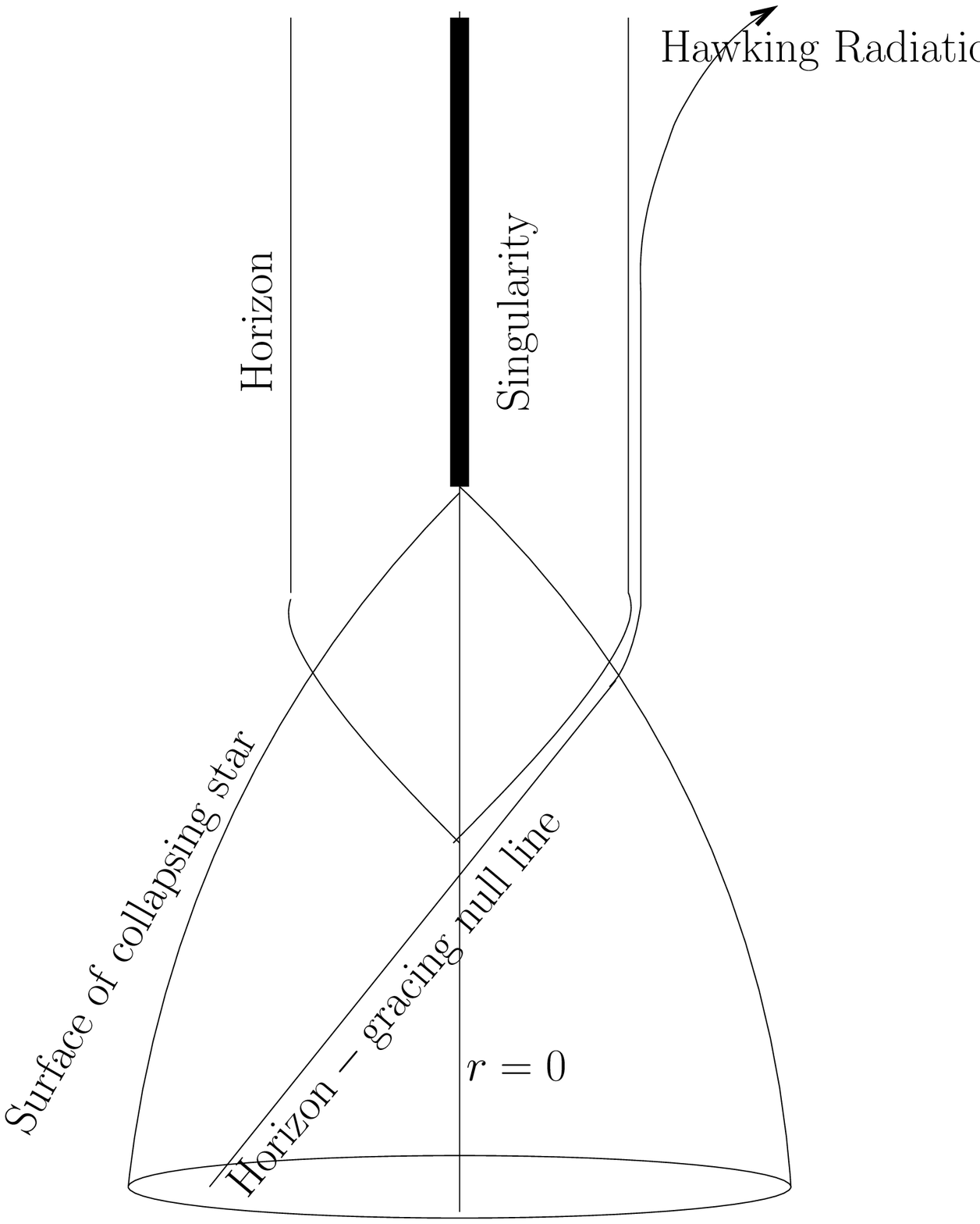}
\caption{Caricature of gravitational collapse. Hawking radiation can be thought of as the almost-trapped geodesics that emerge from near the horizon at arbitrarily late times.}
\label{curves}
\end{center}
\end{figure}
They emerge after being stuck at the horizon for an arbitrarily long time and therefore started out at ${\cal I}^-$ at very high energy. But at high energy, these modes can be treated as particles traversing a geodesic\footnote{Note that one can define a scaling limit where the Klein-Gordon equation reduces to the geodesic equation: this is essentially the WKB approximation. See \cite{Liu, Krishnan} for a recent detailed discussion of this in a similar context.} and one can work with a WKB approximation for the wave equation. This is called the ``geometric optics" approximation, and one can essentially ignore much of the details of the collapse. Solving this frontally via some clever approximations is what Hawking originally did \cite{Hawking}: the details are discussed also in Birrell\&Davies.

But since we have the historical advantage over Hawking, we will use a somewhat more elegant approach to the problem using a path originally suggested by Unruh. Unruh observed that the collapsing star problem can be re-interpreted as a different problem in the full Kruskal geometry, which is much cleaner. The idea is to identify a state in the past half of the Kruskal geometry region I (see Appendix) that most captures the far past vacuum of a collapsing star and then work with it in the Kruskal geometry, instead of in the collapsing black hole geometry. Note that the future half of the Kruskal region I is identical to that of the future part of the collapsing star.

So the outstanding question becomes: which are the positive energy modes that define this {\em Unruh vacuum} of the Kruskal region I? Since we need a complete set and since we are working with massless fields, we need positive energy modes on both ${\cal I}^-$ and ${\cal H}^-$. Note that for massless particles, it is evident that one can think of $\cH^- \cup \cI^-$ (or for that matter $\cH^+ \cup \cI^+$) as global Cauchy surfaces for region I, see figure.
Both Kruskal and collapsing star share ${\cal I}^-$, and the choice of positive energy modes there is clear: one should work with $\sim e^{-i\omega v}$ which are just the positive energy modes with respect to $t$ at ${\cal I}^-$. What about ${\cal H}^-$? A priori, one could work with the positive energy modes of the Killing vector $\partial_t$ there as well, but note that this choice is no longer unique because Kruskal time is also a Killing vector there. Indeed, Unruh's answer is that one should work with positive energy modes with respect to Kruskal time: at the past horizon ${\cal H}^-$ these positive energy modes are of the form $\sim e^{-i\omega U}$. The philosophy being that in the geometric optics approximation, the notion of positivity should be fixed using the affine parameter along the geodesic, and at the horizon the affine parameter is precisely the Kruskal coordinate (see Appendix).

The positive energy modes above defines the Unruh vacuum. We emphasize again that the reason to consider this specific choice is that it is the natural choice when trying to simulate the past vacuum of a collapsing star using Kruskal. Our aim now is to Bogolubov analyze these positive frequency modes in terms of a natural choice of positive energy modes in the future region ${\cal I}^+ \cup {\cal H}^+$. The natural positive energy modes at ${\cal I}^+$ is again with respect to $t$ and take the form $\sim e^{-i\omega u}$. Since we are interested in what the asymptotic observer at late times in the black hole's history is likely to see, we can trace over the Hilbert space of modes on ${\cal H}^+$ (they are inaccessible to the asymptotic observer), and the specific choice of vacuum there won't matter. But for concreteness in computations, we will take the basis to be $\sim e^{-i\omega v}$.

To compute the Bogolubov transformation, it is easiest to work by analogy with the example of the Rindler wedge. To do this, we first introduce modes also in regions other than I in the Kruskal manifold. The idea here is that using this, we can construct the Bogolubov transformation by a version of the analytic continuation argument we used for Rindler.
\bea
u^{I,\cH^-}_{k} \sim \left\{ \begin{array}{ll}
                    e^{-i\omega_k u}
&\quad \mbox{I}\\
          0 & \quad \mbox{II}
                \end{array}\right. \\
u^{II,\cH^-}_{k} \sim \left\{ \begin{array}{ll}
                    0
&\quad \mbox{I}\\
          e^{+i\omega_k u} & \quad \mbox{II}
                \end{array}\right.
\eea
The modes in region II, namely $u^{II,\cH^-}_{k}$, is something that we have introduced by hand to make the analogy with Rindler. Note that this will not affect experiments done in region I. 
Now, note that the positive energy modes $\sim e^{-i\omega U}$ at ${\cal H}^-$ can also be defined as the modes that are analytic in the lower half $U$ plane. The relation between $u$ and $U$ is the same as the relation between Rindler and Minkowski coordinates (up to a factor of $4M$ in the exponent), so we can adapt the Bogolubov transformation that we wrote down before to this case. The result is that instead of working with $\sim e^{-i\omega U}$, we can work with
\bea
U_k^1=\frac{1}{(1-e^{-\beta\omega_k})^{1/2}}(u^{I,\cH^-}_k+e^{-\beta \omega_k/2}u^{{II,\cH^-}*}_k), \\
U_k^{2}=\frac{1}{(1-e^{-\beta\omega_k})^{1/2}}(u^{II,\cH^-}_k+e^{-\beta \omega_k/2}u^{{I,\cH^-}*}_k).
\eea
We have introduced $\beta \equiv 8 \pi M$. The notation is a bit awkward because of the different modes involved, so let us clarify that the $*$ stands for complex conjugation as before. In any event, the past vacuum is therefore defined by the modes spanned by $U_k^1, U_k^2$ and $U_k^{\cI^-} \sim e^{-i\omega v}$. Here $U_k^{\cI^-}$  is the positive energy basis on ${\cal I}^-$ as we discussed.

We will take the future vacuum to be defined by the modes $V^{\cI^+}_k\sim e^{-i\omega u}$ at the future null infinity, the modes at the future horizon that we will trace over $V^{\cH^+}_k\sim e^{-i\omega v}$, and the extra set of modes that we introduced in region II of $\cH^-$, namely $u^{II,\cH^-}_{k}$. The last two are arbitrary choices, the first one because of the tracing over and the second one because it is defined in a region inaccessible to the asymptotic observer at future null infinity. Our task then is to find the Bogolubov transformations relating $(U_k^1, U_k^2, U_k^{\cI^-})$ and $(V^{\cI^+}_k, V^{\cH^+}_k, u^{II,\cH^-}_{k})$.

At this stage, we note that $u^{I,\cH^-}_k, U_k^{\cI^-}, V^{\cI^+}_k, V^{\cH^+}_k$ are not all independent. This is because the full solution of the Schrodinger problem we discussed in the last section is fully fixed by two pieces of data; so the other two can be determined in terms of them. Note  that the situation is entirely analogous to the situation we considered at the end of the previous subsection. In terms of the reflection and transmission coefficients, we can write
\bea
u^{I,\cH^-}_k=t_{\omega_k} V^{\cI^+}_k+r_{\omega_k}V^{\cH^+}_k, \ \ U_k^{\cI^-}=t_{\omega_k}^*V^{\cH^+}_k-r_{\omega_k}^* V^{\cI^+}_k.
\eea
Using all these the final form of the Bogolubov transformations is
\bea
U_k^1=\frac{1}{(1-e^{-\beta\omega_k})^{1/2}}(t_{\omega_k} V^{\cI^+}_k+r_{\omega_k}V^{\cH^+}_k+e^{-\beta\omega_k/2}u^{{II,\cH^-}*}_k), \\
U_k^{2}=\frac{1}{(1-e^{-\beta\omega_k})^{1/2}}(u^{II,\cH^-}_k+e^{-\beta \omega_k/2}(t_{\omega_k}^* V^{\cI^+*}_k+r_{\omega_k}^*V^{\cH^+*}_k)), \\
U_k^{\cI^-}=t_{\omega_k}^*V^{\cH^+}_k-r_{\omega_k}^* V^{\cI^+}_k, \hspace{1.45in}
\eea
and their complex conjugates. Now it is straightforward to write down the ``S-matrix"
following exactly the same steps as in Rindler (except the $\alpha$ and $\beta$ matrices are a bit bigger)
\bea
|0\rangle_{\rm past}=\prod_{\omega_k}(1-e^{-\beta\omega_k})^{1/2}\exp\Big[e^{-\beta\omega_k/2}a_{\omega_k}^{II, \cH^-\dagger}(t_{\omega_k}a_{\omega_k}^{\cI^+ \dagger}+r_{\omega_k}a_{\omega_k}^{\cH^+\dagger})\Big]|0\rangle_{\rm future}
\eea
The $a$'s are the creation/annihilation operators associated with the appropriate basis mode. We can take traces successively on the Hilbert space associated to $\cH^-$ in region II of Kruskal as well as over $\cH^+$ to get the density matrix as seen by an observer in region I. After a little computation, the result takes the form
\bea
\rho=\prod_{\omega_k}
(1-e^{-\beta \omega_k})\sum_{n} \frac{e^{-n\beta \omega_k}|t_{\omega_k}|^{2n}}{(1-|r_{\omega_k}|^{2}e^{-\beta \omega_k})^{n+1}}|n\rangle\langle n|
\eea
The states $|n\rangle$ in the final expression are to be understood as states in the Hilbert space associated to $\cI^+$.
When the reflection coefficient is zero, we see that the density matrix is the normalized thermal density matrix in a canonical ensemble at temperature $\beta=8\pi M$. This result, when re-interpreted in terms of the original collapsing geometry is essentially  Hawking's original result: the observer at late times sees the black hole to be a thermal bath at the Hawking temperature $T_H=1/8\pi M$. Note that the reflection coefficient captures the fact that there is backscattering off of the geometry. One can determine these coefficients numerically (for example) by solving the wave equation in the Schrodinger form we presented before. These corrections away from perfect thermality are sometimes called greybody factors: the ideal thermal case being associated with a {\em black} body.

Note that in all of these results we have suppressed the quantum numbers of the $S^2$ harmonics, there is an implicit summation over them as well. We worked with massless fields in this discussion, more work is required to analyze massive fields: the major conceptual difference being that we cannot take (for example) ${\cal H}^- \cup {\cal I}^-$ as a Cauchy surface, because Cauchy data for the evolution can come from past timelike infinity (which looks like a point - a vertex - in the Penrose diagram) as well.

\subsection{The Trinity of Black Hole Vacua}
\label{UBHH}

In the last section we associated to the Kruskal geometry a vacuum state called the {\bf Unruh vacuum}. This vacuum is defined by the modes which are positive frequency with respect to $t$ on $\cI^-$ and positive frequency with respect to $U$ on $\cH^-$. The reason behind interest in it was that this was the vacuum that worked as a substitute for the collapsing geometry in the simpler setting of Kruskal. Apart from the Unruh vacuum, there are two other vacua which are sometimes considered in the context of Kruskal geometry in discussions of black hole radiance. One of them is the so-called Boulware vacuum while the other is called the Hartle-Hawking vacuum. We discuss each of these in turn.

{\bf Boulware vacuum:} To describe a (past) vacuum, we need to impose a notion of positive energy modes on $\cH^- \cup \cI^-$. As we discussed in the previous subsection, the natural notion of positive energy on $\cI^-$ is that of positive energy with respect to the Killing vector associated to translations in the Schwarzschild time $t$. But ambiguity arises in choosing the positive energy modes at the (past) horizon $\cH^-$ because there, the Kruskal time coordinate is also Killing. Unruh vacuum was the choice where positive energy modes at $\cH^-$ are defined with respect to Kruskal time. Boulware vacuum on the other hand corresponds to choosing positive energy with respect to $t$ even on $\cH^-$.

In a sense, Boulware vacuum state is the naive choice of past vacuum on Kruskal manifold. We will see in a later section that the stress tensor diverges at the horizon, when evaluated on the Boulware vacuum state.

{\bf Hartle-Hawking vacuum:} Both Boulware and Unruh vacuum state were defined by prescribing positivity of modes on $\cH^- \cup \cI^-$. This makes sense because the surface $\cH^- \cup \cI^-$ is a Cauchy surface (at least for massless fields). The definition of Hartle-Hawking vacuum \cite{Hartle1} on the other hand is based on defining positive energy modes on $\cH^- \cup \cH^+$. It is defined as the vacuum state described by choosing positivity with respect to Kruskal time on $\cH^- \cup \cH^+$: said another way, we demand analyticity in the lower half plane for $U$ on $\cH^-$ and for $V$ on $\cH^+$. It is not a priori obvious that this is a consistent choice for defining the 1-particle Hilbert space, so we give an explanation why this is so.

The basic observation is that even though providing Cauchy data on $\cH^-$ is not enough to determine the future in region I of Kruskal, it {\em is} enough to determine what happens inside a box enclosing the black hole with Dirichlet boundary conditions at the boundary. Moreover, if we choose a basis of solutions of Klein-Gordon equation which are positive frequency with respect to $U$ on $\cH^-$, then then they will evolve into positive frequency solutions with respect to $V$ on $\cH^+$. 
Therefore the Hartle-Hawking state captures is the vacuum state appropriate for a black hole inside a reflecting box.

There is yet another reason to take the Hartle-Hawking state as an interesting vacuum for Schwarzschild black holes. As described in the previous section, one can start with the Green function in the Euclidean geometry. See also the discussion in section (\ref{EQG}). The precise form of the Euclidean metric is (\ref{ESch}) with the identification $\tau \sim \tau + \beta(=8 \pi M)$. One can write down the unique Euclidean Green function that falls off at infinity as a solution to
\bea
\nabla^2_E G_E (x,x') =\frac{1}{\sqrt{g_E}}\delta^4_E(x-x').
\eea
This Euclidean Green function is well-defined on the imaginary time axis ($\tau$). Note that according to the discussion in the previous section one can now {\em define} a Lorentzian Green function by analytically continuing it to the real axis. The various ways of taking the boundary value of this analytically continued function on the real axis will be interpreted as Wightman Green functions, Feynman Green functions, etc. The question then becomes: is there a Lorentzian state where we can compute (Lorentzian) Green functions which can reproduce these analytically continued Green functions coming from the Euclidean domain? The answer to this question is also the Hartle-Hawking state. To see this we need to note that the analytically continued Green function is analytic in $t$ in, for example, the strip (see discussion in section \ref{Green})
\bea
-8\pi M < {\rm Im} t <0.
\eea
Using the definition of the Kruskal coordinates $U$ and $V$ in terms of $r$ and $t$ from the appendix this can be shown to translate to the statement that the Green function is analytic in the lower half-$U$ plane when $V=0$ (and similarly on lower half-$V$ plane for $U=0$). Remembering that $U=0$ defines $\cH^+$ and $V=0$ defines $\cH^-$, we have showed that the Green function satisfies precisely the definition of the Hartle-Hawking Green function.

The way to think of the Hartle-Hawking state is as the state which captures the black hole being in equilibrium with radiation at the Hawking temperature. We will come back to the question of whether this equilibrium is a stable one.

\subsection{Black Hole Thermodynamics}

In our discussions in the previous sections, one of the things to notice is that we were working with a fixed gravitational background (i.e., geometry) in which quantum fields propagate. From this set-up, we came to the conclusion that there is a natural notion of thermality and temperature that emerge. In fact historically, the first suggestion that black holes have a connection  with thermodynamics came not from quantum field theory in curved spacetime, but from classical general relativity. The equations of motion of general relativity in black hole backgrounds is formally identical to the laws of thermodynamics as demonstrated by Bardeen, Carter and Hawking \cite{Bardeen}. The fact that the dynamics of black hole horizons is captured by laws identical to the three laws of thermodynamics is what is usually called black hole thermodynamics.  It is important to remember that these laws follow from {\em classical} general relativity applied to black hole spacetimes.

The {\em zeroth law of black hole thermodynamics} is the statement that the surface gravity of a black hole is a constant over the horizon. Surface gravity on a Killing horizon is defined by $\kappa$ in the relation
\bea
k^a\nabla_a k^b=\kappa k^b
\eea
where $k^a$ is the Killing vector that defines the horizon\footnote{A Killing horizon is a null hypersurface which is spanned by Killing vectors: that is, the surface is generated by the orbits of the isometry corresponding to the Killing vector. The horizons we consider will always be Killing horizons.}. The equation should be evaluated at the horizon, and for asymptotically flat spacetimes a normalization convention for the Killing vector has to chosen so that $k^ak_a\rightarrow -1$ as $r\rightarrow \infty$ and $\kappa$ is chosen positive. The basic point is that surface gravity is a purely geometrical quantity, computable from the metric.
For the Schwarzschild black hole, the timelike Killing vector is the one that defines the (Killing) event horizon, so $k=\partial_t=(1,0,0,0)$. Plugging this into the equation above and using the Schwarzschild metric, one finds that $\kappa=1/4 M$, and is manifestly constant at the horizon. The fact that it is a constant even for black holes in any dimensions and in any theory is a less trivial fact, however. Zeroth law is interesting because it makes surface gravity a quantity analogous to temperature, for a body in equilibrium: it is constant all over it. In fact, in the context of quantum field theory in curved spacetime, as we discussed in the last section, the temperature that we calculated (for the Schwarzschild black hole) was $1/8\pi M$ and is precisely the surface gravity of the black hole, up to a numerical factor. This result is more general. The temperature one associates to a black hole via quantum arguments\footnote{The derivation of Hawking temperature via Bogolubov transformations that we saw in the previous subsections was for the Schwarzschild black hole. But we will see a derivation of temperature that is applicable for more general static black holes in the section on Euclidean quantum gravity.} is always its surface gravity. So in this sense, the fact that surface gravity is constant over the horizon makes the analogy with temperature a very natural one.

The {\em first law of black hole thermodynamics} is a statement based on the asymptotic charges of the black hole, and basically captures the dynamics of black holes as stipulated by general relativity. For a rotating black hole with charge, it says that
\bea
dM=\frac{\kappa}{8 \pi} dA+\Omega dJ+\Phi dQ
\eea
where $M$ is the mass of the black hole, $\kappa$ is as before the surface gravity, $A$ is the horizon area, $J$ is the asymptotic angular momentum, $\Omega$ is the angular velocity of the horizon, $\Phi$ is the chemical potential for the charge at the horizon and $Q$ is the charge. We will not define the various quantities in detail, all we need to know for the present discussion is that they can all be computed from the metric and the geometry, yet the relation has the flavor of the thermodynamics. The charges (mass, angular momentum and electric charge) can be taken as the ADM values in flat spacetime. If one interprets the mass as the internal energy, the surface gravity as the temperature and the horizon area as the entropy, this equation is exactly what one would expect from the first law of thermodynamics as applied to a black hole. One way to think of the first law is to consider it as a statement about processes that take place involving black holes. The statement then would be that in any process involving the black hole, the above quantities of the black hole vary in such a way that the relation above holds true.  The fact that we find direct evidence from QFT in curved space arguments that $\kappa$ is indeed the temperature lends further credence to take this as a thermodynamical relation at face value.

The {\em second law of black hole thermodynamics} is the statement that in any process involving the black hole, its horizon area can never decrease. For example, when black holes merge, the final area will not be less than the sum of the two areas. This ties in perfectly with the interpretation above from the first law that the area is analogous to the entropy, and can therefore never decrease as expected from our usual thermodynamic intuition.

We have not presented detailed derivations of these results because they essentially follow from classical general relativity and our focus in these lectures is on the quantum aspect of things. The original paper of Bardeen, Carter and Hawking is pedagogical, and so is the review by Townsend \cite{Townsend}. Note that as they stand, these three classical laws cannot quite fix the numerical pre-factor relating the surface gravity and the temperature, because one could always satisfy the first (and trivially  second) law by a compensating factor in the definition relating area and entropy. But once we know the Hawking temperature independently from a quantum calculation, we know that
\bea
T=\frac{\kappa}{2 \pi}, \ \ S=\frac{A}{4}.
\eea
The latter expression is often called the Bekenstein-Hawking entropy. Bekenstein was the brave soul who had the courage to speculate that area of the horizon is proportional to the entropy of a black hole: without the malice of hindsight, this is a ``crackpot" claim, because classical black holes are zero-temperature objects and it doesn't make any sense to talk about them thermodynamically. Hawking's computation fixed the proportionality constant to be 1/4. Putting back the dimensionful constants, the equations above take the form
\bea
T=\frac{\hbar c^3 \kappa}{2 \pi G k_B}, \ \ S=\frac{k_B c^3 A}{4 G \hbar}.
\eea
where $k_B$ is Boltzmann's constant.

Note that all three laws follow from classical general relativity. This should be contrasted to our previous subsections in two ways. Firstly, the previous sections deal purely with fixed backgrounds where gravity is non-dynamical. Black hole ``thermodynamics" on the other hand follows from the dynamics of black holes as general relativistic objects. Secondly, the most concrete observation that the surface gravity is the temperature seems to arise from quantum considerations, but surprisingly black hole thermodynamics seems to know about this already at the classical level.

There are a couple of caveats that one needs to remember when thinking about black hole thermodynamics. Firstly, it should be borne in mind that the first law of thermodynamics is usually written down for systems where there is quasi-static exchange of heat between the system and it ambience. The claim then is that the heat change ($\sim\kappa dA$) is equal to the work done (the other terms) plus the internal energy change ($\delta M$). On the other hand, the second law applies to black holes treated as closed systems. This is because the full second law of thermodynamics takes the form $T \delta S -\delta Q\ge 0$ and it reduces to $\delta S \ge 0$ only for closed systems. Another point is that in the presence of Hawking radiation, the full second law has to be modified: this is because radiation can take away entropy from the hole. The entropy of the hole and the radiation together should still be non-decreasing, so the modified second law would be
$ \delta S_{rad}+\delta A/4 \ge 0$. It seems plausible to me that the correct way to think of these relations is in terms of local equilibrium: the horizon is in local thermal equilibrium with its ambience. In flat space, black holes cannot be in stable global equilibrium because they Hawking radiate away. 
Another reason to think of black holes as objects in local equilibrium becomes evident when one considers black holes in higher dimensions where there is the possibility of having multiple horizons \cite{5d}. There, each of the horizons can be associated with a distinct Hawking temperature and there cannot exist a notion of even an unstable equilibrium. On the other hand, when working with local equilibrium, one can associate a local notion of temperature \cite{EKsaturn}. We will have a few words to say about this when we briefly discuss the fuzzball picture for black holes in the next section.

To clarify the claim above, one can consider the specific heat of a black hole. From the first law and by associating the mass to the internal energy ($M \sim U$) along with the result $T=1/8 \pi M$, one can define a specific heat for the black hole. Simply using $C=dU/dT$ we find that the specific heat of the Schwarzschild black hole is negative. 
This means basically that the black hole cannot be in stable equilibrium with radiation even if the radiation happens to be at the Hawking temperature: it heats up by radiating and cools down by absorbing. This suggests that the Hartle-Hawking state for the black hole that we defined in the previous section -which is supposed to capture a black hole  exchanging radiation with a heat bath at the Hawking temperature and being in thermal equilibrium- is strictly speaking not well-defined. This assumes that the black hole is in a big enough box: if the box is small, the net gain in the entropy of the radiation plus hole system can be positive, even if the hole by itself is losing entropy. We will see that in Anti-de Sitter space, large black holes have positive specific heat and therefore the Hartle-Hawking state is well-defined. AdS is like a reflecting box, so this is morally the analogue of putting a small enough box around a black hole.

\subsection{Two Challenges: Entropy and Information}

Classical general relativity comes equipped with no-hair theorems which say that quite often, black hole solutions are fully fixed by a  few asymptotic charges of the solution. This means that when matter collapses to form a black hole, almost all of the  information on the initial data is lost. A specific quantum initial state turns in the end into a hairless black hole and essentially information-less thermal radiation. This violates unitarity, and is called the black hole information paradox.

More basically, one could look at the area law for black hole entropy and ask what are the microstates that contribute to the entropy. Thermodynamic entropy arises from the coarse-graining over microstates with the same macroscopic quantum numbers, so the black hole must be comprised of microstates. The challenge in understanding black hole entropy is to understand what precisely these microstates are.

It seems evident that answers to either of these questions will require us to move beyond not just classical gravity, but also QFT in curved space. 
With the emergence of string theory, there has been substantial progress in understanding the microstates of black holes that contribute to its entropy. The thing to note is that because of the no-hair theorems, black holes are completely specified by the global charges they carry\footnote{We have focussed our attention on only one charge, namely the mass of the black hole. But in principle it can carry other gauge charges like electric charge.}. The idea is that in string theory objects carrying these charges have more than one description: when the string coupling\footnote{\label{FN!}What we mean by string coupling here is the effective {\em open} string coupling, which goes as $g_s N$ where $g_s$ is the closed string coupling and $N$ is the number of branes (i.e., the number of Chan-Paton flavors as seen from the string worldsheet). At large $g_sN$ the branes are heavy, backreact, and form a black hole.} is large, they are described as black hole solutions of certain supergravity theories, whereas when the string coupling is small they are captured by certain D-brane states. Because of supersymmetry, we do not expect the degeneracy of the object to change as we tune the string coupling. This means that the entropy of the black hole should be captured by counting the degeneracy of D-brane states with the same charges. This can be done in a controlled way, and the remarkable result is that the entropy is precisely reproduced by the D-brane degeneracy \cite{SW}.

The trouble is that these computations apply only for certain supersymmetric black holes, and supersymmetric black holes are, among other things, extremal and therefore their Hawking temperature is zero. (The entropy is still non-zero.) It is possible to go infinitesimally away from supersymmetry in a scaling limit\footnote{In fact in this limit, the string computation agrees both with the entropy and the Hawking spectrum!}, but what would be fully satisfying is if one could get rid of using supersymmetry as a crutch altogether. The reason for this is not merely because black holes in astrophysics are non-supersymmetric: there is nothing wrong in assuming that a black hole is more symmetric than it is, in order to make an idealization. After all, we are used to working with frictionless inclined planes and spherical cows in physics. The trouble here is that supersymmetry has a way of restricting 
dynamics, and it is precisely this restriction that we are using (namely the protected nature of the degeneracy, i.e., BPS index) in order to count the microstates. So the solution to the idealized problem is entirely silent about the solution to the not-so-idealized problem. Despite this, it is an undeniable fact that the understanding of black holes in string theory is a tremendous step forward in our understanding of black holes in quantum gravity, and indeed 
is very strong evidence that string theory is a consistent quantum theory of (some) gravity.

The information paradox is a harder problem, because it seems to require a detailed understanding of dynamics and is not merely a counting problem. But with the advent of the AdS/CFT correspondence, we believe that we have an in-principle understanding of information flow for black holes, at least when they live in asymptotically AdS spacetimes. The claim of the AdS/CFT correspondence is that quantum gravity with asymptotically anti-de Sitter boundary conditions is described by a quantum gauge theory, but the precise map between local AdS physics and the CFT is diabolically subtle. But since the CFT is unitary, we know that any process that happens in the bulk of AdS {\em should} have a unitary description and therefore information cannot be lost. So string theory (or AdS/CFT) suggests that quantum mechanics is saved. The details of this, however, are not understood in any detail. One suggestion due to Mathur \cite{Mathur}, is that black holes should in fact be thought of as a  D-brane fuzz, and that classical gravity would have to be discarded already at the scale of the horizon, long before the singularity. The analogy is with a burning lump of coal: one does not loose information when the coal burns, it is just that to retrieve information from the end products (i.e., the precise state of all the CO$_2$ and water vapor molecules), is too much work. This analogy with the lump of coal also gives a rationale for the local equilibrium picture of the horizon that we advocated in the previous section. For large enough black holes, the horizon has very low curvature and the temperature is low, so the claim of the fuzzball proposal (that general relativity has to be abandoned already at the horizon) is a bold claim. It will be very interesting to see where these ideas lead.

\section{Euclidean Quantum Gravity}
\label{EQG}

We will view Euclidean quantum gravity as a semi-classical approximation to quantum gravity, whatever quantum gravity might ultimately be. The idea is that one defines a (semi-classical) partition function for gravity in the canonical ensemble via a ``path" integral over Riemannian metrics. Specifically, the canonical partition function for gravity at temperature $T=1/\beta$ is defined as
\bea
Z(\beta)=\int D[g] e^{-I[g]}
\eea
where the integral is over Riemannian metrics which satisfy certain asymptotic fall-offs\footnote{One thinks of this fall-off as part of the definition of the quantization.}, and $I$ is the (gravitational) action. The $\beta$ shows up in the integral as the asymptotic periodicity that needs to be satisfied by all the geometries in the integral. For example, if we are concerned with quantization in an asymptotically flat context,  then one decrees that all metrics have to have a compact isometry direction with $\beta$ as the proper size of the asymptotic periodicity in that direction. We will also be interested in asymptotically AdS quantizations and we will describe them later.

As is clear from the discussion above, asymptotics is an important ingredient in this construction. This is made especially clear because Einstein-Hilbert action contains second derivatives (unlike for example the usual scalar or gauge field action) and to define a variational principle in the standard way\footnote{That is, one would like to have a gravitational variational problem where we hold the boundary value of the metric fixed, while not constraining the normal derivative of the metric at the boundary. This is analogous to the variational principle of a classical Newtonian particle where we hold its position at initial and final time to be fixed, while allowing the velocities at these points to be unconstrained.}, one needs to add a boundary piece to the action to cancel pieces from the bulk. The requisite boundary piece is called the Gibbons-Hawking-York term and since we are working with {\em Euclidean} theory, the total action takes (in $d$ bulk dimensions) the form
\bea
I[g]=-\frac{1}{16 \pi G}\int_{\cal M} \sqrt{g}\ R\ d^d x - \frac{1}{8 \pi G} \int_{\partial {\cal M}} \sqrt{h}\ K d^{d-1} x. \label{ES}
\eea
The $K$ is the extrinsic curvature of the boundary and $h_{ab}$ is the induced metric on the boundary. If we were to follow the standard notion of path integration, we would like to integrate over all field configurations satisfying certain boundary conditions. Here it would be the metric (and possibly topology) configurations with fixed induced metric on the boundary. One of the reasons why Euclidean quantum gravity is not well-defined is because an integral of this form is {\em not} well-defined because the integral is unbounded from below. This can be seen (we skip the details, see \cite{GHP}) by splitting the integral over the conformal class and a Weyl factor: the integral over the Weyl factor is already unbounded from below.

But since we expect that metric should be a meaningful concept in quantum gravity at least semi-classically, we will press on. Our viewpoint will merely be that the saddle point of the path integral above with the least action (and the boundary conditions described above) should be the dominant contribution to the partition function:
\bea
Z(\beta) \approx e^{-I_{\rm saddle}}
\eea
Therefore from the standard relation between entropy and the partition function in the canonical ensemble, we get
\bea
S\equiv \ln Z -\beta\frac{\partial \ln Z}{\partial \beta}=\beta\frac{\partial  I_{\rm saddle}}{\partial \beta}-I_{\rm saddle}. \label{EUentropy}
\eea
This can be used to evaluate the entropy and Hawking temperature of the black hole, as we demonstrate next.

\subsection{Temperature and Entropy, Redux}

It is clear that when we put flat space at finite temperature using the above approach, the Euclidean solution is a saddle for any temperature. The Euclidean flat space metric is
\bea
ds^2=d\tau^2+dr^2+r^2 d\Omega^2
\eea
and it is clear that for any periodicity of $\tau$,  the metric is a solution of Einstein's equation and is regular everywhere. So it is an acceptable saddle. Note also that the periodicity in $\tau$ can be factored out of the definition of the action: $I=\int d\tau d^{d-1}x L=\beta \times\int d^{d-1}x L=\beta \times ({\rm stuff})$. Because of the linear dependence of the action on $\beta$ we immediately find that the entropy computed via (\ref{EUentropy}) is zero. So flat space is a saddle at any temperature, and none of these saddles have any entropy.

On the contrary, when the spacetime under consideration is a (static) black hole, the discussion gets more interesting. Consider the Euclideanize the Schwarzschild metric
\bea
ds^2=(1-2M/r)d\tau^2+\frac{dr^2}{1-2M/r}+r^2 d\Omega^2, \label{ESch}
\eea
In Euclidean quantum gravity, we demand that the saddle geometry be regular. This is a strong constraint because of the presence of (what used to be, in the Lorentzian section) the horizon. For large $r$ the space is asymptotically flat and we want to set $\tau \sim \tau +\beta$. But for generic choices of $\beta$ we would have a conical singularity at the $r=2M$. This can be seen by expanding the Euclidean metric around the horizon as $r=2M+y$
\bea
ds^2 \approx \frac{y}{2M} d\tau^2+\frac{2M}{y} dy^2=d\rho^2+ \rho^2\frac{d\tau^2}{16 M^2}
\eea
where we suppress the sphere part and introduced $\rho =\sqrt{8My}$. If $\frac{\tau}{4M} \sim \frac{\tau}{4M}+2 \pi$, then the last form can be thought of as the metric of the 2-plane in polar coordinates and can be regular at $\rho=0$: otherwise, there will be a conical deficit and the geometry is not regular. This forces the periodicity of $\tau \sim \tau +8\pi M$ and we end up with the Hawking temperature $T=1/8\pi M$, which is the value we found in the previous sections by other means.

This approach of demanding the regularity of the Euclidean section  is a remarkably simple and useful tool for computing the Hawking temperature when the black hole is static. The interesting fact here is that unlike the QFT in curved space computation, this relied only on the classical geometry. Even in the QFT computation, the dynamics of gravity never showed up: the thermal Hawking spectrum arose when we propagated quantum fields on a fixed background. Both these computations therefore seem to be capturing something about (quantum fields living on) geometry that is even more primitive than gravity itself. In this sense, Hawking radiation is hardly a step towards the quantum nature of gravity itself. Of course, in a consistent quantum theory of gravity, we have to have a consistent and comprehensible explanation for both black hole microstates and Hawking radiation, and string theory seems to be able to do this when it is under control.

Note however that we needed to have a precise idea about what the asymptotic conditions are, in order to unambiguously fix the norm of the time-translation Killing vector in the asymptotic region. Choosing the asymptotic condition (flat, AdS, etc.) fixes this. In the asymptotically flat example we could have multiplied the $\tau$ by a constant factor if the fall-off of the $d\tau^2$ piece $(1+O(r^{-1}))$ could be re-scaled by $1/({\rm the \ same \ factor})$. This would have resulted in a corresponding rescaling in the Hawking temperature since the periodicity of $\tau$ is what ultimately determines it. But since the fall-off $1+O(r^{-1})$ is part of the definition of asymptotic flatness, this cannot be.  The message again is that the physics depends on the asymptotics.

Along the same lines, we can also determine the entropy of the black hole using the Euclidean approach via the formula (\ref{EUentropy}). The  point is that the Euclidean metric looks basically like a cigar (again we suppress the trivial sphere).
\begin{figure}
\begin{center}
\includegraphics[
height=0.3\textheight
]{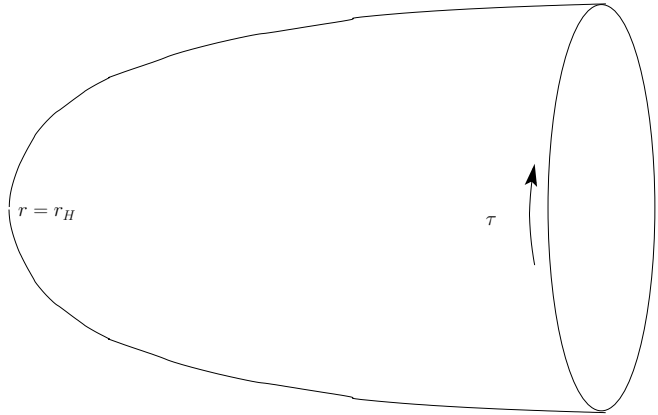}
\caption{Euclidean (static) black hole is fully regular, periodic in Euclidean time $\tau$, and covers only the region outside the horizon. The $r$ and $\tau$ coordinates together make it look like a cigar and therefore it has the topology of $\IR^2$ (times the compact directions, which in four dimensions is the sphere $S^2$ coming from $\theta$ and $\phi$).}
\label{curves}
\end{center}
\end{figure}
The tip of the cigar tapers off smoothly at $r=2M$. Our task is to compute the action for this geometry. Even though the Euclidean space does not have a time direction, to have a natural Hamiltonian interpretation in the Lorentzian version, we do the integrals along constant $\tau$ slices. This means that there is a contribution from the asymptotic boundary as well as the horizon in the integration (i.e., the latter also needs to be treated as a boundary). When we compute the entropy using the result, the bulk term plus the asymptotic boundary does not contribute (by the same argument based on linearity in $\beta$ that we used for flat space). So we need to keep track only of the boundary contribution arising from the horizon. Close to the horizon, the metric is
\bea
ds^2\approx d\rho^2+ \rho^2 \kappa^2 d\tau^2+r_+^2 d\Omega^2 \label{NHmetric}
\eea
where we have expressed it in terms of surface gravity to emphasize that the result we are about to get is general and not tied to Schwarzschild. Using the definition of the extrinsic curvature, the boundary piece can therefore be written as
\bea
\frac{1}{8\pi  G}\int_{\partial M}\sqrt{h} K=-\frac{1}{8\pi G}\frac{\partial}{\partial n}\int_{\partial M} \sqrt{h}=-\frac{1}{8\pi G}\frac{\partial}{\partial \epsilon}2 \pi \epsilon A= -\frac{A}{4G}.
\eea
In the second last step we have evaluated the boundary term at $\rho=r_+ + \epsilon$, where in the limit of vanishing $\epsilon$ we can use the metric (\ref{NHmetric}) and evaluate the integral. The area $A$ naturally arises from the geometry in this derivation. Plugging this into the entropy formula, we immediately find
\bea
S=\beta\frac{\partial  I}{\partial \beta}-I=0+\frac{A}{4G}=\frac{A}{4G},
\eea
which is the universal Bekenstein-Hawking entropy.

\subsection{An Application from Anti-de Sitter Space: Hawking-Page Transition}

One of the most successful applications of Euclidean quantum gravity is in the context of Anti-de Sitter space, where many of its expectations are borne out by expectations in the a dual quantum field theory via the AdS/CFT correspondence.  We will have more to say about AdS/CFT later, but for now, we apply the Euclidean partition function to asymptotically $AdS_5$ spacetimes. The gravitational part of the action is
\bea
I=-\frac{1}{16\pi G_N} \int d^5 x \sqrt{g}  \Big(R + \frac{12}{L^2}\Big).
\eea
$G_N$ is the five-dimensional Newton's constant and $L$ is the AdS scale, which captures the cosmological constant. The ``vacuum" solution in this theory will be empty AdS space, and we will always define the relevant quantities (like the on-shell action) after a suitable subtraction with AdS as the datum. When one does this, it turns out that the boundary piece does not contribute, so we will not keep track of it in the discussion of this section. We will make some of these questions more precise using holographic renormalization in a later section.

In the global coordinates, the anti-de Sitter vacuum is given by
\bea
ds^2  = \left(1+\frac{r^2}{L^2}\right) d\tau^2 + \left(1+\frac{r^2}{L^2}\right)^{-1} dr^2 + r^2 d\Omega_3^2 ,
\eea
where $d\Omega_3^2$ is the metric on $S^3$.  The asymptotics are at $r \rightarrow \infty$ like in flat space, but in AdS the conformal boundary is timelike as opposed to null. This can be seen by drawing its Penrose diagram (exercise).  From the metric above it is clear that the topology of the boundary is $S^1\times S^3$.

As we have seen before, in Euclidean quantum gravity the prescription is to sum over all geometries with the asymptotics (``boundary") fixed. Another solution that has the same asymptotic structure is  AdS-Schwarzschild black hole with metric given by
\bea
ds^2 = \left(1+\frac{r^2}{L^2} -\frac{8 G M}{3 \pi r^2}\right) d\tau^2 + \left(1+\frac{r^2}{L^2} -\frac{8 G M}{3 \pi r^2}\right)^{-1} dr^2+ r^2 d\Omega_3^2 ,
\eea
where $M$ is the mass of the black hole. In the Euclidean signature, the space-time is restricted to the region $r \geq r_+$ for regularity as in the flat space case, where $r_+$ is the (Lorentzian) horizon described the bigger root of $g_{\tau \tau}=0$. By computation analogous to that in the previous section, regularity further fixes the Hawking temperature of the hole to be
\bea
T\equiv \frac{1}{\beta} =\frac{2r_+^2+L^2}{2\pi L^2 r_+}.
\label{betaAdSBH}
\eea

The outstanding question is which of the two geometries is the dominant contribution to the partition function\footnote{We consider only spherically symmetric solutions. Rotating black holes etc., are possible, but they are more natural in a grand canonical ensemble where we allow a chemical potential for the angular momentum. There are further subtleties in the Euclidean quantum gravity approach there due to the fact that the spacetime is not stationary. Here we focus only on static situations where only the canonical, fixed temperature, ensemble shows up.}. To determine this we should use the classical action.  It is easily checked that on a solution of the equations of motion (which is basically Einstein's equation with AdS cosmological constant), the action takes the form
\bea
I = \frac{1}{2\pi G_N L^2} \int d^5 x \sqrt{g}.
\eea
This naively diverges because of the AdS warp factor, so we cut-off the integral at a radial coordinate $R$.
The difference of the two actions\footnote{As mentioned before, we treat empty AdS as the vacuum.} is proportional to
\bea
I\equiv I_{BH}-I_{AdS} \sim \int_0^{\beta} d\tau\int_{r_+}^R dr \  r^3 -
\int_0^{\beta'} d\tau \int_0^R dr \ r^3.
\eea
We want to compare the two objects at the same temperature, but since there is a cut-off involved on two {\em separate} spacetimes we match the proper circumference of the Euclidean time direction at $R$. This determines $\beta'$ via
\bea
\sqrt{g_{\tau \tau}^{BH}(R)} \ \beta=\sqrt{g_{\tau \tau}^{AdS}(R)}\ \beta'
\eea
Reinstating the constants, the datum-subtracted action of the black hole in the limit when the cut-off is taken to infinity is
\bea
I =  \frac{2 \pi^2(L^2 r_+^3 - r_+^5)}{4G_N(4 r_+^2 + 2 L^2)}.
\eea
So the black hole is dominant at high temperatures, or more precisely when $1/\beta \equiv T \geq \frac{3}{2 \pi L}$. (This is the same thing as $r_+ \geq L$.). This phase transition is called the Hawking-Page transition \cite{HawkPage}: in AdS, black holes are the dominant phase at temperatures higher that the AdS scale.

The black hole we considered here is often called the ``big" black hole in the literature. The point is that for a given Hawking temperature, there are two black holes possible in AdS corresponding to two solutions of $r_+$ as function of $\beta$ in (\ref{betaAdSBH}). This ``small" black hole is never the dominant contribution to the canonical ensemble. One can check that the large black hole has positive specific heat\footnote{This can be done using the expression we obtained for the Hawking temperature (\ref{betaAdSBH}), together with the fact that $r_+$ is the bigger root of the $g_{\tau\tau}=0$ equation which contains the mass of the black hole. Note that $C \sim dM/dT$.} while the small black hole has negative specific heat. The latter is therefore unstable. In particular, large Schwarzschild black holes in AdS can have a well-defined Hartle-Hawking state.

\subsection{Wave Function of the Universe}

Path integrals define propagators in ordinary quantum mechanics. A propagator is nothing but the amplitude to find a particle at location $x'$ at time $t'$ if it was at position $x$ at time $t$:
\bea
\langle x',t'|x,t\rangle=\int Dx[t] \exp (iS[x(t)]),
\eea
where we integrate over all histories that start at $x$ at $t$ and end at $x'$ at $t'$, weighted by $\exp (iS[x(t))$ for that history. Inspired by Euclidean quantum gravity, one might propose a similar construction for spacetime geometries. The idea is to foliate a spacetime with spatial slices and then to define a propagator for evolution from the 3-geometry on  one slice to the 3-geometry on another via
\bea
\langle h_{ij}'|h_{ij}\rangle = \int Dg \exp [iS[g]].
\eea
Here $h$ stands for a 3-metric on a slice and where the sum is over all 4-metrics that connect between the two 3-geometries. At least when the universe is spatially closed, the 3-metric fully fixes the information on the 3-surface. Note that unlike in the quantum mechanics example where there was a time parameter $t$, here we do not fix proper time on the 3-surfaces. Rather it is implicitly present in the integral over the 4-geometries. Fixing proper time as well, would involve intermediate measurements between the two slices, which we do not wish to specify.

More generally, one might define a wave function for a 3-geometry by
\bea
\Psi[h_{ij}]=\int Dg \exp[i S[g]]
\eea
where now the integral is over 4-geometries that end on the 3-geometry $h_{ij}$ and over what 4-geometries we integrate is part of the prescription of the wave function. A specific proposal for a ground state wave function for the Universe was put forward  by Hartle and Hawking \cite{Hartle2}: they suggested that the integral should be over all compact 4-geometries bounded by the 3-geometry $h_{ij}$. One can think of this as the amplitude for producing $h_{ij}$ by starting with a trivial 3-geometry, that is nothing (i.e, a point).
\begin{figure}
\begin{center}
\includegraphics[
height=0.4\textheight
]{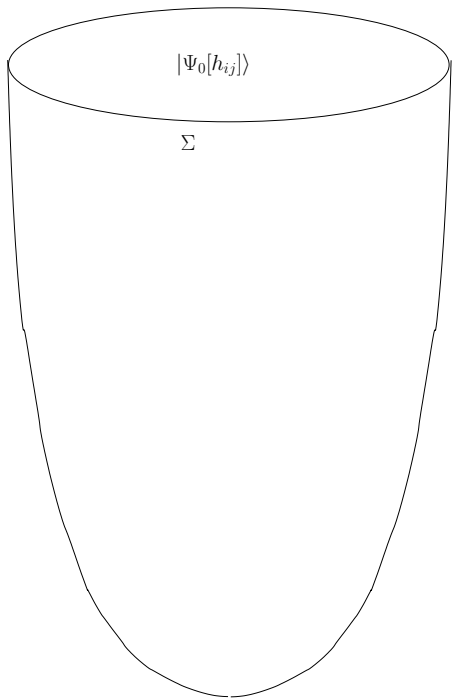}
\caption{Evolution of the wave function from a trivial 3-geometry. Even though we draw only one 4-geometry between initial and final states, the path integral is over all such
4-geometries.}
\label{curves}
\end{center}
\end{figure}
In Wick rotated notation, one can write the ground state wave function of the Universe as
\bea
\Psi_0[h_{ij}]=N \int Dg \exp[-I[g]]. \label{grwf}
\eea
where $I$ is now the Euclidean action and $N$ is a normalization constant (which we had suppressed so far). We can interpret this as the ground state wave function of the Universe in analogy with ordinary quantum mechanics: in the latter, consider the amplitude
\bea
\langle x,0|0,t' \rangle.
\eea
This can be written in two ways. As discussed already, we can write this as a path integral
\bea
\langle x,0|0,t' \rangle=\int D[x] \exp (iS[x(t)]).
\eea
But by inserting a complete energy basis we can also write it as
\bea
\langle x,0|0,t' \rangle=\sum_n\langle x,0|n\rangle \langle n|0,t'\rangle=\sum_n \psi_n(x) \psi_n(0)^* \exp (iE_n t')
\eea
We have used the definition of the coordinate space Schrodinger picture wave function: $\psi_n(x)=\langle x,0|n\rangle$. The time evolution of such a wave function is fixed as usual by $\psi_n(x,t)=\psi_n(x)e^{-i E_n t}$.
Wick rotating $t'$ to $-i \tau$  and considering the limit $\tau \rightarrow -\infty$, this last expression reduces (assuming there is no zero point energy and up to a constant normalization) to $\psi_0(x)$, while the path integral goes over to the Euclidean path integral
\bea
\int D[x] \exp (-I[x(t)]).
\eea
Since the two should be identical, this means that this last expression is a representation of the ground state wave function. Equation (\ref{grwf}) is the gravity version of this statement, so it is natural to interpret the left hand side as the ground state wave function of the Universe\footnote{This wave function is often called the Hartle-Hawking wave function (or state). But since the phrase has already been used in another context (cf., subsection \ref{UBHH}) we will try to minimize this usage.}.

\subsection{Wheeler-DeWitt Equation}

Just like the wave function of quantum mechanics satisfies the (time-dependent) Schrodinger equation, the wave function of the Universe satisfies a functional differential equation. This is the Wheeler-DeWitt equation. To motivate the latter, we first present a perspective on how the Schrodinger equation emerges starting from the path integral.

Since the Schrodinger equation is about the time evolution of the wave function, what we would like is to write down the wave function at a location $y$ at a time $t+\epsilon$, assuming that we know what the wave function is for any position at $t$. Using the usual time slicing definition of the path integral, we integrate the wave function $\psi(x,t)$ over all $x$ on the slice $t$ and then also integrate over all paths leading from each $x$ at $t$ to $y$ at $t+\epsilon$:
\bea
\psi(y,t+\epsilon) = \int_{-\infty}^\infty dx\;\;\psi(x,t)\int_{x(t)=x}^{x(t+\epsilon)=y} Dx(t) \exp \ {i\int_t^{t+\epsilon} (m\dot{x}^2/2 - V(x)) dt} \label{SchrPath}
\qquad
\eea
We take the usual kinetic-minus-potential form for the action. From this point, the emergence of Schrodinger equation is essentially automatic: since the derivative piece blows up for large differences between $x$ and $y$, the path integral gets most of its contribution from $x$'s close to $y$. Making this approximation and evaluating the integral approximately, one can reproduce the time dependent Schrodinger equation in the form
\bea
i\frac{\partial \psi}{\partial t}=\Big(-\frac{\nabla^2}{2m}+V(x)\Big) \psi \equiv H \psi.
\eea
We will not do it in detail: it is elementary and only the origin of the equation (\ref{SchrPath}) is interesting for our purposes.

To derive an analogue for the wave function of the Universe, one might hope to proceed similarly. But there is one big conceptual difference. This is because in quantum mechanics time is a parameter. But in a diffeomorphism invariant theory, time is merely a crutch, a coordinate that is really not there in the full quantum theory. In fact we will see that the Wheeler-DeWitt equation is effectively the statement that the Hamiltonian annihilates the physical states, instead of evolving them. It should also be noted that since we are using a functional integral approach which is not really well-defined for gravity, the ``derivation" below should be taken with a grain of salt.

First we write down the action that goes into the path integral in a form that treats the 3-geometry in a special way. This is the ADM decomposition, where one thinks of the spacetime as spatial slices at constant $t$ and writes the metric as
\bea
ds^2=-(N^2-N_iN^i)dt^2+2 N_i dx^i dt+h_{ij}dx^idx^j,
\eea
where $h_{ij}$ is the three metric and the indices are raised and lowered using it. $N$ is called the lapse, $N_i$ is the shift. The Einstein action, including the boundary piece is the Lorentzian version of (\ref{ES}),
\bea
S=\frac{1}{16 \pi G}\int_{\cal M} \sqrt{g}\ (R-2 \Lambda)\ d^4 x + \frac{1}{8 \pi G} \int_{\partial {\cal M}} \sqrt{h}\ K d^{3} x,
\eea
where we have included a cosmological constant and specialized to four dimensions, but ignored matter terms because they add nothing substantial in our present discussion. For the ADM metric, this action takes the form
\bea
S=\frac{1}{16 \pi G}\int d^4 x\sqrt{h}N(K_{ij}K^{ij}-K^2+{}^{(3)}R-2 \Lambda)
\eea
where $K_{ij}$ is the extrinsic curvature and ${}^{(3)}R$ is the Ricci scalar of the 3-metric $h$.

Translation in field-space is a symmetry of the path integral (if the measure is invariant). This is usually interpreted as the equation of motion for the field, being satisfied as an operator equation. For the lapse field $N$, this gives
\bea
0=\int Dg \frac{\delta S}{\delta N} \exp (iS[g]).\label{varact}
\eea
Classically,
\bea
\frac{\delta S}{\delta N}\equiv\sqrt{h}(K^2- K_{ij}K^{ij}+{}^{(3)}R-2 \Lambda)=0
\eea
is the Hamiltonian constraint of general relativity. Using the fact that $K_{ij}$ involves only first derivatives of $h_{ij}$ we can invert for it using the canonically conjugate momenta $\pi_{ij}$ of $h_{ij}$. That is, we can solve
\bea
\pi_{ij}=-\sqrt{h} (K_{ij}-h_{ij}K),
\eea
for $K_{ij}$ in terms of $\pi_{ij}$:
\bea
K_{ij}=-\frac{1}{\sqrt{h}}\Big(\pi_{ij}-\frac{h_{ij}}{2}\pi\Big).
\eea
Furthermore, noting that one can write these canonical momenta as functional derivatives acting on the path integral via
\bea
-i\frac{\delta}{\delta h_{ij}}\int Dg \exp(iS[g])=\int Dg \ \pi^{ij}\exp(iS[g]),
\eea
we can finally rewrite (\ref{varact}) as
\bea
\Big(G_{ijkl}\frac{\delta^2}{\delta h_{ij} \delta h_{kl}}+\sqrt{h}({}^{(3)}R-2 \Lambda)\Big) \Psi[h_{ij}]=0, \label{WdW}
\eea
interpreted as an operator equation, where by defining
\bea
G_{ijkl}=\frac{1}{2\sqrt{h}}(h_{ij}h_{kl}+h_{il}h_{jk}-h_{ij}h_{kl}),
\eea
we have picked a specific operator ordering prescription. Here $\Psi[h_{ij}]=\int Dg \exp (iS[g])$. Equation (\ref{WdW}) is the Wheeler-DeWitt equation and is the analogue of the Schrodinger equation in the gravitational context. Schematically it is  often written as
\bea
\hat H |\psi \rangle=0
\eea
to emphasize that it is the quantum version of the Hamiltonian constraint. It emphasizes the problem of time in quantum gravity, in that time evolution in the canonical sense is non-existent. We have suppressed matter terms in this discussion, but they are straightforward to add. To make the WDW equation tractable, one often works with a minisuperspace approximation, where an ansatz is adopted for the metric degrees of freedom. An example is cosmology where the scale factor of an FRW ansatz is treated as the only metric degree of freedom in the quantum theory.

\section{Backreaction and Renormalization}

Once we have quantized the fields, we can go ahead and compute composite operators involving them using the standard principles of quantum field theory. Because of the short-distance divergences inherent in QFT, this will require us to do renormalization. For scalar fields in curved spacetime, the interesting composite operator to compute is the stress tensor which one can think of as a source term for the expectation value form of the Einstein equation (cf. Introduction). The stress tensor can also be used to compute physically interesting quantities like particle fluxes in curved backgrounds. Eg. the Hawking flux from a black hole horizon.

The stress tensor is an operator that is quadratic in the fields, and therefore requires renormalization to define it. Doing this will be our goal in this section. We start first by making some comments about the precise nature of the divergences.

\subsection{Overview: Why and What}

The necessity to renormalize is not a phenomenon restricted to curved space. In flat spacetime, if one computes the stress tensor for a scalar field, one sees divergences, this is essentially what the zero point energy is. In flat space, we just subtract this and things work out fine. But in curved space, such a simple subtraction scheme is not obvious. In fact, spacetimes which can be reduced to flat space limit by tuning some parameters, often have divergences in the stress tensor on top of those that are visible in the flat space limit. This means that we need to have a more general approach towards regularizing and renormalizing stress tensors in curved spacetime.

The basic trouble is that a product of operator-valued distributions (i.e., fields) need not be another operator valued distribution. The simplest example is the two point function between fields at spacetime points $x$ and $y$. When $x$ approaches $y$ the 2-pt. function blows up. What one does when dealing with such infinities is to (1) define a parameter called the regulator so that for finite values of the regulator the operator is finite, but in the limit where the regulator is taken to infinity\footnote{Our mental picture of the regulator is as a cut-off in a momentum space integral and therefore we talk about the limit where it is taken to infinity. But there can be various kinds of regulators - like dimensional, point-splitting, lattice etc. - where the divergence arises not when the regulator is taken to infinity, but to some other limit.} it goes back to the original divergent quantity, (2) subtract a part of the regulated operator in such a way that the remainder stays finite when we take the regulator to infinity.  The way one implements these subtractions is usually by the addition of ``counter-terms" which cancel the divergences. The various renormalization schemes correspond to the addition of counter-terms that differ in their finite parts: the infinities need of course to be canceled in any consistent choice of counter-terms.

In renormalizing the stress-tensor, potentially there are ambiguities. But Wald, elaborating on the work of Christensen, has showed that demanding certain reasonable conditions essentially fixes all these ambiguities (see section 6.6 of Birrell\&Davies for a detailed discussion). We will demand that the counter-terms to be local c-number subtractions. Locality is a natural condition to impose because the divergences arise from short-distance effects which have nothing to do with long-distance phenomena. The demand for c-number subtractions is by analogy with usual field theory renormalization in flat space: we expect the counter-terms to be independent of the state in which $\langle  T_{\mu\nu}^{\rm ren}\rangle$ is evaluated\footnote{Of course, this does not say anything about the background in which the renormalization is being defined. Indeed, the counter-terms will depend on the background, but this dependence will be local.}. Here ``${\rm ren}$" stands for the renormalized stress tensor. A further condition is that at any finite value of the cut-off, we also expect that the divergent parts are conserved tensors since the stress tensor is covariantly conserved.  These conditions essentially fix the form of the subtractions uniquely by dimensional analysis.  If $\epsilon$ is the short-distance length scale, the fact that the stress tensor has dimensions of (length)$^{-4}$ together with dimensional analysis and the  balancing of tensor indices  enables us to write the expectation-value-form of the Einstein equations as
\bea
R_{\mu \nu}-\frac{1}{2} g_{\mu\nu} R +\Lambda_B g_{\mu \nu}=-8 \pi  G_B\langle  T_{\mu\nu}^{\rm reg}\rangle \equiv \hspace{0.5in} \nonumber\\
=-8 \pi  G_B\Big[\langle  T_{\mu\nu}^{\rm ren}\rangle+\alpha_{\rm reg} g_{\mu\nu}+\beta_{\rm reg} G_{\mu\nu}+\gamma_{\rm reg} H_{\mu\nu}^{(1)}+\delta_{\rm reg} H_{\mu\nu}^{(2)}\Big], \label{renEin}
\eea
with $H_{\mu\nu}^{(1)}$ and $H_{\mu\nu}^{(2)}$ are certain specific dimension four operators defined in terms of the Riemann tensor, the metric and their covariant derivatives. The explicit formulas for the $H$'s can be found in eqns. (6.53-6.54) of Birrell\&Davies. Here $\alpha, ..., \delta$ are the regulator dependent divergent terms, the subscript $B$ stands for bare parameters, and the subscript $R$ for renormalized objects. The basic idea here is that because of symmetries and dimensional analysis,  the only allowed divergences are of the form $1/\epsilon^4$, $1/\epsilon^2$ and $\log \epsilon$, whose coefficients must be proportional to covariantly conserved tensors of dimension zero, two and four respectively. These conserved tensors are precisely $g_{\mu\nu}$, $G_{\mu\nu}$ and the two dimension four operators listed above. In general dimensions, there is one more operator at this order that one could write down, but in four dimensions (which is what we will restrict our attention to) the fact that the Gauss-Bonnet term
\bea
\sqrt{-g}(R_{\mu\nu\rho\sigma}R^{\mu\nu\rho\sigma}-4 R_{\mu\nu}R^{\mu\nu}+R^2)
\eea
is metric-independent (it's integral computes the Euler number and is a topological invariant of the spacetime) means that its variation with respect to the metric vanishes. This can be used to express the third operator in terms of $H^{(1)}$ and $H^{(2)}$, so we will not need it. By moving things to the LHS and dividing by (bare and regulator-dependent) constants, one can bring (\ref{renEin}) to the form
\bea
G_{\mu\nu}+\Lambda_{R}g_{\mu\nu}+\alpha_R H_{\mu\nu}^{(1)}+ \beta_R H_{\mu\nu}^{(2)}=-8 \pi G_{R} \langle T_{\mu\nu}^{\rm ren}\rangle \label{RenormEinstein}
\eea
Everything in this equation is finite. Note that the four-derivative terms were not there originally in the Einstein equations, but they are induced by the divergences: the correct way to think of the renormalized values of their couplings is by introducing bare couplings $\alpha_B$ etc. for them in the first place, and then thinking of the renormalized values as arising as finite combinations of the bare values and the regulator dependent couplings together. 
Of course, since we know that standard general relativity is remarkably successful experimentally, these renormalized couplings are extremely tiny.

\subsection{Stress Tensor Renormalization: The Setup}
\label{srs}

In this section we will present some of the details of stress tensor renormalization that was sketched in the last section. Our aim will be to describe the renormalization procedure and demonstrate the emergence of (\ref{RenormEinstein}), but {\em not} to compute the renormalized stress tensor appearing on the right hand side of (\ref{RenormEinstein}) for any specific spacetime. Both tasks are complicated, but the first one will be useful to us in deriving the trace anomaly in the next subsection. For the second task, which is often not doable in a closed form in any but the simplest backgrounds, we will merely refer the reader to section 6.4 of Birrell\&Davies \cite{Birrell} and the references there.

We start with the generating function for Green's functions defined for the matter part of the action:
\bea
Z[J]=\int D[\phi] \exp\Big(iS_m[\phi]+i\int d^nx J(x)\phi(x)\Big)
\eea
Here $S_m$ is understood to be the matter action, appropriate to the background geometry, and we have kept the dimension $n$ general because we will be using dimensional regularization. Note incidentally, that the matter part of the action can also contain fluctuations of the metric because we will be working only up to one loop level (see the discussion in the Introduction). The viability of a functional integral of the above form rests on the assumption that there are suitable vacuum states
\bea
|0, {\rm in}\rangle \ \ {\rm and} \ \ |0, {\rm out}\rangle
\eea
in the system so that $Z[0]$ is the vacuum-to-vacuum propagator:
\bea
Z[0]=\langle 0, {\rm out}|0, {\rm in}\rangle_{J=0}.
\eea
Note that because of the essential non-uniqueness  of vacuum states in curved spacetime, this quantity is not necessarily equal to unity. Varying the generating function with respect to the field, we obtain {\em Schwinger's variational principle}:
\bea
\delta Z[0]=i\int D[\phi]\delta S_m \ \exp iS_m[\phi]=i \langle 0, {\rm out}|\delta S_m|0, {\rm in}\rangle_{J=0}.
\eea
By analogy with the classical definition of the stress tensor as
\bea
T_{\mu\nu}=\frac{2}{\sqrt{-g}}\frac{\delta S_m}{\delta g^{\mu\nu}},
\eea
we may now define the quantum stress tensor by
\bea
\langle T_{\mu\nu} \rangle\equiv \frac{\langle 0, {\rm out}|\delta S_m/\delta g_{\mu\nu}|0, {\rm in}\rangle_{J=0}}{\langle 0, {\rm out}|0, {\rm in}\rangle_{J=0}}=\frac{2}{\sqrt{-g}}\frac{\delta W}{\delta g^{\mu\nu}},\label{unrenT}
\eea
where $W$ is given by
\bea
W\equiv-i \ln Z[0].
\eea

So in order to define a systematic renormalization procedure for the stress tensor we need to systematically identify the divergences in $W$, or equivalently, $Z[0]$. For a free field theory, which is the only case we consider, one can explicitly evaluate the (Gaussian) path integral that defines $Z[0]$. This is a standard text-book result in flat spacetime (see \cite{Peskin, Ramond} or \cite{Birrell}) and the derivation is essentially unchanged in curved space, so we will only quote the result:
\bea
W=-\frac{i}{2}{\rm tr}[\ln(-G_F)] \label{trlnform}
\eea
Computing this object is therefore what needs to be accomplished in order to describe the renormalization of the stress tensor.

Here $-G_F$ is to be thought of as an operator (matrix) with continuous row and column indices $x$ and $y$, which is forced to be diagonal via a delta function:
\bea
G_F(x,y)=\frac{1}{(\nabla^2_x+m^2+\xi R-i\epsilon)}\frac{\delta^n(x-y)}{\sqrt{-g(y)}} \equiv K_{xy}^{-1}.
\eea
Last line is the definition of the Kernel $K$ which we will useful.
It is very convenient to think of the matrix as acting on the bra-ket basis defined by $|x \rangle$ so that
\bea
\langle x' |x \rangle = \frac{\delta^n(x-y)}{\sqrt{-g(y)}}, \ \ , \langle x' |G_F|x \rangle=G_F(x, x'),\ \ ,
{\rm Tr} M\equiv \int d^n x \sqrt{-g(x)} \langle x |M|x \rangle.
\eea
for any operator $M$.

To compute $G_F$ in general backgrounds, we need some technology that has close connections with something called the ``Heat Kernel". It is a widely useful approach for dealing with the spectrum of the Laplace operator and therefore is a recurrent theme in theoretical physics, so we will take a digression to discuss it.

\subsection{An Aside on the Heat Kernel}

The aim is to get a handle\footnote{We want to diagonalize the Laplacian so we can do manipulations with it. Diagonalization is the same problem as computing the spectrum of eigenvalues.} on the Laplacian (or more precisely the d'Alembertian, since we are in Lorentzian signature). For this, the heat kernel approach considers an auxiliary problem. To compute the spectrum of $\nabla^2_x$, we consider instead an associated heat equation:
\bea
\partial_s {\cal K}(s,x,x')+\nabla^2_x {\cal K}(s,x,x')=0.
\eea
The solution ${\cal K}$ will be called the Heat Kernel. Often, the practical advantage in introducing the extra structure of the heat equation is simply that it provides us with an extra variable $s$, which when inside integral expressions, can often be interchanged with other integrals. A formal solution for the heat equation can be written as
\bea
{\cal K}(s,x,x')=\langle x| e^{-s \nabla^2_x}| x'\rangle. \label{formalHK}
\eea
Note also that the formal expression also implies that we are tacitly assuming an initial condition:
\bea
{\cal K}(0,x,x')=\frac{\delta^n(x-x')}{\sqrt{-g(x')}}.
\eea
It is possible to write down the explicit expression for the heat kernel in some cases. The canonical example that is easily verified, is the heat kernel for flat Euclidean space $\IR^d$,
\bea
{\cal K}(s,x,x')=\frac{1}{(4 \pi s)^{d/2}}e^{-|{\bf x}-{\bf x}'|^2/4 s},
\eea
which is a solution with the correct initial condition in $s$ and solves the heat equation for all $s>0$. This is basically all we will need to know about the heat kernel, per se. In the following, we can in principle avoid mentioning the heat kernel altogether: but the language is closely related, hence the above digression.

Now, we will compute an expression that is convenient for extracting divergences in our ``${\rm Tr} \ln$" expression for the generating function $W$ (which contains the Green function for the d'Alembertian). Since we are interested in short-distance divergences, we want to evaluate $W$ when $x \rightarrow x'$. On a general spacetime, the metric around $x'$ can be taken as
\bea
g_{\mu\nu}(x)=\eta_{\mu\nu}+\frac{1}{3}R_{\mu\alpha\nu\beta}y^\alpha y^\beta+...
\eea
where $y$ are the Riemann normal coordinates of the point corresponding to $x$, around $x'$ as the origin. Note that $R_{\mu\alpha\nu\beta}$ and all higher coefficients are evaluated at $x'$, i.e., $y=0$. In these coordinates, one can solve the wave equation\footnote{For minimal coupling, $\xi=0$, and for the so-called conformal coupling in 4 dimensions, $\xi=1/6$. We will keep $\xi$ general for convenience when discussing the trace anomaly in a later section.}
\bea
(\nabla^2_x +m^2+\xi R)G_F(x,x')=-\sqrt{-g(x)}\delta^n(x-x')
\eea
as an expansion. The idea is that Riemann normal coordinates are a sort of Taylor expansion around the point $x'$ and we can write the wave operator as a derivative expansion and then solve it iteratively. Such an expansion is called an adiabatic expansion. We will not do it in detail, the basic idea is the same as in any perturbation expansion, see eg. Ramond \cite{Ramond}\footnote{The authoritative reference on the applications of the heat kernel to quantum field theory and gravity is Bryce DeWitt's \cite{DeWittQFT}. Be warned, however, that it is a stiff book.}. We will merely write down the final result, and to do that it is useful to define
\bea
{\cal G}_F(k) =
\int d^n ke^{i\eta^{\alpha \beta}k_\alpha y_\beta}(-g(x))^{1/4}G_F(x,x').
\eea
Note that this Fourier transformation is useful and meaningful only because we are working in a derivative expansion in Riemann normal coordinates around $y=0$: $y$'s are really curvilinear coordinates. Using this momentum space as an intermediate crutch, after tedious computation, we can show that the adiabatic expansion for the Green function is
\bea
(-g(x))^{1/4}G_F(x,x')\approx \hspace{4.5in}\nonumber \\
\approx\frac{1}{(2\pi)^n}\int d^n ke^{-i\eta^{\alpha \beta}k_\alpha y_\beta} 
\Big\{a_0(x,x')-a_1(x,x')\frac{\partial}{\partial m^2}+a_2(x,x')\frac{\partial^2}{(\partial m^2)^2}+ ...\Big\}\frac{1}{k^2-m^2+i\epsilon} 
\eea
where dots represent higher adiabatic terms which can in principle be systematically computed. The DeWitt coefficients (also called Heat Kernel coefficients or DeWitt-Seeley-Gilkey coefficients) are defined as
\bea
a_0(x,x')=1,\ \ a_1(x,x')=(1/6-\xi)R+..., \ \ a_2(x,x')=..., \ \ ...
\eea
where again dots represent higher order adiabatics which are in principle computable. They are all geometrical quantities and they are expressed as expansions in $y$, with coefficients evaluated at $y=0$. The $i \epsilon$ prescription has been introduced as a specific choice of contour: we are working with Feynman Green functions. Other contours can be chosen and will yield other Green functions.

Up to this point, we have merely found a derivative expansion for the Green function. Now, using the trick of writing 
\bea
\frac{1}{k^2-m^2+i\epsilon} =-i\int_0^\infty ds \exp is (k^2-m^2+i\epsilon),\label{trick1}
\eea
we can exchange the $k$ and $s$ integration and  perform the former explicitly. The result (again after some work), when written in a general (as opposed to Riemann normal) coordinates is
\bea
G_F(x,x')=-i \frac{\Delta(x,x')^{1/2}}{(4 \pi)^{n/2}}\int_0^\infty i ds \ (is)^{-n/2}\exp\Big(-im^2 s+\frac{\sigma}{2is}\Big) F(x,x',s), \label{DWS}
\eea
where the Van Vleck-Morette determinant is defined via
\bea
\Delta(x,x')=-\frac{{\rm det}(\partial_\mu\partial_\nu\sigma(x,x'))}{\sqrt{g(x)g(x')}},
\eea
with $\sigma(x,x')$ being half the proper distance ($=\frac{1}{2}g_{\mu\nu}y^\mu y^\nu$ in Riemann coordinates). We use the adiabatic expansion
\bea
F(x,x',s)\approx  a_0(x,x')+a_1(x,x') is+a_2(x,x')(is)^2+... \label{DPS}
\eea
The above ``DeWitt-Schwinger" representation of the Green function is what we will use in order to compute the divergences and perform renormalization.

\subsection{Stress Tensor Renormalization: The Result}

Going back to the $G_F$ and its inverse $K$ defined in subsection \ref{srs}, we have
\bea
\langle x| G_F | x'\rangle=\langle x| K^{-1}| x'\rangle=-i\int_0^\infty \langle x| \exp(-iK s)| x'\rangle \ ds \label{compare}
\eea
where we have used a trick analogous to the one used in (\ref{trick1}) to write the last step. Comparing (\ref{compare}) with (\ref{DWS}) gives us an expression for $ \langle x| \exp(-iK s)| x'\rangle$ in terms of the DeWitt expansion. To compute the generating functional $W$, what we need is not $\langle x| G_F | x'\rangle$, but $\langle x|\ln(- G_F )| x'\rangle$ (after tracing/integrating over the coordinates after setting $x=x'$). In the compact operator notation, this latter object can be written as
\bea
\ln (-G_F)=-\ln K= \int_0^\infty \exp(-iK s) \ \frac{ds}{s}.
\eea
The final expression is divergent in the lower limit, but if we introduce a lower cutoff and then take it to zero after the computation, one finds that this divergence is a metric independent constant that is present also in flat spacetime. So we will discard it and consider only the finite part, which is $-\ln K$. Plugging in the DeWitt expansion for $\exp(-iK s)$ that we mentioned above, we can immediately write this as
\bea
\langle x| \ln (-G_F)| x'\rangle=-\int_{m^2}^\infty G_F(x,x') dm^2.
\eea
where the DeWitt form is understood on the right hand side: in particular, the structure of the DeWitt expansion is of such a form that the integral over $m^2$ brings down the necessary factor of $1/s$.

Putting these ingredients together and using (\ref{trlnform}) we can finally write
\bea
W \equiv \int d^n x \sqrt{-g} \ L_{\rm eff}=\int d^n x \sqrt{-g}\Big(\frac{i}{2}\lim_{x\rightarrow x'}\int_{m^2}^{\infty} dm^2 G_F (x,x')\Big),
\eea
where again the Green function on the right hand side is understood to be in the DeWitt expansion form. This is the one-loop effective action.

Renormalization is necessary because the expression for $W$ we have written above contains divergences: they arise from the lower end of the integral over $s$ in the DeWitt expansion when $x \rightarrow x'$ and the exponent vanishes. This limit (sometimes called the ``coincidence limit") corresponds to short distances and it is natural that we have UV divergences. The divergences arise only from a finite number of low order (in powers of $s$) terms in  (\ref{DPS}). For generic dimension $n$ the first $n/2+1$ terms in the DeWitt expansion are divergent.

To regularize the divergences, we will use dimensional regularization and take $n$ to be complex. Then the coincidence limit can be taken without blow ups and the integrals can be performed to yield simple gamma functions. The result is
\bea
L_{\rm eff} \approx \sum_{j=0}^{\infty}a_j(x,x)(m/\mu)^{n-4}\frac{(m^2)^{2-j}}{2(4 \pi)^{n/2}}\Gamma(j-n/2). \label{DWgamma}
\eea
At this stage we have revealed our ultimate interest in 4 dimensions by fixing the mass dimenion of the effective Lagrangian to be 4: to enforce this when $n$ is analytically continued, we have introduced the arbitrary mass scale $\mu$ as usual in dimensional regularization.

When $n \rightarrow 4$, the divergences emerge as poles of the gamma function for $j=0,\ 1, \ 2$. Expanding the expression above in $(n-4)$, the divergent part of $L_{\rm eff}$ can therefore be separated. The explicit form of the dimensionally regulated divergences are easy to write down from our previous expressions, but we will not need it. Instead we only need the schematic expressions
\bea
L_{\rm div}={\rm div}_0 \times a_0(x,x)+{\rm div}_1 \times a_1(x,x)+{\rm div}_2 \times a_2(x,x).
\eea
Here the ${\rm div}$'s stand for terms containing $\frac{1}{n-4}$ which diverge as $n\rightarrow 4$. The $a_j$ are the coincidence limits of the DeWitt coefficients introduced earlier, and are purely geometric, well-defined, scalar quantities defined at $x$ (constructed from various curvatures and their derivatives). The expressions for $a_0$ and $a_1$ are simple,
\bea
a_0(x,x)=1, \ \ \ a_1(x,x)=(1/6-\xi) R,
\eea
while the expression for $a_2$ is nasty:
\bea
a_2(x,x)=\frac{R_{\mu\nu\rho\sigma}R^{\mu\nu\rho\sigma}}{180}
-\frac{R_{\mu\nu}R^{\mu\nu}}{180}+\frac{(1/6-\xi)^2R^2}{2}-\frac{(1/5-\xi)\nabla^2 R}{6}.
\eea
Note that the last term gives rise to a total derivative and often will not play a role in discussions of the action. The structure of these divergences  (except for the coefficient of $a_2$) is precisely of the form that was already present in the gravitational part of the original action. So we can treat the original coupling constants present in the ``bare" gravitational action as getting renormalized due to these (regulated) one loop divergences resulting in a finite (experimentally accessible) final result. The divergence associated with $a_0$ gets absorbed into the cosmological constant and the one associated to $a_1$ renormalizes $G_N$.  It is crucial for the result of this to be  finite, that the bare couplings themselves have compensating divergences in them.

The $a_2$ term is not present in the standard gravitational action. But this is not particularly problematic because what we measure are the renormalized values. So we can assume the existence of a piece corresponding to $a_2$ in the bare action, and then take the bare value to precisely\footnote{Or at least, up to experimental limits.} cancel the UV divergence associated to $a_2$, resulting in a zero (to within experiment) net result.

Note that we are treating only matter (including metric fluctuations) quantum mechanically at one loop, as discussed in the introduction. The conclusion is that the one loop corrections from the matter part, correct the gravitational part of the action, but these corrections can be thought of as corrections to the cosmological constant and Newton's constant, together with the couplings of the new terms arising from $a_2$. As a result we have a renormalized action that is finite. The final message is this: We started out with classical gravity and quantum matter
\bea
S=S_g+W.
\eea
At one loop, the quantum matter has divergences, but these can be absorbed into appropriately defined couplings of the gravitational part, with the result
\bea
S=S_{g, {\rm ren}}+W_{\rm ren}, \ \ {\rm where} \ \ W_{\rm ren}=W-W_{\rm div}, \ \ {\rm and}\ \ S_{g, {\rm ren}}=S_g+W_{\rm div}. \label{asif}
\eea
In this renormalized expression, all the couplings are physical, finite and renormalized. Both $S_{g, {\rm ren}}$ and $W_{\rm ren}$ are finite. The variation of this action written in terms of renormalized quantities therefore takes the form
\bea
G_{\mu\nu}+\Lambda_{R}g_{\mu\nu}+\alpha_R H_{\mu\nu}^{(1)}+ \beta_R H_{\mu\nu}^{(2)}=-8 \pi G_{R} \langle T_{\mu\nu}^{\rm ren}\rangle
\eea
where
\bea
\langle T_{\mu\nu}^{\rm ren}\rangle =\frac{2}{\sqrt{-g}}\frac{\delta W_{\rm ren}}{\delta g^{\mu\nu}}\equiv \frac{\langle 0, {\rm out}|T_{\mu\nu}|0, {\rm in}\rangle_{J=0, {\rm ren}}}{\langle 0, {\rm out}|0, {\rm in}\rangle_{J=0}}. \label{renT}
\eea
is the renormalized version of (\ref{unrenT}).
Note also that the variation of the renormalized gravitational action contains terms coming from variation of $a_2$, which contains Riemann$^2$, Ricci$^2$ and Ricci-Scalar$^2$ (ignoring a total derivative term that does not contribute to the equations of motion). As discussed before, the topological (in four dimensions) Gauss-Bonnet term is built up precisely from a linear combination of these three quantities, so the resultant pieces that show up in the equations of motion can be taken to be two (as opposed to three) independent couplings that are the coefficients of $H^{(1)}_{\mu\nu}$ and $H^{(2)}_{\mu\nu}$. The explicit forms of these two quantities involve curvature tensors and their derivatives: they can be straightforwardly (but complicated-ly) computed by varying the curvature scalars, they can be found in the references, and they will not be important in our future discussion. They contain four-derivatives (i.e., they are adiabatic order 4) and are to be thought of as higher derivative corrections to the Einstein equations.

Another important thing to be observed here is that the renormalized stress tensor as it appears in the above expression is not really an expectation value, rather it is evaluated between two different states, one in the past and one in the future. This was enough for computing the divergences of the stress tensor: one can show by expanding (for example) the out-vacuum in terms of the in-basis, that the divergent part of a genuine stress tensor expectation value like $\langle 0, {\rm out}|T_{\mu\nu}|0, {\rm out}\rangle$ is the same as in (\ref{renT}).
This ties in with the expectation that the short-distance divergences are state independent. They are also independent of the global properties of the background geometry as illustrated by the fact that the divergences were proportional to locally computable curvature scalars.

\subsection{Conformal Anomaly}

A conformal (or Weyl) transformation is a rescaling of the metric that can be position-dependent. An action that remains invariant under a conformal transformation is called conformally invariant. Since the metric measures distances, an action of this form is to be thought of as scale invariant. Note that a mass term typically breaks conformal invariance of the action. A massless scalar has a chance of being conformally invariant. In fact, a minimally coupled scalar field in a curved geometry is {\em not} conformally invariant even if it is massless, but it can be checked that in four dimensions an action of the form
\bea
S=\int d^4 x \sqrt{-g} (\nabla_a \phi \nabla^a \phi +\xi R \phi^2)
\eea
is in fact conformally invariant when $\xi=1/6$. This can be explicitly checked by computing the variations of the Christoffel symbols and the Ricci scalar under a conformal transformation
\bea
g_{\mu\nu}\rightarrow \Omega^2(x) g_{\mu\nu},
\eea
and noting that the variations can be compensated if the scalar transforms under conformal re-scalings as
\bea
\phi(x) \rightarrow \phi(x)/\Omega(x).
\eea
A theory of this form is called a conformally coupled scalar field theory. The stress tensor for the action above can be computed as
\bea
T_{\mu\nu}=\frac{2}{\sqrt{-g}}\frac{\delta S}{\delta g^{\mu\nu}}.
\eea
It is easy to check directly that this stress tensor is traceless $T^{\mu}_{\mu}=0$. Tracelessness of the classical stress tensor is in fact a general consequence of conformal invariance. To see this, note that the stress tensor is defined by
\bea
\delta S \sim \int d^4 x \sqrt{-g} T_{\mu\nu}\delta g^{\mu\nu}
\eea
For an infinitesimal conformal transformation, $\delta g^{\mu\nu} = (2 \delta \omega) g_{\mu\nu}$. (Note that $g'_{\mu\nu}=g_{\mu\nu}+\delta g_{\mu\nu}=(1+2\delta \omega)g_{\mu\nu}\approx e^{2\delta \omega}g_{\mu\nu}\equiv \Omega^2 g_{\mu\nu}$.) Plugging this into the expression above, we see that the variation of the action vanishes when $T_{\mu}^{\mu}=0$.

Note that all of this is at the level of the classical theory. The question of tracelessness of the quantum stress tensor is more interesting. This is because in the quantum theory, one is forced to introduce a scale when regulating the divergences. (In dimensional regularization, this was the scale $\mu$ that we introduced.) This scale can result in a breaking of scale invariance and the quantum stress tensor need not be conformally invariant. This phenomenon is called the conformal (Weyl/Trace) anomaly. 
The idea is that a conformal transformation, since it affects the metric and therefore changes distances, necessitates a rescaling of the cutoff:
\bea
g_{\mu\nu} \rightarrow (1+2 \delta \omega) g_{\mu\nu}, \ \ \mu \rightarrow (1+\delta \omega) \mu.
\eea
The trace of the quantum stress tensor arising from renormalization, when it is non-vanishing, is called the trace anomaly.

With this general argument in hand, now we go on to see this in detail building on our previous computation of the stress tensor divergences. We will look at a conformally invariant classical theory and see how loop corrections can destroy it. More practically, what we will assume is that the regulated, but unrenormalized effective action (i.e., the $W$) is conformally invariant.

Since the mass term breaks conformal invariance already classically, we will focus on the massless limit. As mentioned already, a conformal scalar in four dimensions corresponds to $\xi=1/6$. The generalization of this for arbitrary $n$ follows from a similar computation and will be important when we dimensionally regulate:
\bea
\xi(n)=\frac{n-2}{4(n-1)}.
\eea
Our starting point is (\ref{DWgamma}), which means that our ultimate interest is in four dimensions. From the structure of the expressions, since we will be looking at $n\sim 4$ and since we are interested in the limit $m^2 \rightarrow 0$, it is clear that for $j>2$, there are divergences. Since these divergences arise from $m^2 \rightarrow 0$, they are best interpreted as IR divergences, and we are interested here in the UV behavior: so these divergences do not concern us. On the contrary, the cases $j=0, \ 1, \ 2$ have no IR divergences: in fact $j=0, \ 1$ vanish without drama. So the only case one has to pay attention to is $j=2$. This leads to the UV divergent part of the effective action (remember that $L_{\rm div}$ is the divergent part of $L_{\rm eff}$):
\bea
W_{\rm div}=\int d^n x \ \sqrt{-g} L_{\rm div}= (m/\mu)^{n-4}\frac{1}{2 (4\pi)^{n/2}}\Gamma(2-n/2)\int d^n x \ \sqrt{-g} a_2(x,x).
\eea
We will show that this regulated/divergent effective action is conformally invariant {\em when} $n=4$. Note that in working with $a_2$, the total derivative piece can be omitted in the action, and that the $R^2$ piece (because of a conspiracy due to the fact that its coefficient contains $\xi=\xi(n)$) vanishes in the $n \rightarrow 4$ limit. This means that we only need to keep track of the Riemann$^2$ and Ricci$^2$ pieces. Rearranging them and taking the $n \rightarrow 4$ limit, one ends up with
\bea
W_{\rm div}=(m/\mu)^{n-4}\frac{1}{2 (4\pi)^{n/2}}\Gamma(2-n/2)\int d^n x \ \sqrt{-g}\ \Big(\frac{1}{120} F-\frac{1}{360}G\Big)+{\cal O}(n-4), \  \ {\rm where} \nonumber \\
F\equiv R_{\mu\nu\rho\sigma}R^{\mu\nu\rho\sigma}-2R_{\mu\nu}R^{\mu\nu}+\frac{R^2}{3}, \ \ G \equiv  R_{\mu\nu\rho\sigma}R^{\mu\nu\rho\sigma}-4R_{\mu\nu}R^{\mu\nu}+R^2. \hspace{0.5in}
\eea
In four dimensions when $n=4$, two special things happen. The first piece, $F$, becomes the square of Weyl curvature $C_{\mu\nu\rho\sigma}$ and therefore it is conformally invariant. The second piece is the Gausss-Bonnet term which is topological in four dimensions and therefore that is also conformally invariant. The net result is that $W_{\rm div}$ is conformally invariant in four dimensions\footnote{Of course, since we are in the strict $n=4$ limit, the regulator has been removed and the coefficient is infinite.}.

But this does not mean that the quantities computed with it away from $n=4$ will be conformally invariant, if one goes back to $n=4$ after the computation. This is the operational origin of the conformal anomaly. Using the expression above we can compute the divergence in the stress tensor as
\bea
\langle T_{\mu \nu}\rangle_{\rm div}=\frac{2}{\sqrt{-g}}\frac{\delta W_{\rm div}}{\delta g^{\mu\nu}},
\eea
and its trace in the $n\rightarrow 4$ limit can be found to be
\bea
\langle T_{\mu}^{\ \mu}\rangle_{\rm div}=\frac{1}{16 \pi^2}\Big(\frac{F-\frac{2}{3}\nabla^2 R}{120}-\frac{G}{360}\Big)=\frac{a_2}{16 \pi^2}.
\eea
Now, immediately from the conformal invariance of $W$, and the relation $W_{\rm ren}=W-W_{\rm div}$ (see \ref{asif}), we see that
\bea
\langle T_{\mu}^{\ \mu}\rangle_{\rm ren}=-\frac{a_2}{16 \pi^2},
\eea
which is the celebrated trace anomaly. The computation using the DeWitt expansion can be readily extended to other dimensions, and one finds that the anomaly exists only in even dimensions. This is because $L_{\rm eff}$ as defined by a generalization of (\ref{DWgamma}) to generic dimensions has only IR divergences in odd dimensions. An especially simple result is obtained in two dimensions (note that $\xi(n=2)=0$):
\bea
\langle T_{\mu}^{\ \mu}\rangle_{\rm ren}= -\frac{R}{24 \pi},
\eea
a result which is of importance in obtaining the correct spacetime dimensionality in (critical) string theories.

\subsection{Renormalized Stress Tensor On Rindler and Schwarzschild}

One of the uses of the stress tensor is that in the absence of a canonical particle concept that is meaningful everywhere, it provides us with other useful local  observables. Particles are merely a choice of positive energy modes chosen globally, and it is not meaningful to ask questions about (eg.) Hawking decay using the notion of particle emission localized at the horizon. On the other hand, one can compute the flux of the stress tensor at the horizon and the answer is physical. We will find that the flux into the horizon is negative, indicating that the hole is indeed loosing energy-momentum due to Hawking radiation.

Our computation in this section will not be the most thorough, but it will capture this basic point. While we have developed the technology for extracting and renormalizing the divergences, we have not developed a systematic approach to computing the (remaining, finite) stress tensor expectation values in specific states (built on specific vacua in specific spacetimes). As we mentioned before, this is a messy subject where the pay-off is not worth the investment. Besides, in the cases of interest to us, we will be able to get the essence of the answer without too much work. We start with Rindler as the usual prototype for black holes and then work by analogy for Schwarzschild.

There are two natural vacua in Rindler, corresponding to the Rindler modes and the Minkowski modes. Without doing any computations, we can proceed by physical arguments: we will declare\footnote{Of course, this needs to be checked by explicit computations of the renormalized stress tensor. It works.} that the expectation value of the renormalized stress tensor is zero in the Minkowski vacuum, because we expect no backreaction on flat space. Now, we know from our discussion of Rindler space that the Rindler observer sees a thermal distribution of quanta in the Minkowski vacuum, and sees nothing in the Rindler vacuum\footnote{The latter is the definition of the Rindler {\em vacuum}.}. In the language of the stress tensor this means that the difference between the energy densities in the two vacua, as measured in the comoving frame of the Rindler observer should be given by the Planckian energy density at the Rindler temperature, $T_R=\frac{1}{2\pi\xi}$: 
\bea
{}_M\langle 0|T^{\rm ren}_{00}|0\rangle_M-{}_R\langle 0 |T^{\rm ren}_{00}|0\rangle_R=\int \frac{d \omega}{2 \pi^2} \frac{\omega^3}{\exp\frac{\omega}{T_R}-1}=\frac{1}{480 \pi^2 \xi^4}.
\eea
The divergent parts are state independent, so the subtraction gets rid of them and this equation is in fact legitimate even if one works with the unrenormalized stress tensors. We are using the notations from our section on Rindler, and it is to be emphasized that the stress tensor expectation value is in a specific frame, namely the one comoving with the Rindler observer (i.e., the $T_{00}$ components are evaluated by the Rindler observer accelerating at the uniform rate $1/\xi$). The Planck/Debye integral is evaluated above via
\bea
I(n) \equiv \int_0^\infty \frac{x^n}{e^{x}-1} dx=\int_0^\infty \frac{x^n e^{-x}}{1-e^{-x}} dx=\int_0^\infty x^n \sum_{k=1}^{\infty} e^{-k x} dx=\sum_{k=1}^{\infty} \frac{1}{k^{n+1}}\int_0^\infty y^n e^{-y} dy\nonumber \\
=\Gamma(n+1)\sum_{k=1}^\infty \frac{1}{k^{n+1}}=\Gamma(n+1)\zeta(n+1), \hspace{1.75in}
\eea
where in one step we have expanded $1/(1-e^{-x})$ in a power series in $e^{-x}$, changed variables to $y=k x$ in another and then used the definitions of the Gamma function and Riemann zeta function. The value relevant for us is $n=3$ in which case $\zeta(4)=\pi^4/90$.

The previous result, together with the fact that  the stress tensor vanishes in Minkowski vacuum  means that the stress tensor as evaluated in the Rindler vacuum, is {\em negative}. This effect is called vacuum polarization. It is because of the presence of the compensation due to vacuum polarization that the Minkowski vacuum, despite being seen as full of hot thermal radiation, has vanishing stress energy to the Rindler observer. Note that as we get closer to the Rindler horizon ($\xi=0$), the vacuum polarization diverges.

Now we are ready to proceed to the black hole case by analogy. One general observation before we begin: as we observed in the discussion on Rindler, the Rindler coordinate system is analogous to Schwarzschild while the Minkowski coordinates are analogous to Kruskal. The natural vacuum analogous to Rindler vacuum therefore  is  Boulware, and the vacuum analogous to Minkowski is Hartle-Hawking. We can make the analogous argument as before regarding the difference between the stress tensors (with the temperature replaced by the appropriate black hole temperature). This quantity is again divergent at the horizon. If one pushes the analogy further and declares that the stress tensor in the HH vacuum is finite like it was in the Minkowski vacuum, then we come to the conclusion that the (renormalized!) stress tensor diverges in the Boulware vacuum at the horizon. These conclusions can be checked by detailed computations of the renormalized stress tensor in the black hole background.

The divergence in the Boulware state is one of the reasons why it is often considered as ill-defined. The Hartle-Hawking vacuum is expected to capture the hole  being in equilibrium with radiation, so incoming and outgoing fluxes balance, so that there is no net flux. In the next section, we will take another point of view on the divergence in the Boulware state.

\subsection{Digression: Hawking Radiation from Gravitational Anomaly}

One of the remarkable things about Hawking radiation is that there are many alternate ways to derive it, demonstrating the robustness of the result. We have seen two ways of doing this already: one was Hawking's original derivation based on wave scattering in a collapsing background (with the input from Unruh which translated it to an equivalent computation in the cleaner Kruskal geometry). This computation was as ``physical" as things could be. The trouble with this derivation is that in the collapsing context, the radiance arose from modes that were stuck close to the horizon for an indefinite amount of time. Because of the horizon red-shift, this translates to modes that started out at extremely high energies (the so-called ``trans-Planckian" modes) in the asymptotic past. It is not so clear to what extent these ultra high energy modes can be reliably used without doing a full quantization of gravity. We also presented another derivation inspired by Euclidean quantum gravity, which was based on the Euclidean extension. While this looks quite elegant, unlike the Hawking computation it does not offer us any physical mechanism for the origin of the radiation.

Instead of clarifying these issues (which are not fully resolved even after forty years from the original discovery of Hawking radiation), in this section we will present a derivation of Hawking radiation from another angle that demonstrates its robustness in yet another way. 
This computation is interesting because it is a direct implementation of the Wilsonian effective field theory philosophy of looking at Hawking radiation via purely low energy experiments, thereby avoiding some of the worries about transPlanckain modes. 

The starting point is the observation that each partial wave mode of a scalar propagating in the black hole background can be re-interpreted as a scalar propagating in the $1+1$ dimensional $(r-t)$ plane, in the near-horizon limit. This involves writing out the scalar action expanded in spherical harmonics and re-interpreting the resulting equations in terms of 1+1 dimensions: this is straightforward, so we will skip the details and refer the reader to Robinson's PhD thesis \cite{phd}. The result is that one effectively has free massless 1+1 dimensional scalar fields, which are labeled by the angular momentum quantum numbers, but are otherwise degenerate. Now, we can define a notion of energy that is well-defined outside the horizon using the Killing vector $\partial/\partial t$. We wish to think of the scalar field theory in terms of an effective field theory where the energy scale is determined by this notion of energy. This is a perfectly reasonable thing to try outside the horizon, but it runs into trouble at the horizon because as we saw in the previous section, the vacuum associated to the $\partial/\partial t$ modes is the Boulware vacuum, and the stress tensor  on this vacuum diverges at the horizon because of the pile-up of outgoing modes.

In the language of effective field theory, this means that there should be a resolution to this problem in the UV which we are just not able to see from the IR scalar field theory. This means that in a theory of near-horizon scalar fields, one has effectively integrated out the troublesome outgoing high energy modes and one is left with an effectively chiral theory\footnote{When there are left moving and right moving degrees of freedom and the theory depends only on either left or right degrees of freedom, we have a chiral theory. In the present case, the modes are functions of $t-r$ and $t+r$ because they are solutions of the massless two dimensional wave equation whose general solution is $f(t+r)+g(t-r)$. The fact that the outgoing modes are too high energy and are inaccessible to low energy experiments means that it should be describable purely in terms of modes that depend only  on $(t+r)$, and therefore we have a chiral theory.}. In other words, an effective low energy experiment at the horizon should be describable only using the degrees of freedom that are accessible to the experiment, and since outgoing modes are too high energy, our effective field theory will contain only ingoing modes and therefore will be chiral.

A crucial ingredient enters the discussion at this point: a chiral scalar theory in 1+1 dimensions suffers from gravitational anomalies. A gravitational anomaly is an inconsistency that shows up in a quantum field theory, when it is coupled to gravity. That is, a theory with gravitational anomalies is fine as a standalone theory, but it cannot arise as a limit of a full theory of the Universe which also contains gravity. Since the full UV complete theory in the black hole background is a consistent theory of nature, it cannot have gravitational anomalies. But the low energy effective theory where we have integrated out the outgoing modes is chiral, and a 1+1 dimensional chiral scalar field is known to be gravitationally anomalous. What gives?

To understand how this problem is fixed we first need to know what the anomaly looks like. Gravitational anomaly is an anomaly arising from a breakdown of diffeomorphism invariance that arises when the theory is coupled to gravity. The way in which it shows up is as the violation of energy-momentum conservation. (Note that diff. invariance is directly responsible for the conservation of energy-momentum in a Lagrangian that contains the metric.) The precise form of the violation of energy-momentum conservation in 1+1 d chiral theories can be looked up \cite{Gaume, Wilczek1}:
\bea
\nabla_{\mu}T_{\nu}^{\mu} = \frac{1}{\sqrt{-g}}
\partial_{\mu}N^{\mu}_{\nu}\equiv A_{\nu},
\label{ganomaly}
\eea
where
\bea
N^{\mu}_{\nu}=\frac{1}{96\pi}\varepsilon^{\beta\mu}\partial_\alpha
\Gamma^{\alpha}_{\nu\beta}.
\label{Nmunu}
\eea
Here $\varepsilon^{\mu\nu}$ is the fully anti-symmetric tensor
($\varepsilon^{01}=1$), and $\Gamma^{\alpha}_{\nu\beta}$ is the
Christoffel symbol on the $(1+1)$-dimensional spacetime whose metric is
\bea
ds^2=-f dt^2+\frac{dr^2}{f}.
\eea
For Schwarzschild, $f=(1-2M/r)$, but nothing really prevents us from keeping it more general. The diffeomorphism anomaly can be avoided (since we know it must be fixed in the low energy theory somehow since the full theory is consistent), if there is a flux of energy-momentum at the horizon that precisely compensates the $A^\nu$. It turns out that this flux is exactly the flux that one expects from Hawking decay of the black hole. This is a simple example of what can be referred to as {\em anomaly inflow}.

Now we proceed to see this in detail following closely the presentation in \cite{Wilczek1}. As explained above, we want to set outgoing modes to zero as a condition at the horizon. We do this by imposing this condition on a slab of width $2 \epsilon$ straddling $r=r_H$ where $r_H$ is the horizon, i.e., $f(r_H)=0$. Now, in any effective theory descending from a consistent UV theory, we expect that the variation of the effective action under diffeomorphisms is zero. That is, we expect that the full energy-momentum tensor, arising from the variation of the effective action, is indeed conserved. The full stress tensor can be split into three pieces: one piece each on either side of the $2 \epsilon$-slab and one on the slab. So we write the energy-momentum tensor as
\bea
T^{\mu}_{\nu}=T^{\mu}_{{\bf I}\,\nu}\Theta_{+} +
T^{\mu}_{{\bf O}\,\nu}\Theta_{-} + T^{\mu}_{{\bf X}\,\nu}H,
\eea
where $\Theta_{\pm}=\Theta(\pm(r-r_H)-\epsilon)$ are step-functions on either side of the slab and $H=1-\Theta_{+}-\Theta_{-}$ is the ``top-hat"function that is non-zero only on the slab. The indices ${\bf I}$ and ${\bf O}$ stand for inside and outside the horizon-straddling slab, and ${\bf X}$ stands for the on-the-slab piece which is chiral\footnote{$X$ is to be read as (capital) $\chi$, for chiral.}. The quantities $T^{\mu}_{{\bf I}\,\nu}$ and $T^{\mu}_{{\bf O}\,\nu}$ are covariantly conserved inside and outside the horizon, respectively. However,
$T^{\mu}_{{\bf X}\,\nu}$  is the anomalous chiral piece, and is not covariantly conserved. Imposing the condition that the variation of the full effective action vanishes under a diffeomorphism $x^{\mu} \rightarrow x^{\mu} + \lambda^{\mu}(x)$, (i.e., the full energy momentum tensor is conserved) yields
\bea
0&=&\int d^{2}x
\sqrt{-g}\lambda^{\nu}\nabla_{\mu}\left\{T^{\mu}_{{\bf I}\,\nu}\Theta_{+}
+ T^{\mu}_{{\bf O}\,\nu}\Theta_{-} + T^{\mu}_{{\bf X}\,\nu}H\right\}
\nn \\
&=&\int d^{2}x
\sqrt{-g}\lambda^{t}\left\{\partial_{r}\left(N^{r}_{t}H\right)+
\left(T^{r}_{{\bf O}\,t}-T^{r}_{{\bf X}\,t}+N^{r}_{t}\right)\partial_r\Theta_{+}
+
\left(T^{r}_{{\bf I}\,t}-T^{r}_{{\bf X}\,t}+N^{r}_{t}\right)\partial_r\Theta_{-}\right\}\nn\\
&&+\int d^{2}x \sqrt{-g}\lambda^{r}
\left\{\left(T^{r}_{{\bf O}\,r}-T^{r}_{{\bf X}\,r}\right)\partial_r\Theta_{+}
+
\left(T^{r}_{{\bf I}\,r}-T^{r}_{{\bf X}\,r}\right)\partial_r\Theta_{-}\right\}.
\eea
This equality should be understood as holding in the $\epsilon \rightarrow 0$ limit.

Purely on general grounds by inspection, an energy-momentum tensor that satisfies equation (\ref{ganomaly}) and is independent of $t$ can be solved as
\bea
T^{t}_{t}=-\frac{(K+Q)}{f}-\frac{B(r)}{f}-\frac{I(r)}{f}+T^{\alpha}_{\alpha}(r), \hspace{0.5in}\\
T^{r}_{r}=\frac{(K+Q)}{f}+\frac{B(r)}{f}+\frac{I(r)}{f}, \ \
T^{r}_{t}=-K+C(r)=-f^{2}T^{t}_{r},
\eea
where we have defined
\bea
B(r)= \int^{r}_{r_H}f(x)A_{r}(x)dx, \ \ C(r)= \int^{r}_{r_H}A_{t}(x)dx, \ \
I(r)= \frac{1}{2}\int^{r}_{r_H}T^{\alpha}_{\alpha}(x)f'(x)dx.
\eea
The constants $K$, $Q$, and the trace $T^\alpha_\alpha(r)$ are
undetermined. These expressions hold for the chiral piece because we used the anomalous equation (\ref{ganomaly}), but note that the same symmetry arguments also fix the form of the outside and inside pieces, when we set $A^{\mu}$ to zero. Taking derivatives of the $\Theta$ functions and expanding for small $\epsilon$, we get
\bea
0= \int
d^{2}x\lambda^{t}\left[\left(K_{{\bf O}}-K_{{\bf I}}\right)\delta\left( r-r_H
\right)
- \epsilon\left(K_{{\bf O}}+K_{{\bf I}}-2K_{{\bf X}}-2N^{r}_{t}\right)\partial_r\delta\left( r-r_H \right)
\right]+\nn\hspace{0.5in}\\
-\hspace{-0.05in} \int d^{2}x\lambda^{r}\left[\hspace{-0.05in}\left(\frac{K_{{\bf O}} +Q_{{\bf O}}+K_{{\bf I}}+Q_{{\bf I}}-2K_{{\bf X}}-2Q_{{\bf X}}}{f}\right)\hspace{-0.05in}
-\hspace{-0.05in}
\epsilon\hspace{-0.05in}\left(\frac{K_{{\bf O}}
+Q_{{\bf O}}-K_{{\bf I}}-Q_{{\bf I}}}{f}\right)\partial_r\delta\left(r-r_H\right)
\right]+\ldots \nn
\label{huh}
\eea
The ellipses stand for higher order in $\epsilon$ terms, but they don't add any new conditions. Since the parameters $\lambda^{t}$ and $\lambda^{r}$ are
independent, each of the four terms in square brackets in Equation
(\ref{huh}) must vanish simultaneously. Note however that because of the delta functions etc., this needs to happen only {\em at} the horizon. These conditions yield
\bea
K_{{\bf O}}=K_{{\bf I}}=K_{{\bf X}}+\Phi,\ \ Q_{{\bf O}}=Q_{{\bf I}}=Q_{{\bf X}}-\Phi,
\eea
where
\bea
\Phi=N^{r}_{t}\Big|_{r_H}=\frac{\kappa^2}{48\pi}= \frac{\pi}{12}
T_H^2 \label{flux}
\eea
and $T_H$ is the Hawking temperature
\bea
T_H=\frac{\kappa}{2\pi}.
\label{THawk}
\eea
Equation (\ref{flux}) is exactly the Stefan's law flux that would result from a thermal distribution at $T_H$ in two dimensions. What we have shown is that the flux $\Phi$ necessary to make the theory diff. invariant at the quantum level, is consistent with a thermality assumption at the Hawking Temperature. Even though we have not strictly {\em derived} the thermal distribution, this consistency is quite remarkable and has been applied to various kinds of black objects in numerous theories in various dimensions successfully.

Before leaving this section, we pause to note that there is in fact yet another derivation of Hawking temperature and radiation (which we will not discuss in detail). This comes from string theory. The idea is to model (closed string) Hawking radiation as arising from the joining of certain open string fluctuations of D-branes \cite{DasMathur, Malda12}. The result reproduces the temperature, entropy and the Hawking spectrum (up to sub-leading corrections). The trouble is that the D-brane picture of black holes is in control only in certain nearly supersymmetric cases. But it is indeed interesting that (1) Hawking radiation is robust even in string theory to a first approximation, and (2) that dynamical phenomena like Hawking decay (and not just entropy counting) can be accomplished from a microscopic theory of gravity.

\section{Excerpts from AdS/CFT}
\label{adscft}

The holographic duality between (certain) non-gravitational quantum field theories and gravity is arguably the deepest insight that we have acquired in the last quarter century about the laws of nature. In this section we will explore a specific aspect of this conjectured relationship: we will consider the duality between conformal field theories (which arise for example as RG fixed points of quantum field theories) and (quantum) gravity in asymptotically anti-de Sitter spaces\footnote{The asymptotically anti-de Sitter boundary conditions means morally that in the gravitational path integral we allow only field configurations that satisfy certain specific fall-offs at spacelike infinity.}.

\subsection{Gauge-String Duality and Holography}

The most concrete realization of holographic duality is in string theory. The claim there is that type IIB string theory (which is believed to be the UV completion of Einstein gravity coupled to certain specific matter fields), on spaces that are asymptotically $AdS_5 \times S^5$ is identical as a quantum theory to the specific gauge theory called ${\cal N}=4$ Supersymmetric Yang-Mills theory in four dimensions with gauge group $SU(N)$. The original argument for this duality due to Maldacena goes as follows\footnote{The following few paragraphs require some familiarity with string theory.}.

Flat ten-dimensional spacetime is expected to be an exact background of IIB string theory. Start with a stack of parallel D3-branes in it. At low energies one expects that the system will be described by the low energy effective action for the bulk spacetime, the low energy effective action for the branes, and the low energy effective action for the coupling between the two. We know that the bulk action will be type IIB supergravity with higher derivative corrections which are suppressed at low energies. This follows from a closed string computation. An analogous open string computation shows that the low energy D3-brane action is the ${\cal N}=4$ supersymmetric Yang-Mills theory plus higher derivative terms which are suppressed at low energies.  The leading order interaction between the brane and the bulk is dictated by covariantizing the brane action with respect to the bulk metric. In the deep IR, everything except the freely propagating bulk gravitons and the ${\cal N}=4$ $SU(N)$ supersymmetric Yang-Mills theory on the bulk are frozen out. Essentially, what we are doing is to take the $\alpha' \rightarrow 0$ limit and keeping only the pieces in the action that remain.

Now there is another viewpoint on this low energy limit, where one first dials up $g_s N$ (where $g_s$ is the string coupling and $N$ is the number of branes, see footnote \ref{FN!}.) first so that the branes are heavy and their backreaction leads to the $D3$-brane supergravity solution:
\bea
ds^2 = f^{-1/2}(-dt^2 + dx^2_1+ dx^2_2+ dx^2_3) + f^{1/2}(dr^2 + r^2d \Omega^2_5) ,\hspace{0.3in}  \\
F_5 = (1 + \star)dt \wedge dx_1 \wedge dx_2 \wedge dx_3 \wedge df^{-1} , \ \
f = 1 +{R^4 \over r^4} , \ \ R^4 \equiv 4\pi g_s\alpha'^2 N .
\eea
$F_5$ is the self-dual five-form of IIB string theory and $\star$ stands for the Hodge dual. Note that string perturbation theory is well-defined precisely in the opposite limit, where $g_sN\ll 1$, i.e., small number of strings interacting weakly.  Maldacena's proposal is that the low energy limit from the supergravity viewpoint corresponds to taking a near-horizon limit of this geometry. This limit for the above metric basically corresponds to taking $r$ to be small, so that the 1 in the warp factor can be dropped. The result is the $AdS_5 \times S^5$ metric in Poincare coordinates:
\bea
ds^2=\frac{r^2}{R^2}(-dt^2+dx^2_1+ dx^2_2+ dx^2_3)+\frac{R^2}{r^2}dr^2+ R^2d \Omega^2_5. \label{AdS5S5}
\eea
There is also $N$ units of five-form flux threading the $S^5$.
Taking the near-horizon limit as a low energy limit is reasonable because one can measure the energy of an excitation as measured by an observer at infinity, and near-horizon fluctuations are redshifted by the warp factor in the metric, and therefore are low-energy\footnote{The systematic scaling limit we are taking is to let $r \rightarrow 0$ while holding $u=r/\alpha'$ fixed. To compensate for an overall factor of $\alpha'$ that shows up, we measure things in terms of it: note that $\alpha'$ is dimensionful. In writing down the $AdS_5 \times S^5$ metric above, we assume that such rescalings have been done: i.e., the $r$ in the near horizon geometry should not be directly compared to the $r$ in the D3-brane supergravity solution.}. That is, the claim is that the low energy limit corresponds to IIB string theory on the the near horizon limit of D3-brane solution, plus freely propagating gravitons in the asymptotic region. 
The original system that we started with is protected by supersymmetry, so the value of the effective string coupling is a modulus and we should in principle be able to go back and forth between this system and the system that we had at the end of the last paragraph, by tuning the coupling. This means that
\bea
\text{Free gravitons in 10 D + ${\cal N}=4$ $SU(N)$ supersymmetric Yang-Mills theory}=  \hspace{0.5in}\nonumber \\
=\text{Free gravitons in 10 D + Type IIB string theory on $AdS_5 \times S^5$ geometry with } \hspace{0.3in} \nonumber\\  \text{$N$ units of RR-flux}.
\eea
This motivates the conjecture that type IIB string theory on $AdS_5 \times S^5$ is the same thing as ${\cal N}=4$ $SU(N)$ supersymmetric Yang-Mills theory, and is called the AdS/CFT correspondence. Note that in making this conjecture, we are making the assumption that tuning the coupling commutes with the low energy limit. The simplest check of the conjecture is that the symmetries should match on both sides. The conformal group of ${\cal N}=4$ SYM is $SO(4,2)$ of 4D Minkowski space and these are precisely the isometries of $AdS_5$. The $SO(6)=SU(4)$ isometries of the $S^5$ are realized as the $SU(4)$ R-symmetries of ${\cal N}=4$ SYM theory. In fact, it is possible to show that the full supergroup $PSU(2,2|4)$ of global symmetries (of which the R-symmetry and the conformal symmetry are a part) is realized on both sides.

How are the parameters of the gauge theory and string theory related? Note that there are two dimensionless quantities on the string side: string coupling $g_s$ and the flux $N$. This $N$ captures the length scale on $AdS_5 \times S^5$ via the IIB equations of motion (it is really the five-form flux through the $S^5$). The supergravity equations enforce the condition
\bea
{R^4 \over \alpha'^2} = g_s N \ .
\eea
Similarly, the gauge coupling can be determined by doing an open string computation with the D-brane boundary conditions and the result is that
\bea
g^2_{YM}=4 \pi g_s.
\eea
On the gauge theory side also there are two dimensionless quantities, namely the gauge coupling $g^2_{YM}$ and the rank of the gauge group $N$. Since ${\cal N}=4$ Yang-Mills is expected to be an exactly conformal theory even in the full quantum theory, $g^2_{YM}$ is a dimensionless parameter. Together  then, the relation between the gauge theory and string theory is completely fixed by the two {\em Maldacena relations}
\bea
4 \pi g_s={\lambda \over N}, \ \ {R^2 \over \alpha'} =\sqrt{\lambda}\ .
\eea
where $\lambda$ is the 't Hooft coupling defined as $\lambda=g^2_{YM} N$. Inspired by the large $N$-limit where the gauge theory is expected to have a stringy description, we have expressed the two independent parameters of the gauge theory in terms of the 't Hooft coupling and gauge group rank (as opposed to the gauge coupling and the gauge group rank).

From the string theory perspective, the natural ``small" parameters in which one could potentially do perturbation theory are $g_s$ and $\alpha'$ (or $\alpha'/R^2$). For fixed $\lambda$ (which is the same as fixed radius of curvature of the AdS space in units of $\alpha'$), string tree level  $g_s=0$ corresponds to the strict large $N$ limit on the gauge theory. In this limit, worldsheet $\alpha'$ perturbation theory is highly non-perturbative (in  $\lambda$) from the gauge theory point of view because of the second Maldacena relation above. 
In this sense AdS/CFT is a strong-weak duality. This makes the duality hard to check, but also means that one can learn a lot about the strong coupling dynamics of one theory by studying the weak-coupling dynamics of the other. Large radius of curvature at infinite $N$ is the same as working in the supergravity approximation on the string side.


So far what we have done amounts to motivating the conjectured duality between a specific string theory on a certain ten dimensional spacetime and a particular conformal gauge theory in four dimensional spacetime. Note that this claim says nothing about the specific map between the two theories and it is not clear how the observables are related. The remarkable fact is that this duality is in fact a {\em holographic} duality, as we now discuss. The idea of holography belongs to 't Hooft\footnote{It is quite remarkable that two of the key ingredients of AdS/CFT - the observation that gauge theories in the large-$N$ limit have a stringy (and therefore gravitational) description, and the speculations about holography - both come from 't Hooft.} and Susskind. The schematic idea is that since black hole entropy scales like the area instead of the volume (and since black holes carry the maximum entropy in a given volume) it might be reasonable that the entire dynamics of a region of spacetime can be captured by its ``boundary". Immediately after Maldacena's observation that there is a duality between gauge theory and string theory, Witten \cite{WittenI} (and Gubser-Klebanov-Polyakov \cite{GKP}) realized that this is in fact a holographic duality and gave a precise prescription (``GKP-W") for mapping the observables on either side. 
In fact the duality is a generic statement about gravitational theories in asymptotically AdS spacetimes and conformal quantum theories on the timelike boundary of that AdS space. This of course is a direct realization of holography: a non-gravitational quantum field theory on the boundary of AdS describes gravity in AdS!

The GKP-W map between gauge theory and string theory is best written using the partition functions on both sides. Before we begin, it is worthwhile emphasizing that the $S^5$ part of the metric (\ref{AdS5S5}) is negligible when we are looking at the boundary\footnote{Really, the conformal boundary, which is 4D Minkoswki space.}, where $r \rightarrow \infty$. So the ``boundary" here is really four-dimensional. Therefore what we have is really a duality between a certain gravitational theory with asymptotically $AdS_5$ boundary conditions and a gauge theory in four-dimensions. Now lets describe the AdS/CFT map. One motivation for the map is that the expectation value of the dilaton (which captures $g_s$) in the string theory is dual to the gauge coupling by Maldacena's first relation. The string coupling of a background is most naturally defined by the asymptotic (``boundary") value of the dilaton.
If one thinks of the gauge coupling as the source for the gauge theory Lagrangian\footnote{i.e., we think of the gauge coupling as the source $J$ coupling to the operator $O$ which happens to be the Lagrangian .}, the GKP-W map below can be thought of as the natural generalization of this, for operators other than the Lagrangian. We denote by $O_i(x)$ the gauge-invariant local operators\footnote{Strictly speaking this map is for ``single trace operators" $O_i$, but we will not get into the details.} in the gauge theory, with scaling dimension $\Delta_i$, and by $\phi_i(x, r)$ the corresponding string theory fields in AdS, then
\begin{itemize}
\item  the mass of the field and the dimension of the dual operator are related by
\bea
m^2_{\phi_i}=\Delta_i(\Delta_i-4)/R^2
\eea
where $R$ is the radius of the AdS space, and
\item the map between the CFT and string partition functions can be written as
\bea
\langle e^{i\int d^4 x J_i(x) O_i(x)} \rangle_{CFT}=Z_{IIB}[\lim_{r\rightarrow \infty}(\phi_i(x,r)r^{4-\Delta_i})=J_i(x)]
\eea
\end{itemize}
The CFT partition function is the standard partition function familiar from quantum field theory with the source given by $J_i(x)$. The claim is that this partition function is the same as the string theory partition function on AdS with the boundary value of the field $\phi_i(x,r)$ (after compensating for the fall-off) given by  $J_i(x)$. This map is strictly speaking defined in Euclidean AdS spaces, and it says that the boundary values of bulk fields are the sources of the dual operators in the gauge theory.  Note that the structure of the map does not depend crucially on the specific string theory construction we used to derive the Maldacena duality. In fact the  ingredients here are the fact that we are working with a quantum gravity on asymptotically AdS backgrounds and that the dual theory is a CFT. The partition function of the quantum gravity then would be captured by the partition function of the gauge theory according to a  map of the form above. By differentiating the generating functions above, one can obtain correlation functions as familiar in quantum field theory.

Note that in the above map between partition functions, the left hand side (the CFT side) is rather well-defined. We can imagine defining it precisely in terms of a path integral which can (at least in principle) be made sense of operationally on the lattice. But this is not true about the gravity/string side, because we don't know what {\em precisely} it means to have a non-perturbative definition of string theory. In certain limits (like the supergravity limit of the string theory), we have more control.

Before we leave the AdS/CFT map, let us mention one more remarkable fact. Note that both theories in the AdS/CFT duality are gauge theories. On the gravity side the gauge invariance is diffeomorphism invariance, while on the gauge theory side it is the $SU(N)$ gauge invariance of the Yang-Mills theory. The claim that the two theories are the same is a statement about the full quantum theories on either side\footnote{It is in this sense, that AdS/CFT is a quantum theory of gravity: we know that the theory contains semi-classical gravity and that the full quantum theory is ${\cal N}$=4 SYM theory. This means, by definition, that   ${\cal N}$=4 SYM is the quantization of a certain gravity theory (IIB supergravity on asymptotically AdS spacetimes) that we were seeking. We emphasize that asymptotically AdS condition is a boundary condition for the fields one should integrate over in the theory and something of this form should be there in any quantum theory.}. 
Gauge invariance is merely a redundancy 
which one mods out (``solves") by the time one gets to the quantum theory. So the claim of the AdS/CFT duality amounts to the statement that once one solves the diffeomorphisms on the gravity side, the operator algebra that one is left with is identical to the operator algebra that one gets by modding out by the $SU(N)$ gauge invariance of the ${\cal N}=4$ SYM theory. Note that observables on the gravity side in a fully quantum sense are diff. invariant objects: spacetime coordinate is meaningless in a diff. invariant theory\footnote{One might think that, say, the Ricci scalar which is often called a diff. invariant object might be acceptable as an observable in the quantum theory. This is not so, because the argument of a scalar function is the spacetime coordinate. The spacetime coordinate in a diff. invariant theory is entirely analogous to the gauge index in a Yang-Mills theory and must be traced over, i.e. integrated over, to get honest-to-god observables.}. Only integrals over all geometry are meaningful observables (operators) in the quantum theory, at least if one believes the semi-classical picture that diffeomorphisms should be modded out. The amazing thing is that this picture implies that the radial direction in AdS is completely invisible in the gauge theory picture, and that locality in the bulk can be made manifest only by introducing diff. invariance back in. Bulk diffeomorphisms are the price that one must pay to manifest the bulk locality of the classical description.

Substantial evidence for the AdS/CFT correspondence has accumulated over the years: this includes both detailed tests of kinematics and dynamics (where the dynamics is under control), as well as very suggestive qualitative evidence that arises from the natural way in which AdS/CFT blends with our understanding of black holes, quantum gauge theories (including supersymmetric gauge theories) and gravity. As mentioned in the introduction, many of the recent applications of AdS/CFT to condensed matter physics and the quark-gluon plasma can in fact be taken as evidence that black hole thermodynamics fits naturally with holography. Since our major interest is in black holes, this particular line in the holographic dictionary is what we develop in what follows.

\subsection{Euclidean Black Holes: Deconfinement at Strong Coupling}

The attempt to understand black holes by working in AdS is often pooh-poohed by some. The argument being that AdS is vastly different from flat space (it has a curvature scale, the asymptotics are different, it is not globally hyperbolic\footnote{This last one is inessential when one imposes reflective boundary conditions at the boundary.}, etc.) and therefore one should not celebrate any kind of understanding that one gleans by studying black holes in AdS. While noble, this argument thoroughly misses the point. The reason to study black holes in AdS is not because of some morphological similarity with (astrophysical) black holes in flat space. The point is that black holes in AdS exhibit the {\em same} puzzles that the black holes in flat space exhibit. Apparently the fact that AdS is different from flat space is {\em not} enough to alleviate these problems. Yet, AdS black holes possibly admit a solution to those puzzles in terms of the dual gauge theory. This is the reason to study them as theoretically interesting objects: AdS is potentially a context where the paradoxes of black hole physics have {\em some} solution. It should also be kept in mind that Hawking radiation, entropy, information, etc. are all {\em theoretical} puzzles with essentially no chance of being experimentally accessible in the near-future, and if one does not believe in the robustness of theoretical physics they can all be ignored. The fact that AdS/CFT ties up beautifully with the old calculations of Hawking and Page should be taken as a remarkable piece of evidence that the successful threads in theoretical physics are indeed the longest. This particular connection is what we will discuss in this section.

In a previous section we saw that Euclidean quantum gravity in the canonical ensemble on asymptotically Anti-de Sitter space exhibit a phase transition. When the temperature is above that determined by the AdS scale, the large AdS black hole is the dominant contribution to the partition function. The transition between the empty AdS phase and the AdS-black hole phase was called the Hawking-Page transition.

In light of the AdS/CFT correspondence, we expect that this should have an interpretation in the dual gauge theory. The moral being that gravity captures the features of the gauge theory at strong coupling. It was shown by Witten in a remarkable paper that indeed, the Hawking-Page transition is dual to a phase transition between confining and deconfined phases of the gauge theory.

To describe a phase transition, one needs an order parameter: a quantity that is different in the two phases. One of the order parameters for the gauge theory deconfinement is the action itself.
In a large-$N$ Yang-Mills theory, the action scales like $O(N^0)$ in the confined phase and $O(N^2)$
in the deconfined phase. This is because the individual color degrees of freedom of a gauge singlet do not count separately in the confined phase, but they do in the deconfined phase. The action is composed of traces of adjoint matrices, and they have $O(N^2)$ individual elements at large $N$.  In the gravity side, the AdS phase is the datum and is always subtracted out and has zero action, which is $O(N^0)$. The black hole phase has an action proportional to $1/G_N$, as we saw in the subsection on Hawking-Page transition.  By dimensional reduction from the 10D IIB theory, this is proportional to $1/{g_s^2}$ and at large but fixed $\lambda$ (which corresponds to working at large but finite AdS radius) this is proportional to $N^2$ according to the Maldacena relations. Therefore the black hole phase corresponds to the deconfined phase.

A more subtle order parameter is the expectation value of the Polyakov loop (i.e., a Wilson loop wrapping the Euclidean time direction), defined by the path ordered exponential
\bea
L(x)\equiv  \frac{1}{N} {\rm Tr} P e^{i \int_0^\beta d\tau A_0(x,\tau)},
\eea
Note that before the tracing ${\rm Tr}$ over the color indices, the Polyakov loop is an $N \times N$ matrix in color space. We start by describing why the Polyakov loop is an order parameter for deconfinement. Note first a feature of the discrete subgroup $\IZ_N$ with elements defined by $e^{2\pi i k/ N}{\cal I}$ ($k=0,1,\dots,N-1$) of the color gauge group: it commutes with all the elements of the gauge group $SU(N)$. Such a subgroup is called the center of the group. Here ${\cal I}$ is the $ N \times N$ unit matrix. Now, under a non-periodic gauge transformation of the form $V_k(x)=e^{2\pi i k \tau/ N \beta}{\cal I}$, the gauge
fields receive a constant shift,
\bea
 A_0 \;\rightarrow\; A_0^k =
  V_k \bigl[ A_0 -(i g)^{-1}\partial_0 \bigr] V_k^\dagger
  = A_0 - \frac{2\pi k}{g N \beta} ,
\label{eq:transformedA}
\eea
so that the Polyakov loop transforms as $L(x) \to e^{2\pi i k/ N} L(x)$.
Because $A_0^k$ still keeps the periodicity in $\tau $, such a non-periodic gauge transformation still forms a symmetry of the action when the theory contains only pure glue. This is called the center symmetry. Note that quark pieces in the action will break the symmetry because it breaks the anti-periodicity of the fermions in the imaginary time direction. A crucial fact is that the expectation value of the Polyakov loop is related to the quark excess free energy $F_q$ in a gluon medium according to
\bea
\langle L(x)\rangle=e^{-\beta F_q}.
\eea
(Note that adding a free quark, in a gauge-invariant language, means adding a semi-infinite Wilson line with a quark on one end.)
In other words, in a confining medium where the free energy of a single quark is infinite, the Polyakov loop is zero, whereas in a deconfined phase where $F_q < \infty $, it is finite. This is the rationale behind using it as an order parameter for confinement.

We want to compute this Polyakov loop at strong gauge coupling (more precisely large 't Hooft coupling) which has a semi-classical description in terms of a classical string in the AdS background, with its boundary ending on the loop.  The expectation value of the loop is calculated from the (regularized) area of the minimal surface ending on the loop
\bea
\langle L \rangle \sim e^{-T_{s}{\cal A}},
\eea
where ${\cal A}$ is the area of the minimal surface in the bulk ending on the loop and $T_{s}$ is the tension of the fundamental string. We want to argue that this object vanishes in the thermal AdS phase and is finite in the black hole phase. The thermal $AdS_5$ geometry has a topology $S^1\times B^4$ ($B^4$ stands for the four-dimensional solid ball). The $S^1$ comes from the thermal circle, whereas the fact that the holographic radial coordinate is semi-infinite, i.e., $r \in [0,\infty)$, means that the rest of the directions together have the form $\IR^+ \times S^3 \sim B^4$. Since the thermal circle stands by itself it is non-contractible. So this topology cannot have a worldsheet with a disk topology that ends on the thermal circle at the boundary. This means that the Polyakov loop expectation value is zero at the closed string tree level, or in the $N \rightarrow \infty$ limit. Thus the thermal AdS geometry corresponds to the confined phase.

On the other hand, for the AdS-Schwarzschild black hole geometry has a topology of  $\IR^2\times S^3$. The $\IR^2$ comes from the previously alluded to fact that the $t$ and $r$ coordinates of a Euclidean black hole together form a cigar, whose topology is that of $\IR^2$. The $S^3$ is simply the 3-sphere in $AdS_5$. Clearly, the $\IR^2$ guarantees that the thermal circle is contractible: there exist discs that end on the boundary and the (regularized) area of the surface gives the Polyakov loop expectation value, which is non-zero. Therefore, the AdS-Schwarzschild black hole geometry corresponds to the deconfined phase as we claimed before.

The punchline, then, is that the Hawking-Page transition is dual, at strong 't Hooft coupling, to the deconfinement transition in the gauge theory. At weak coupling, this computation is silent about the existence of a phase transition. But the weak-coupling regime is precisely where one can do perturbation theory in the ${\cal N}=4$ SYM gauge theory. Indeed, it has been shown by direct perturbative computations that even at weak coupling, the phase transition persists \cite{Minwalla}.



\subsection{Lorentzian Black Holes}

Euclidean AdS black holes are obtained by Wick rotating and periodically identifying the temperature, and the dual interpretation in the gauge theory is that we deal with the theory in thermal equilibrium (at the Hawking temperature of the hole). Equilibrium means that there is no change in the system with time and that there is no dynamics. In particular, the region beyond the horizon is no longer there in the Euclidean picture. Since we expect that AdS/CFT captures the full dynamics of asymptotically AdS spacetimes we should also be able to see the internal structure of black holes inside the horizon from the gauge theory. A proposal for doing this has been put forward by Maldacena, building on some old ideas of Israel and this is what we will sketch in this section.

\begin{figure}
\begin{center}
\includegraphics[
height=0.3\textheight
]{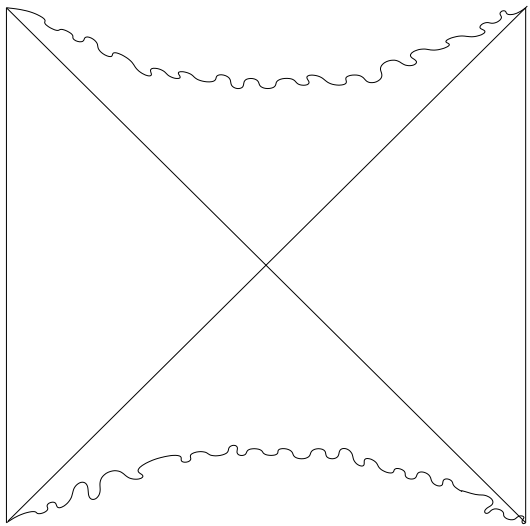}
\caption{Penrose diagram of Kruskal-extended AdS Schwarzschild black hole. One interesting fact is that the conformal structure forces the singularity to be bowed in. This is significant for B1-B2 geodesics that probe the singularity, see \cite{Shenker1, Shenker2, Liu, Rozali, Krishnan}.}
\label{AdSKrusk}
\end{center}
\end{figure}
A basic observation is that a thermal state can be expressed as a specific entangled state between two Hilbert spaces. Note that a thermal state $|0(\beta)\rangle$ is {\em defined} to be a state where the relation
\bea
\langle 0(\beta)| A |0(\beta)\rangle = \frac{1}{Z}{\rm Tr}(e^{-\beta H} A)
\eea
holds for any observable $A$. Typically the observables we will worry about are two-point functions, corresponding to scalar fields in the background spacetime. Then, this definition of thermality is equivalent to KMS periodicity as we showed in an earlier section. The same periodicity is what we see from imposing Euclidean regularity of the black hole. The crucial point is that if one defines
\bea
| 0(\beta)\rangle = \frac{1}{Z[\beta]^{\frac{1}{2}}} \sum_n e^{-\frac{\beta\, E_n}{2} } |n\rangle
\otimes |n\rangle,
\eea
then expectation values of operators acting on only one Hilbert space are thermal, as can be checked by a simple direct computation. This is the entangled description of thermal states: entangled, between the two Hilbert spaces. This prescription is sometimes called {\em thermo field theory}.

Now, we note that large AdS black holes are the best candidates for the statement ``black holes are thermal states". They have positive specific heat and have a well-defined Hartle-Hawking vacuum. Maldacena's proposal is that these two pieces in the tensor product should be thought of as living on the two asymptotic boundaries of the maximally extended Kruskal geometry of the AdS black hole, see figure \ref{AdSKrusk}.
That is one imagines that the two asymptotic regions are associated to the two CFT Hilbert spaces. Putting operators on one CFT (we will call them B1-B1 correlators) gives thermal correlators in the black hole background: $\langle 0(\beta)| A_1(0)B_1(t) | 0(\beta)\rangle$ turns out to the thermal expectation value of the operator $A(0)B(t)$. Subscripts 1 and 2 stand for the Hilbert spaces.

\begin{figure}
\begin{center}
\includegraphics[
height=0.3\textheight
]{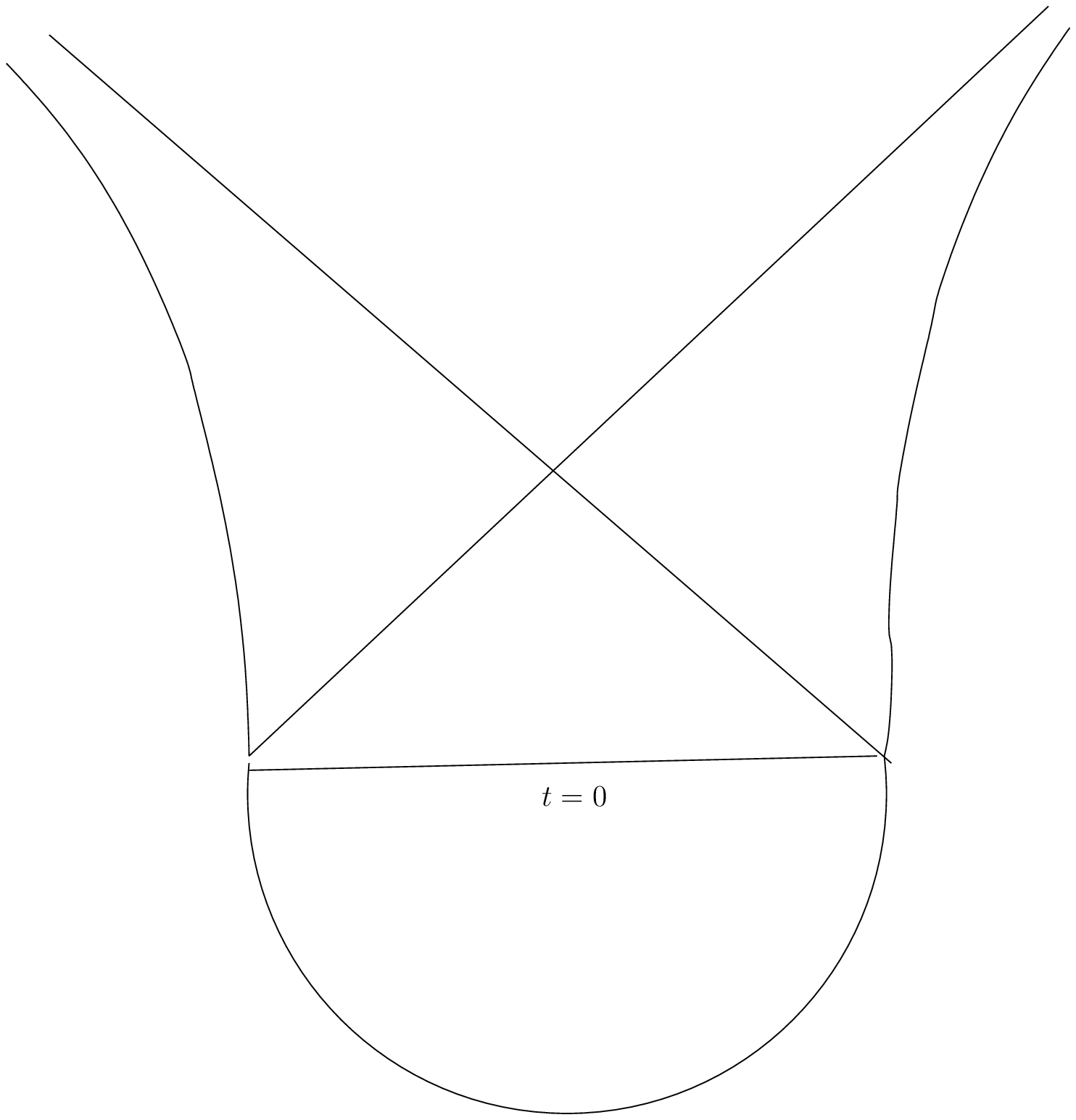}
\caption{The lower half represents the compact Euclidean geometries over which the path integral is done. The result of the path integration is the initial value for the Lorentzian evolution starting at $t=0$. This is the Hartle-Hawking construction. The upper part of the figure represents the Lorentzian (Kruskal) geometry, truncated at $t=0$.}
\label{Malda-Hartle}
\end{center}
\end{figure}

One motivation for this prescription is from the Hartle-Hawking ``wave function of the Universe" idea. There, we can view the Euclidean part of the geometry as giving the initial wave function which we then evolve in Lorentzian signature. The $t=0$ slice of the geometry has a reflection symmetry, so we are allowed to do this. The picture for this is given in figure \ref{Malda-Hartle}. 
From the CFT, this has a natural interpretation. The boundary of the Euclidean black hole is $S^1_{\beta}\times S^{d-1}$. A section of the Euclidean metric intersects the boundary on two disconnected spheres $S^{d-1}\times S^{d-1}$. The Euclidean time direction 
in the lower half of the figure connects these two spheres. The path integral of the boundary CFT is then over the Southern hemisphere which can be written as $I \times S^{d-1}$, where $I $ is an interval of length $\beta/2$. 
The states at $t=0$ are obtained by evolving (in Euclidean time) in the Schrodinger picture for $\beta/4$ in each CFT Hilbert space.  From this perspective, the wave function at $t=0$ is of the entangled form
\bea
| \Psi \rangle = \frac{1}{Z[\beta]^{\frac{1}{2}}} \sum_n e^{-\frac{\beta\, E_n}{2} } |n\rangle
\otimes |n\rangle
\eea
where the sum is over the energy eigen-basis. This wave function is to be considered as the initial condition for Lorentzian evolution of the black hole spacetime, starting at $t=0$.

An analytic continuation in the correlators corresponds to insertion of operators on the two distinct boundaries (``B1-B2" correlators). This is because it can again be checked by direct computation that $\langle 0(\beta)| A_1(0)B_2(t) | 0(\beta)\rangle \equiv \langle 0(\beta)| A_1(0)B_1(-t-i\beta/2) | 0(\beta)\rangle $. So the analyticity properties of the correlator $\langle A(0)B(t) \rangle$ in complex $t$ can be used to move back and forth between correlators within and between the Hilbert spaces.

With this prescription at hand, we can explore the internal structure of black holes by computing correlators in the CFT. The idea is that the dominant contribution to boundary correlators in the large mass limit of the scalar field comes from geodesics connecting the two points. For example, we can focus on (B1-B2) correlators that correspond to (spacelike) geodesics that go inside the horizon and connect the two boundaries. By choosing the geodesic appropriately, we can therefore identify the features in the correlators that correspond to the singularity.
This gives us the boundary signature of the singularity.  The basic strategy is to
\begin{itemize}
\item Define bulk correlators in the Hartle-Hawking vacuum.
\item Take them to the boundary to define boundary correlators and obtain B1-B2  correlators.
\item Identify the correlators that are dominated by geodesics that get close to the singularity, to identify the correlators that capture the features of the singularity.
\end{itemize}
This line of enquiry has been explored in \cite{Shenker1, Shenker2, Liu, Irene2, Rozali, Krishnan} for various black holes, but we will not get into this rather technical topic.




\section*{\bf Acknowledgments}

Thanks to (1) Jarah Evslin and Zoran Skoda (and the pigs) for their sacrifices in the making of a great school, (2) Alessandro Lovato, Eduardo ``The Rock" Conde and especially Daniel Arean for help with transporting the lectures and the lecturer, (3) Sabine Ertl, Malcolm Fairbairn, Gaston Giribet, Matt Kleban, Alex Morisse, Holger Nielsen, Tomas Prochazka, Andrea Prudenziati, Joris Raeymaekers, Javier Tarrio, Dieter Van den Bleeken, John Wang and Ho-Ung Yee for the enjoyable discussions that improved the manuscript and their author. Finally, another round of thanks to Daniel Arean for a careful reading of (parts of) the draft. 

\appendix

\section{Spacetime Geometry and Black Holes}
\label{app}

In the appendix we collect some basic facts about causal structures in general relativity, with emphasis on black holes. Two good references are \cite{Townsend, HawkEllis}.

\subsection{Global Hyperbolicity and the Cauchy Problem}

\begin{figure}
\begin{center}
\includegraphics[
height=0.3\textheight
]{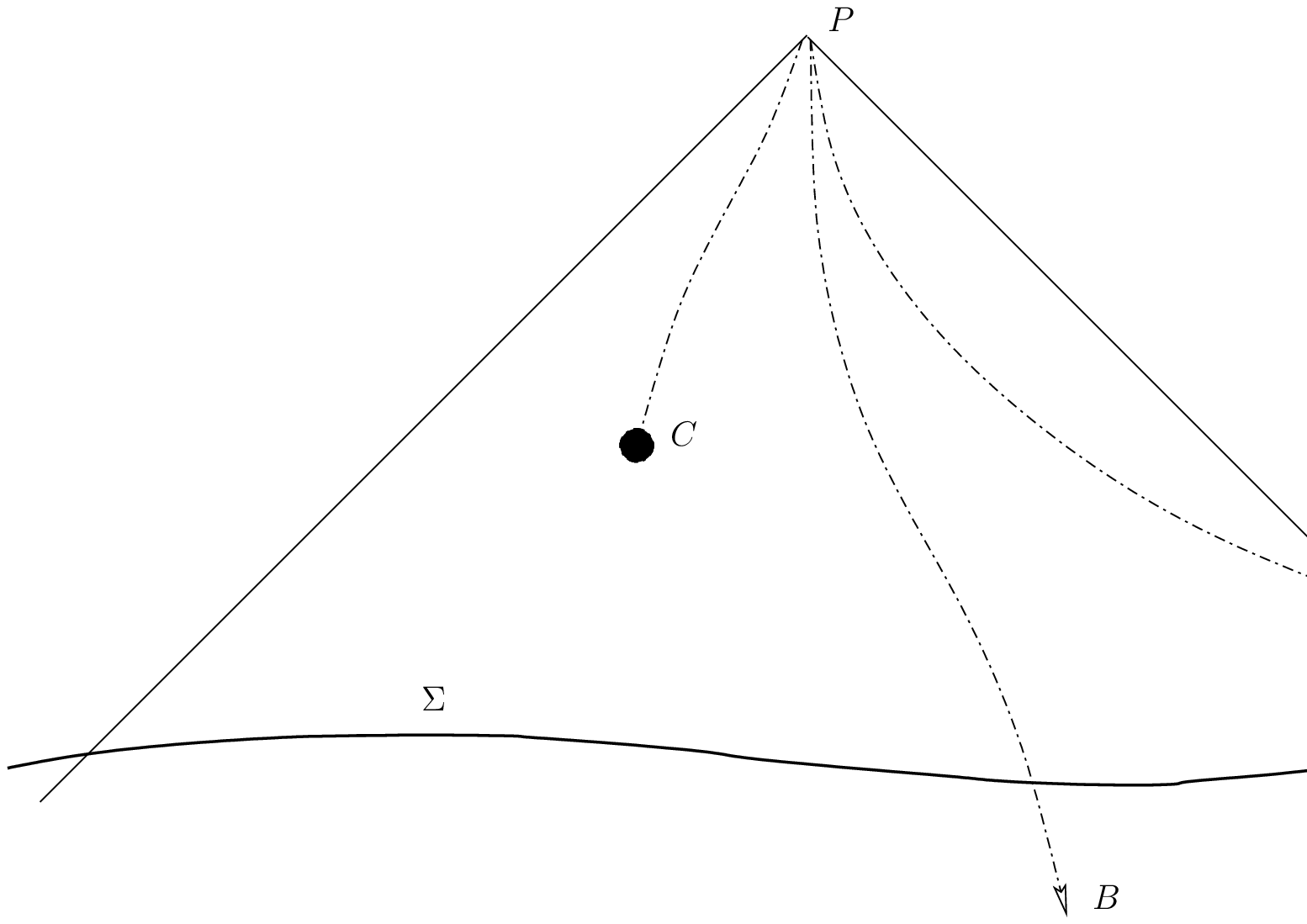}
\caption{A (partial) Cauchy surface $\Sigma$, and various kinds of curves. The past light cone of a point in the future of $\Sigma$ is shown to clarify the nature of these curves. $OA$ is not a causal curve because it gets outside the light cone, $OB$ and $OC$ are causal curves, but $OC$ is not past-inextendible because it can be continued if one chooses to.}
\label{curvey}
\end{center}
\end{figure}

We will start by defining a Cauchy surface. A {\em (partial) Cauchy surface}, $\Sigma$, is a subset of spacetime $M$ which intersects no causal curve (i.e., spacelike or timelike curve) more than once. Roughly, this represents a spatial slice which is an instant of time. But note that this need not be a ``global" spatial slice of the spacetime. Now, a causal curve is said to be {\em past-inextendible}, if it has no past endpoints in $M$. See figure \ref{curvey} for a depiction of these essentially trivial definitions.

With these basic notions at hand, we are ready to define a Cauchy development. The {\em future Cauchy development} of $\Sigma$, ${\cal D}^+(\Sigma)$, is comprised of the set of points $p$ in $M$ such that all past inextendible curves through $p$ intersect $\Sigma$.
\begin{figure}
\begin{center}
\includegraphics[
height=0.3\textheight
]{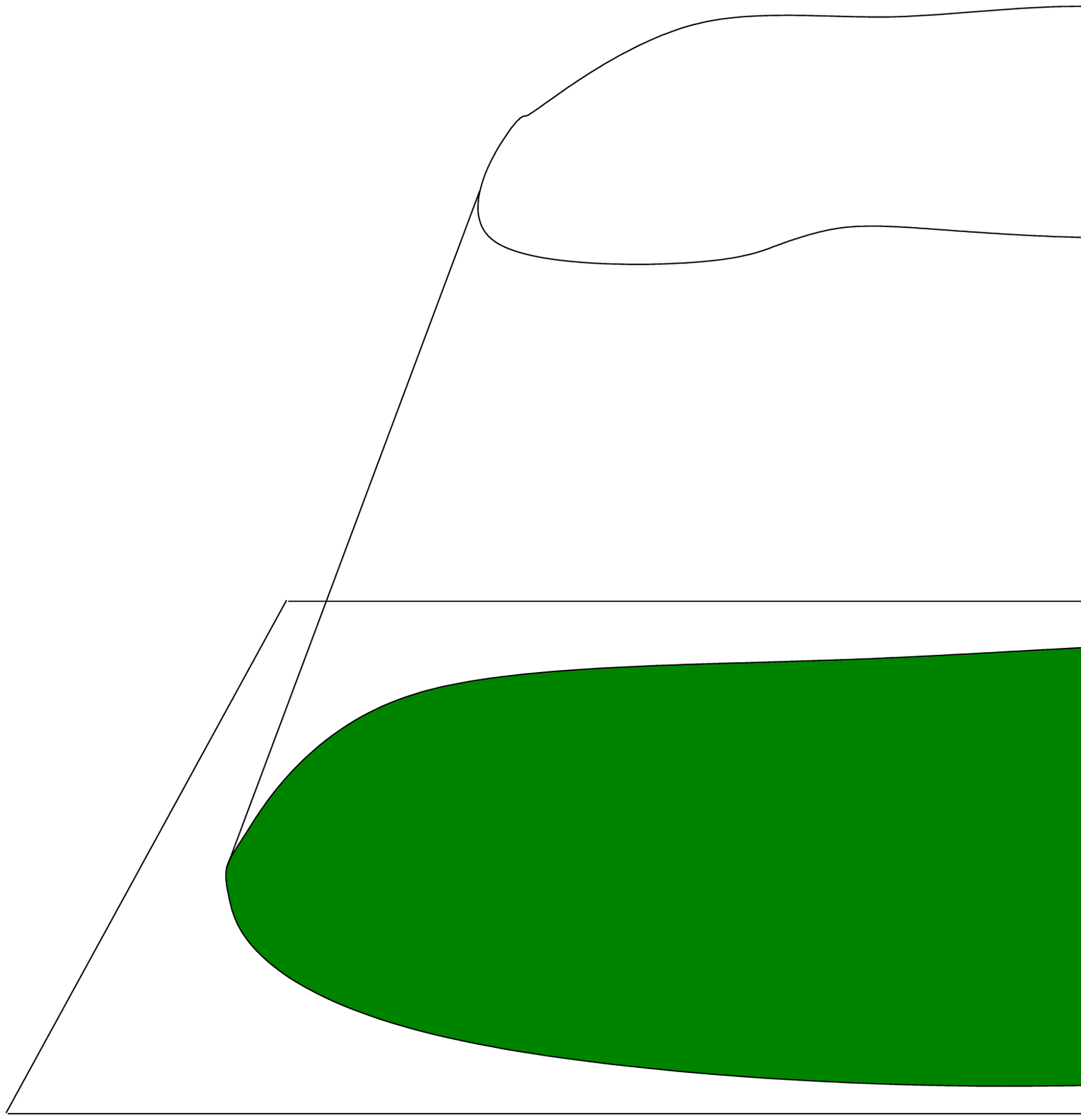}
\caption{A schematic picture of the future Cauchy development of $\Sigma$. The frustum between the parallel planes with null boundaries is a subset of ${\cal D}^+(\Sigma)$. The full ${\cal D}^+(\Sigma)$ goes on into the future. In drawing this figure we have assumed that the spacetime is roughly flat, for illustration. The structure can be more complicated in general, see eg., the Penrose diagram of the rotating (``BTZ") black hole.}
\label{frustum}
\end{center}
\end{figure}
The future Cauchy development is interesting because solutions of hyperbolic PDEs (i.e., wave equations) on ${\cal D}^+(\Sigma)$ are fully determined by data on $\Sigma$. We can have a similar definition for past Cauchy development. The important thing to keep in mind is that the Cauchy development is fixed {\em completely} by the data on the Cauchy surface. This should be contrasted to a future light cone. The frustum ``expands"  for the light cone as we move to the future from $\Sigma$, while it ``shrinks" for the Cauchy development.

We say that $\Sigma$ is a {\em (global) Cauchy surface} iff the future and past Cauchy developments together make up the entire spacetime, i.e., ${\cal D}^+(\Sigma) \cup {\cal D}^-(\Sigma)=M$. A spacetime that admits a global Cauchy surface is called {\em  globally hyperbolic}. This essentially means that the entire spacetime can be determined by providing Cauchy data on some surface\footnote{When we say Cauchy surface, it will generally mean such a {\em global} Cauchy surface in a globally hyperbolic geometry.}. An interesting, and intuitive, fact is that if the spacetime has one Cauchy surface, then it has a Cauchy surface through every point. It is possible to choose a time coordinate such that each $t=$const. slice is Cauchy.

If $M$ is not globally hyperbolic, then $ {\cal D}^+(\Sigma)$ and/or $ {\cal D}^-(\Sigma)$ will have a boundary in $M$. This boundary is the future/past {\em Cauchy horizon}. Flat Minkowski spacetime, collapsing dust, maximally extended Schwarzschild and FRW cosmology are examples of globally hyperbolic spacetimes. On the other hand, the maximal extensions of charged and/or rotating black holes have Cauchy horizons. This is because they have an inner horizon (which is the Cauchy horizon) beyond which, things are not just determined by the initial data on some Cauchy surface, but also by the boundary conditions we put on the timelike singularity that lies beyond. As a typical Penrose diagram of this sort, we provide here the Penrose diagram of a 2+1 dimensional black hole (the BTZ black hole).
\begin{figure}
\begin{center}
\includegraphics[
height=0.7\textheight
]{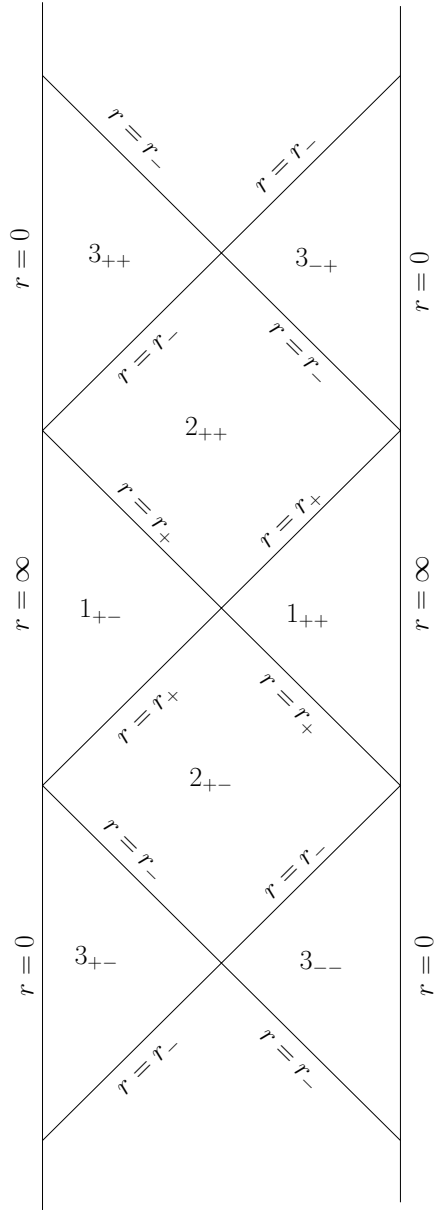}
\caption{This is the Penrose diagram of a rotating 2+1 dimensional black hole in Anti-de Sitter space, the so-called BTZ black hole. The only substantial difference of this black hole with flat space black holes (at the level of the Penrose diagram) is that the boundary is timelike. Here, $r=0$ are timelike singularities, $r=\infty$ are the various asymptotic regions, $r=r_+$ are the outer horizons and $r=r_-$ are the inner (Cauchy) horizons.}
\label{curves}
\end{center}
\end{figure}
The wavy lines denote the timelike singularity, and it is clear that initial conditions provided at the singularity can affect the time evolution in the regions to the future of the Cauchy horizon.

\subsection{Schwarzschild and Kruskal}

One of our aims in the main text is to explore the propagation (and to a lesser extent) backreaction of quantum fields in a static spherically symmetric black hole background.  Such a spacetime -say outside the horizon- is empty, and is a solution of the vacuum Einstein equations. It is described uniquely by Birkhoff's theorem to be the Schwarzschild metric:
\bea
ds^2=-(1-2M/r)dt^2+\frac{dr^2}{1-2M/r}+r^2(d\theta^2 +\sin^2 \theta d\phi^2).
\eea
To understand the causal structure, it is useful to follow the null geodesics $ds^2=0$ in this geometry. Radial null geodesics are captured by $\theta=$const., $\phi=$const. and are solved by
\bea
t=\pm r_* + C_0, \ \ {\rm where} \ \ r_*=\int \frac{dr}{1-2M/r}=r+2M \ln \frac{r-2M}{2M}+C.
\eea
is called the tortoise coordinate. $C$ and $C_0$ are constants of integration and we can choose to absorb the former into the latter. The plus sign corresponds to increase in radius as time flows to the future and therefore capture outgoing geodesics, and the negative sign corresponds ingoing geodesics. Note that $r=2M$ corresponds to $r_*=-\infty$. Since to keep $r_*+t=C_0$ fixed while sending $r_*$ to $-\infty$, we need $t \rightarrow \infty$: 
it takes and infinite amount of coordinate time to get to the the horizon from outside. But despite this, one can show that the affine parameter on a radial null geodesics remains finite as it crosses the horizon.  This can be seen by solving the following equations simultaneously for $\lambda$, the affine parameter:
\bea
du\ dv=0, \  \  \ (1-2M/r)\frac{d (v+u)}{d \lambda}= {\rm const.}
\eea
It is sometimes convenient to use coordinates $(v\equiv t+r_*,  r)$ or $(u\equiv t-r_*, r)$ instead of $(t, r)$ and these are called ingoing and outgoing Eddington-Finklestein coordinates.  In the above we work with $(v, u)$ instead of $(t, r)$. That is, when we write $r$ in the second equation above, we implicitly think of $r$ as a function of $v$ and $u$, i.e., $r\equiv r(v,u)=r(v-u)$.
The first equation is the condition for the radial null geodesic, whereas the second one is the statement that there is a conserved quantity, energy, because the spacetime is static. That is, the second equation comes from the relation that $g_{ab}\xi^a u^b$ is a constant along geodesics whose tangent vectors are $u^a=\frac{dx^a}{d\lambda}$, and $\xi=\frac{\partial}{\partial t}$ is the timelike Killing vector (this is in the Schwarzschild $t$, but we translate it to $u$ and $v$.). Using
\bea
\Big(1-\frac{2M}{r}\Big)=\frac{2M}{r} e^{(v-u)/4M}e^{-r/2M}
\eea
which is an immediate consequence of the definitions of $r_*, u$ and $v$, for infalling (i.e., $v=C_0$) geodesics we can write
\bea
d\lambda=\frac{2M}{r} e^{(C_0-u)/4M}e^{-r/2M}du.
\eea
Near the horizon, this can be approximately integrated to write $\lambda \approx C' -{\tilde C} e^{-u/4M}$ for some constants $C'$ and $\tilde C$, with $\lambda \rightarrow C' $ as we approach the horizon ($r_*=-\infty$). A similar statement holds for outgoing geodesics as well, with the final expression for the affine parameter being $\lambda \approx C' +{\tilde C} e^{v/4M} \rightarrow C' $.

This means basically that (an approximation to) the affine parameters for the outgoing and ingoing null geodesics can work as coordinates that are regular through the horizon. This choice is called Kruskal coordinates\footnote{We did essentially the same construction in Rindler as well.}:
\bea
V=e^{v/4M}, \ \ U=-e^{-u/4M}.
\eea
Note that in Schwarzschild coordinates the horizon looks singular, but it is not really a divergent location for the curvature scalars like $R, R_{ab}R^{ab}$ and $R_{abcd}R^{abcd}$. So the existence of regular coordinates is not surprising: what we have is merely a coordinate singularity, it is not physical. Reinstating the angular parts, the metric now takes the form
\bea
ds^2=-\frac{32 M^3}{r}e^{-r/2M} dU dV+r^2 d\Omega_2^2
\eea
The $r$ is thought of implicitly as a function of $U$ and $V$. This is the Kruskal (sometimes also called Kruskal-Szekeres) extension of the Schwarzschild metric. The extension is maximal, in that the only singular regions where the metric breaks down are genuine curvature singularities ($r=0$).
We started with the exterior Schwarzschild metric to get here. In particular, $r_*$ is real and $r > 2M$. This fixes some choices of signs and square-roots and we have explicitly (for the region $r>2M$)
\bea
U=-\Big({r \over 2M}-1\Big)^{1/2}\exp\Big(\frac{r-t}{4M}\Big), \ \ V=\Big({r \over 2M}-1\Big)^{1/2}\exp\Big(\frac{r+t}{4M}\Big).
\eea
An important observation is that this (the exterior Schwarzschild region) covers only the region $(U<0, V>0)$ in $U-V$ plane. The full extension for all real values of $U$ and $V$ is obviously a solution of Einstein's vacuum equations and is therefore the maximally extended Kruskal ``manifold" (within quotes because it is singular at $r=0$). 
\begin{figure}
\begin{center}
\includegraphics[
height=0.35\textheight
]{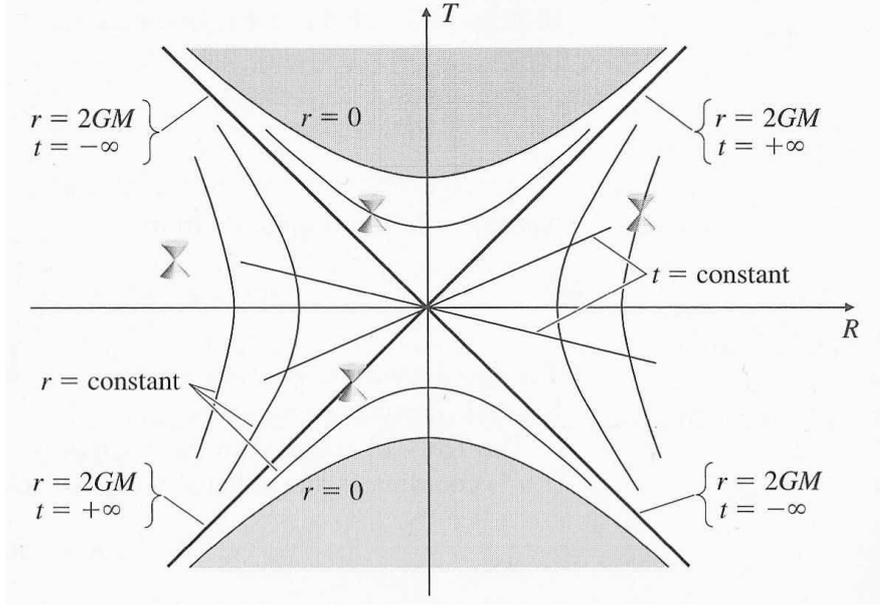}
\caption{The Kruskal manifold. Figure taken from \cite{Kpic}. The coordinates $T$ and $R$ are related to our $U$ and $V$ via $U \sim T-R$ and $V\sim T+R$.}
\label{ord}
\end{center}
\end{figure}
It is often said that in general relativity we take a manifold first and then put a metric on it.  But in practice, this is never what we do. We first look for a physically interesting metric (here the standard Schwarzschild form captures the physically interesting notion of gravitational field outside a spherically symmetric object) and then look for a ``manifold" on which it is maximally well-defined. This manifold is the Kruskal ``manifold".

It is clear from figure \ref{ord} and the coordinate definitions that straight lines through the origin are constant $t$ (Schwarzschild time) slices and $UV$=const. hyperbolas are constant $r$ (Schwarzschild radial coordinate) slices. In the regions II $(U>0, V<0)$, Up $(U>0, V>0)$ and Down $(U<0, V<0)$, we can introduce coordinates $r$ and $t$ as follows, so that the metric still takes the Schwarzschild form there: this is again analogous to the Rindler case that we discussed.
\bea
{\rm II}\ (U>0, V<0): \   U=\Big({r \over 2M}-1\Big)^{1/2}\exp\Big(\frac{r-t}{4M}\Big), \ \ V=-\Big({r \over 2M}-1\Big)^{1/2}\exp\Big(\frac{r+t}{4M}\Big),  \nonumber \\
{\rm U}\ (U>0, V>0): \   U=\Big(1-{r \over 2M}\Big)^{1/2}\exp\Big(\frac{r+t}{4M}\Big), \ \ V=\Big(1-{r \over 2M}\Big)^{1/2}\exp\Big(\frac{r+t}{4M}\Big),  \\
{\rm D}\ (U<0, V<0): \   U=-\Big(1-{r \over 2M}\Big)^{1/2}\exp\Big(\frac{r+t}{4M}\Big), \ \ V=-\Big(1-{r \over 2M}\Big)^{1/2}\exp\Big(\frac{r+t}{4M}\Big),  \nonumber
\eea
Note that because of the flipping of lightcones, the U and D regions are {\em not} static. The $r$-coordinate in this region is the time coordinate. 

\subsection{Penrose Diagrams}

\begin{figure}
\begin{center}
\includegraphics[
height=0.4\textheight
]{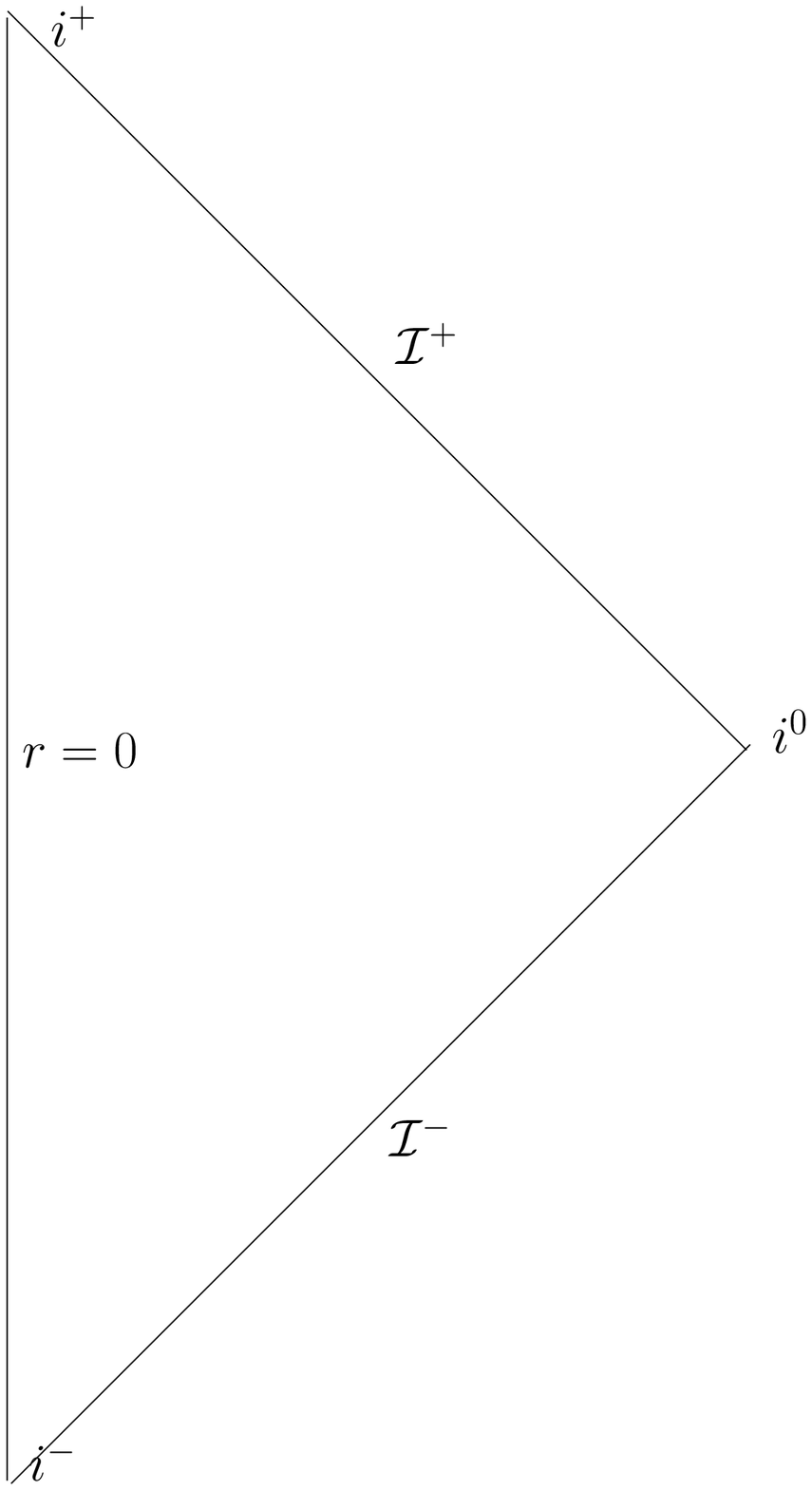}
\caption{Penrose Diagram of Minkowski space.}
\label{minpen}
\end{center}
\end{figure}
Penrose diagrams are useful for understanding the causal structure  of a spacetime. The basic idea is that causal influence is determined by what is inside or outside the lightcone. But to determine the null geodesics of a geometry, we only need to know the metric up to conformal rescalings. So we can do a conformal rescaling to introduce new coordinates such that non-compact coordinate ranges are translated in the new coordinates into compact ranges. If the maximally extended metric can be captured by one coordinate chart, this means that we can depict the entire geometry by a compact region.

The simplest example is flat space. In polar coordinates $t$ and $r$ are non-compact coordinates while $\theta$ and $\phi$ are compact. We will suppress the latter, and in every picture we draw it is understood that there is a sphere at every point that is being suppressed. Clearly this is a trivial suppression only in spherically symmetric spacetimes: in particular, in the case of rotating black holes, the usual Penrose diagram has two {\em non-trivial} suppressed dimensions.

In this two dimensional form, flat space can be written as
\bea
ds^2=du\ dv=\frac{1}{4}\sec^2 (\tilde u/2) \sec^2(\tilde v/2) d \tilde u\ d\tilde v.
\eea
where $u=t-r\equiv\tan(\tilde u/2) \ v=t+r\equiv\tan(\tilde v/2)$, where we have introduced new tilde'd coordinates. In this  form of the metric, one can do a conformal rescaling to get rid of the overall scale factor. It looks like we have ended up with what we started and gained nothing, but this is not so, because now the ranges $\tilde u$ and $\tilde v$  are compact: from $-\pi$ to $\pi$. The original range $r \in (0, \infty)$ and $t \in (-\infty, \infty)$ translates to a triangle in the $(\tilde u, \tilde v)$ plane, shown in figure \ref{minpen}. Really, the figure should have been a square tilted on a vertex, but the fact that incoming rays turn into outgoing rays at $r=0$ means that we only need to consider the $r \in (0, \infty)$ half, which is a triangle.
\begin{figure}
\begin{center}
\includegraphics[
height=0.3\textheight
]{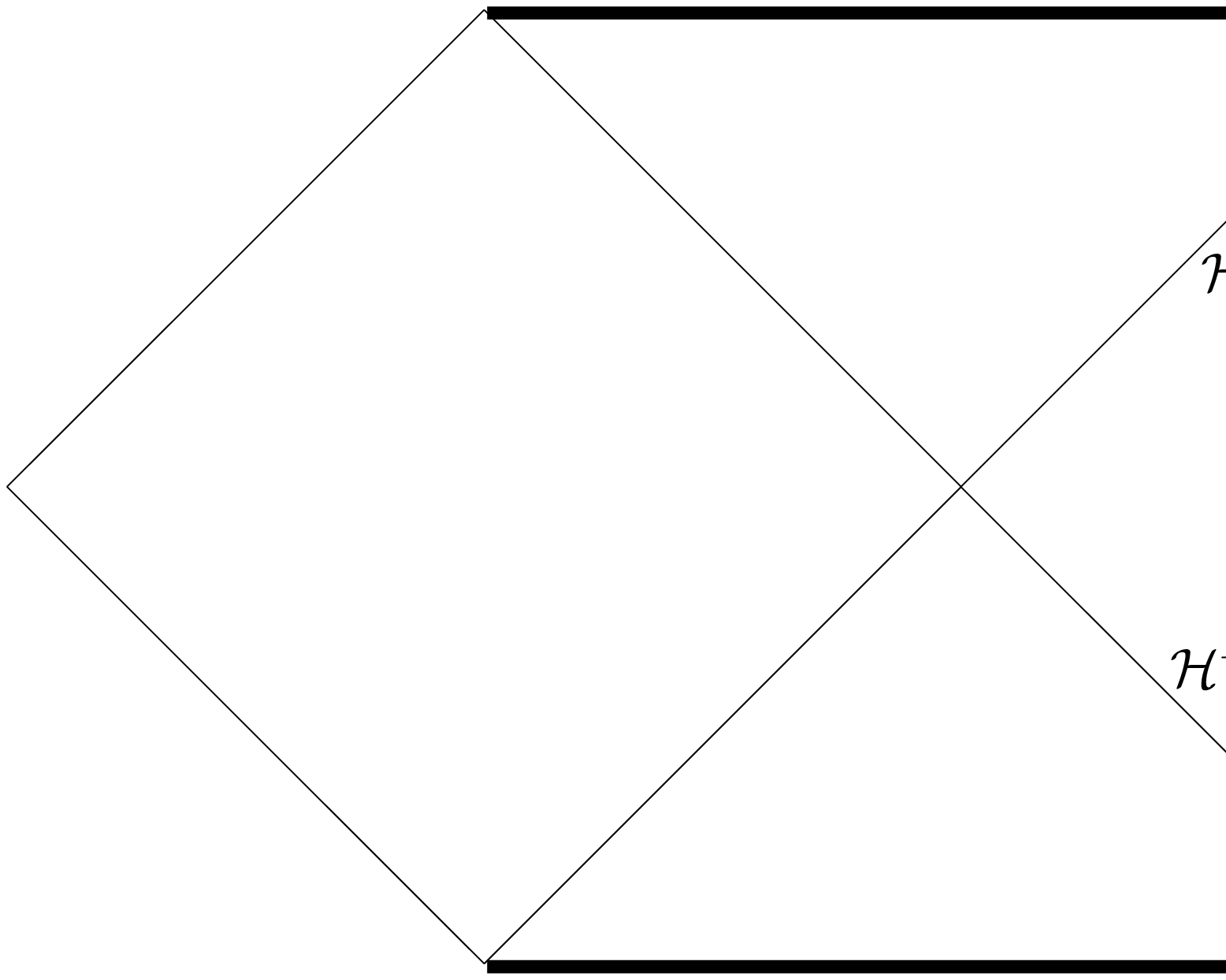}
\caption{Penrose Diagram of Schwarzschild/Kruskal.}
\label{Krupen}
\end{center}
\end{figure}
Note that as in the Kruskal diagram of the previous appendix, we draw the axes for $(\tilde u, \tilde v)$ at 45 degree angles to easily capture the fact that these are light-cone coordinates. The distinguished points on the figure are
\begin{itemize}
\item $i^+$, future timelike infinity. $u=\infty, v=\infty$. On a curve in the geometry that ends here, $t$ increases faster than $r$. Timelike curves end here.
\item $i^-$, past timelike infinity. $u=-\infty, v=-\infty$. On a curve in the geometry that ends here, $t$ decreases faster than $-r$. Timelike curves begin here.
\item ${\cal I}^+$, $v=\infty$, $u=$const. Future null infinity where outgoing null curves end.
\item ${\cal I}^-$, $u=-\infty$, $v=$const. Past null infinity where ingoing null curves begin.
\item $i^0$, spacelike infinity. $u=-\infty, v=\infty$. $r$ increases faster than $t$ for curves that end here.  All (global) Cauchy surfaces in the geometry have an end point here.
\end{itemize}
Note that ingoing curve turns into an outgoing curve at the origin $r=0$.

The discussion above demonstrates that the Penrose diagram reduces the causal structure of the geometry to the fine art of drawing cartoons. Since the structure of the Kruskal metric is identical to that of flat space (when one suppresses the sphere as before) up to a conformal factor, we can use the same transformation as in flat space to compactify the geometry:
\bea
U=\tan(\tilde U/2)\  ,  \ V=\tan(\tilde V/2).
\eea
where the $U$ and $V$ are the Kruskal null coordinates from the last Appendix. Keeping track of the ranges of  the coordinates there, one finds the Penrose diagram shown in the figure. Note that since the radial direction $r$ gets analytically continued beyond $r=0$, there is no boundary at $r=0$ (this is different from the flat space case). So unlike the Minkowski case, here we should have the full square. But because of the future and past singularities, the upper and lower vertices of the square are chopped off. So in effect what we have in Kruskal is a doubled and chopped version of the Minkowski Penrose diagram.

\newpage

%

\end{document}